\documentclass[fleqn,usenatbib]{mnras}

\usepackage{newtxtext,newtxmath}

\usepackage[T1]{fontenc}
\usepackage{ae,aecompl}

\usepackage{graphicx}	
\usepackage{amsmath}	
\usepackage{amssymb}	

\title[THE GALANTE PHOTOMETRIC SYSTEM]{THE GALANTE PHOTOMETRIC SYSTEM}

\author[A. Lorenzo-Guti\'errez et al.]{
\newauthor A. Lorenzo-Guti\'errez,$^{1}$\thanks{E-mail: alorenzo@iaa.es}
E. J. Alfaro$^{1,3}$,
J. Ma\'{\i}z Apell\'aniz$^{2}$,
R. H. Barb\'a$^{4}$,
\newauthor A. Mar\'{\i}n-Franch$^{3,5}$,
A. Ederoclite$^{3,5,6}$,
D. Crist\'obal-Hornillos$^{3,5}$,
J. Varela$^{3,5}$,
\newauthor  H. V\'azquez Rami\'o$^{3,5}$,
J. Cenarro$^{3,5}$,
D. J. Lennon$^{7}$,
and P. Garc\'{\i}a-Lario$^{7}$
\\
$^{1}$Instituto de  Astrof\'{\i}sica de Andaluc\'ia. Glorieta de la Astronom\'{\i}a s/n. E-18\,008 Granada. Spain\\
$^{2}$Centro de Astrobiolog\'{\i}a, CSIC-INTA. Campus ESAC. Camino bajo del castillo s/n. E-28\,692 Villanueva de la Ca\~nada. Spain\\
$^{3}$Unidad Asociada CEFCA-IAA. CSIC. Teruel. Spain\\
$^{4}$Universidad de La Serena. La Serena. Chile\\
$^{5}$Centro de Estudios de F\'{\i}sica del Cosmos de Arag\'on. Teruel. Spain\\
$^{6}$Instituto de Astronomia, Geof\'isica e Ci\^encias Atmosf\'ericas. Universidade de S\~ao Paulo. Brazil\\
$^{7}$European Space Agency. ESAC. Camino bajo del castillo s/n. E-28\,692 Villanueva de la Ca\~nada. Madrid. Spain\\
}

\date{Accepted XXX. Received YYY; in original form ZZZ}

\pubyear{2019}

\begin{document}
\label{firstpage}
\pagerange{\pageref{firstpage}--\pageref{lastpage}}
\maketitle

\begin{abstract}
This paper describes the characterization of the GALANTE photometric system, a seven intermediate- and narrow-band filter system with a wavelength coverage from 3000~\r{A} to 9000~\r{A}. We describe the photometric system presenting the full sensitivity curve as a product of the filter sensitivity, CCD, telescope mirror, and atmospheric transmission curves, as well as some first-  and second-order moments of this sensitivity function. The GALANTE photometric system is composed of four filters from the J-PLUS photometric system, a twelve broad-to-narrow filter system, and three exclusive filters, specifically designed to measure the physical parameters of stars such as effective temperature $T_{\rm eff}$, $\log(g)$, metallicity, colour excess $E(4405-5495)$, and extinction type $R_{5495}$.
Two libraries, the Next Generation Spectral Library (NGSL) and the one presented in \citet{2018A&A...619A.180M}, have been used to determine the transformation equations between the Sloan Digital Sky
Survey (\textit{SDSS}) \textit{ugriz} photometry and the GALANTE photometric system. We will use this transformation to calibrate the zero points of GALANTE images. To this end, a preliminary photometric calibration of GALANTE has been made based on two different \textit{griz} libraries (\textit{SDSS} DR12 and ATLAS All-Sky Stellar Reference Catalog, hereinafter \textit{RefCat2}). A comparison between both zero points is performed leading us to the choice of \textit{RefCat2} as the base catalogue for this calibration, and applied to a field in the Cyg OB2 association.
\end{abstract}




\section{Introduction}
Photometric systems are defined as a set of filters and a detector capable of obtaining information from energy spectral distribution of the objects observed, from which to derive their intrinsic and extrinsic physical
properties. The \textit{UBV} system \citep{1953ApJ...117..313J} based on the spectral response of the human eye ($V$) and the photographic plate ($B$), added a $U$ filter to measure the Balmer jump. This addition made it possible to stablish a first quantitative relationship between photometric colours and spectral properties. The \textit{UBV} system aimed to take advantage of historical catalogues (visible and photographic) to establish a photometric database with the addition of a new filter. Another 3-band system was defined \citep{1946VeGoe...5..159B} with photographic detectors that disregarded the historical catalogues and focused more on the properties of the emitter than on those of the receiver. On paper, the \textit{RGU} system allowed a better stellar classification than the \textit{UBV}, but the fact that it was based on the photographic plate, with a non-linear response, and depended on transformation equations between the \textit{RGU} and the \textit{UBV} greatly limited its development and nowadays it is rarely used. \citet{1960BAN....15...67W} introduced a photometric system with five bands \textit{VBLUW} allowing to measure temperatures for early-type stars. The \textit{W} and \textit{L} bands in the ultraviolet region enabled to estimate the Balmer series in hot stars, in addition with the Johnson's bands. \citet{1966ARA&A...4..433S} developed a new approach to the photometric system based on narrower filters that not only provided information from the stellar continuum, but also allowed a quantitative measurement of some spectral lines. This fact enabled the possibility of estimating parameters such as metallicity ([Fe/H]) for the most common range of temperature we can find in the solar neighborhood. DDO system \citep{1968AJ.....73..313M} is another intermediate five-bands system designed to measure temperature and metallicity in late-type stars with a high precision in comparison with previous photometric systems. Recently, Gaia DR2 has provided a whole-sky optical photometric survey using three very broad bands $G$, $G_{\rm BP}$, and $G_{\rm RP}$ \citep{2018A&A...616A...1G,2018A&A...616A...4E}. The coverage, uniformity, and dynamic range of Gaia photometry will undoubtedly be a golden standard for optical photometry in the future. More specifically, \citet{2018A&A...619A.180M} (hereinafter MAW) have shown that Gaia DR2 photometry can be calibrated without significant systematic biases and with  photometric residuals of one hundredth of a magnitude or better. Since the introduction of the CCD, the development of new photometric systems has halted, reaching its peak with the Sloan Digital Sky Survey (\textit{SDSS}) \citep{1996AJ....111.1748F}, which combined the design of a photometric system in five bands, covering the visible range, with an industrial aspect of the observation. Observational strategy in astronomy has undergone a drastic change and the number of large photometric surveys specifically designed to obtain singular information about types of celestial objects has increased significantly.

Examples of this kind of astronomical projects are J-PLUS and J-PAS which are currently being developed in a new observatory in Javalambre (Teruel, Spain) with 2 telescopes of 80 cm and 250 cm. The main scientific objective of J-PAS is cosmological, trying to measure the spectrum of Baryonic Acoustic Oscillations (BAOs) with photometric redshift estimations \citep{2014arXiv1403.5237B}, while J-PLUS was originally planned to be an auxiliary survey for J-PAS calibration purposes \citep{A.-J.-Cenarro:2018aa}. Given the observational restrictions of these programmes in terms of sky darkness and seeing, we designed a new monitoring programme, which we called GALANTE, focused on the brighter stars of the Galactic disk. This programme would be carried out on those clear nights where J-PLUS were not observable.

The GALANTE project is a photometric survey that will cover the Northern Galactic plane defined by $\delta \geq $ $0^{o}$ and |b| $\leq$ $3^{o}$. This project uses 7 intermediate+narrow band filters aimed at measuring
all stars of that region of the sky with AB magnitudes $\leq $ 17. These 7 filters are 4 J-PLUS filters (F348M, F515N, F660N, and F861M) and 3 new customized filters developed by the GALANTE team (F420N, F450N, and
F665N). The number, width, and effective wavelength of the filters compose an optimal system to accomplish those objectives proposed by \citet{2017edrs.confE..15M}. A Southern version of the GALANTE program has been proposed to be performed with the twin Cerro Tololo 80 cm telescope developed for the S-PLUS project \citep{2018IAUS..334..358S}.

The main goal of this paper is to present the characterization of the 7 GALANTE filters in the optical range, and to stablish the first transformation equations between GALANTE and \textit{SDSS}. The GALANTE project covers
from 3000~\r{A} to 9000~\r{A} using its 7 filters located at the best place to attain our goal: to obtain more precise information about stellar effective temperatures, gravities, metallicities, and also type of
extinction \citep{2013hsa7.conf..657M}.

In Section~\ref{section_phot_syst}, we present the characterization and definition of the GALANTE photometric system. In Section~\ref{section_standards}, we analyse in detail the set of GALANTE standard stars.
Transformation equations between the \textit{SDSS} and the GALANTE photometric system are obtained in Section~\ref{section_trans_eqs}, setting up the comparison between the synthetic magnitudes of 378 stars from the
Next Generation Spectral Library (NGSL) plus 122 stars from a new catalogue from MAW in both photometric systems. Next, in Section~\ref{calibration_using_SDSS} we estimate zero points from the GALANTE photometry using the \textit{SDSS} and \textit{RefCat2}. Finally, in Section~\ref{section_conclusions} we summarize the results of this paper.

\section{THE GALANTE PHOTOMETRIC SYSTEM}
\label{section_phot_syst}

\subsection{DESCRIPTION}

Deriving physical stellar (or galaxy) parameters is the main goal of every photometric system. In our case, we achieve this aim with a new filter set designed to plug some gaps found in previous surveys. The GALANTE
photometric system has been defined using 4 J-PLUS filters \citep{2017hsa9.conf...11C} and 3 purpose-built filters. This composition of intermediate and narrow band filters covers the optical range from
3000~\r{A} to 9000 \r{A}. This filter set has been selected with a direct goal: deriving stellar effective temperatures for hot stars in an
optimal way \citep{MaizSota08,2014A&A...564A..63M,MaizBarb18}. We will also be able to obtain the gravity and metallicity (the latter for stars cooler than 10\,000~K), allowing us to discriminate between giants, supergiants, dwarfs, and between solar or SMC metallicities.

\begin{figure*}
\includegraphics[width=1.0\textwidth]{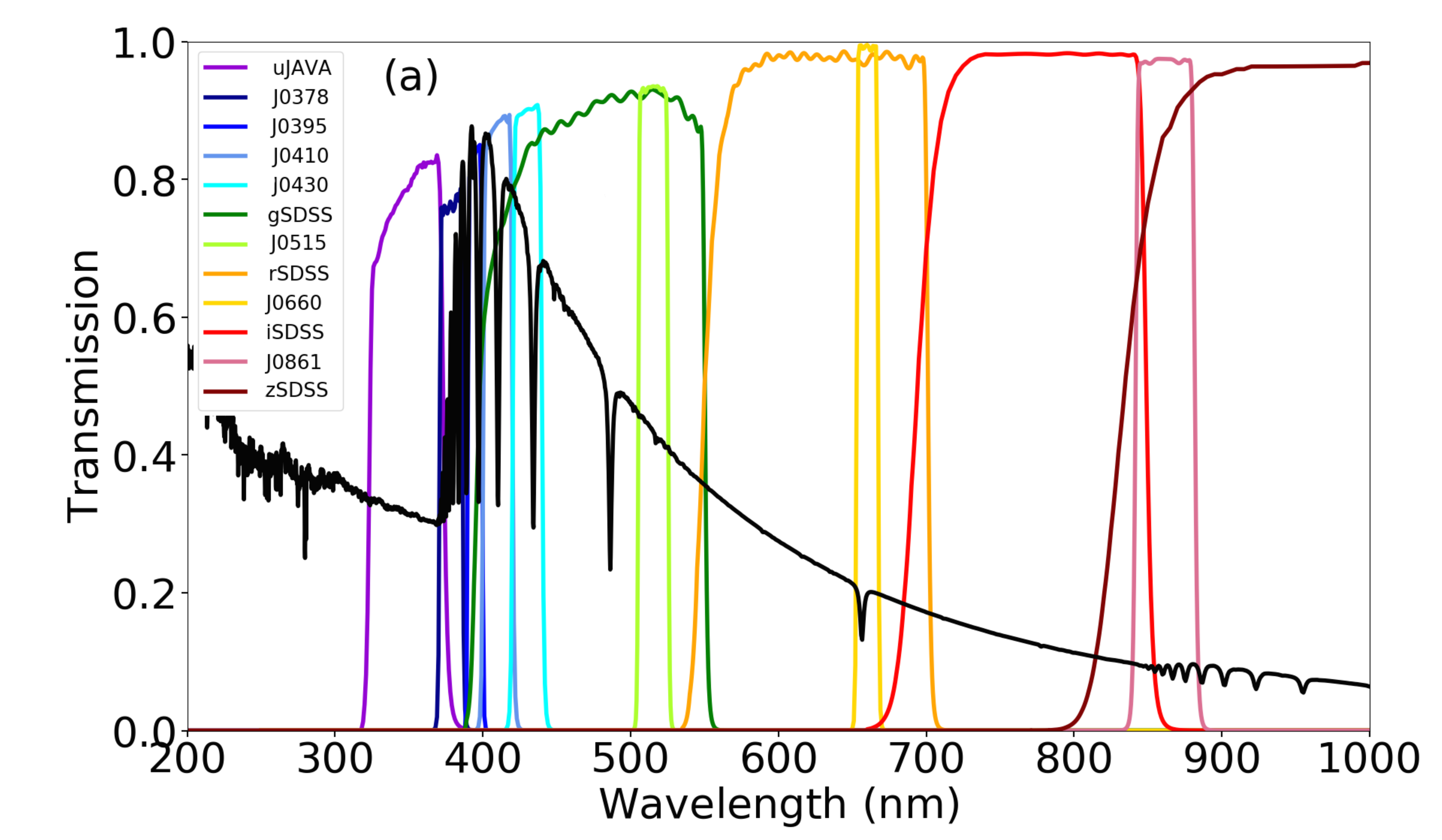}
\includegraphics[width=1.0\textwidth]{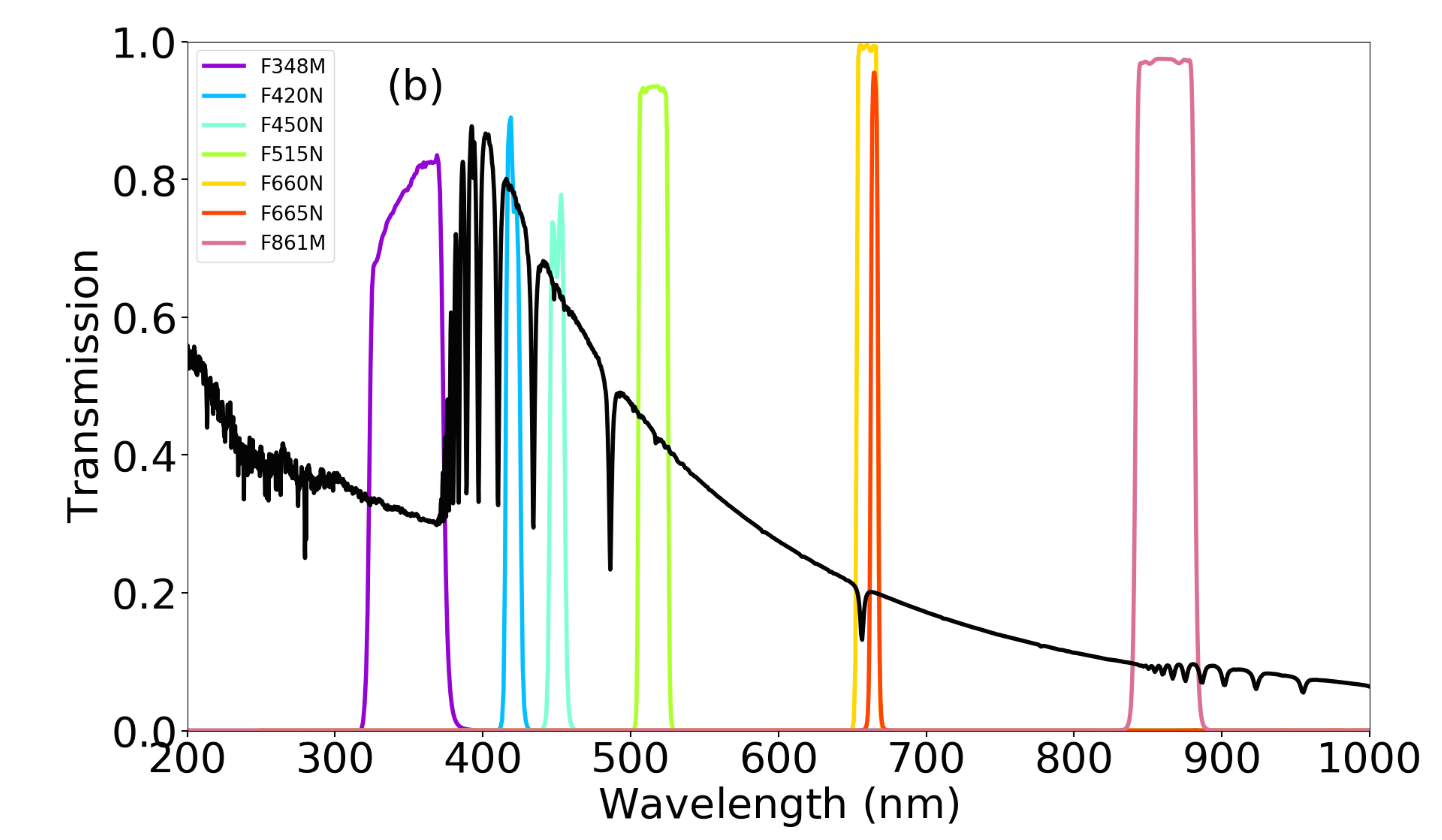}
\centering
\caption{Response functions of the J-PLUS and the GALANTE photometric system filters. In both pictures we only took into account filter transmission curves to visualizate their positions. Figure (a) represents the Vega spectrum superimposed onto J-PLUS filters transmission curves. Figure (b) represents the Vega spectrum superimposed onto GALANTE filters transmission curves. To make the graphic possible, the flux of Vega has been normalized and scaled properly.}
\label{fig:filtros_JPLUS_GALANTE_VEGA}
\end{figure*}

\begin{figure*}
\includegraphics[width=1.0\textwidth]{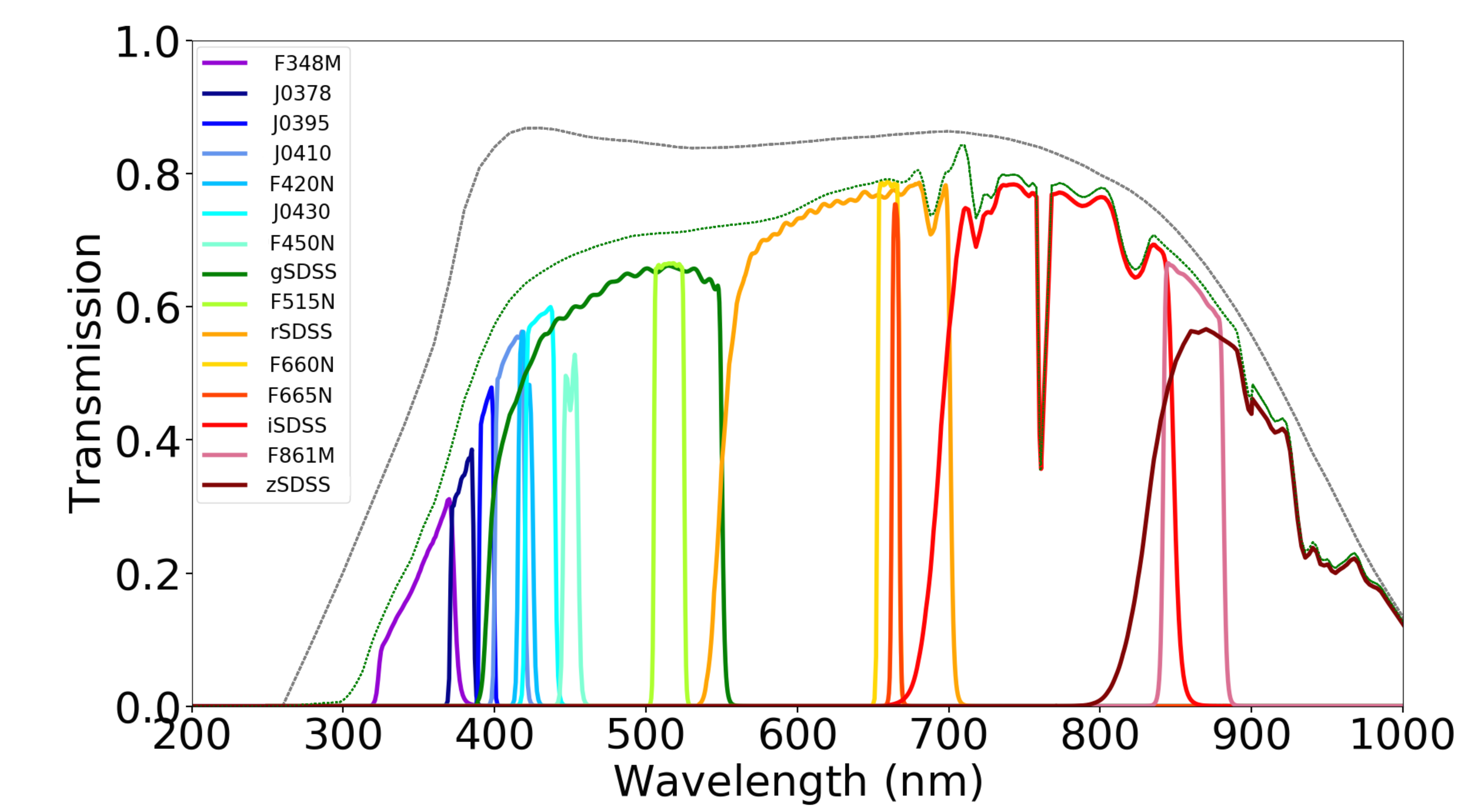}
\centering
\caption{Response functions of the J-PLUS and the GALANTE photometric system filters. In discontinuous grey we represent the response curve of the CCD and mirror, while discontinuous green is the CCD, mirror and atmospheric transmission at 1.3 air mass curve. The \textit{SDSS} response curves are represented by thick colored lines with the other continuous curves are the GALANTE and J-PLUS filter response curves convolved with CCD, mirror and atmospheric transmission at 1.3 air mass.}
\label{fig:filtros_JPLUS_GALANTE_SDSS}
\end{figure*}

\begin{figure}
\includegraphics[scale=1.0]{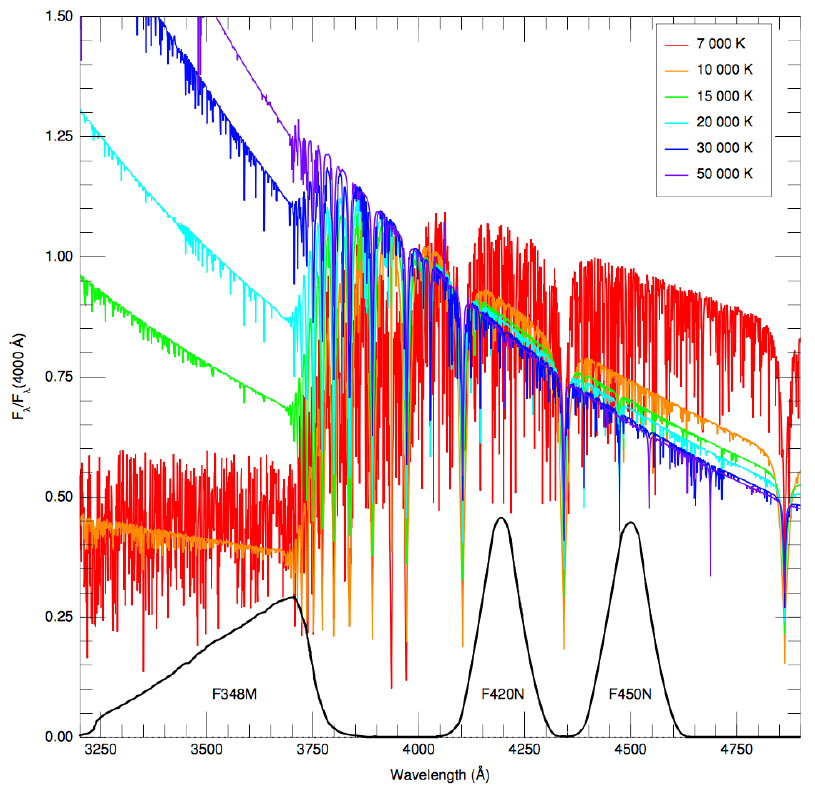}
\centering
\caption{SEDs for main-sequence hot stars normalized to the flux at 4000~\r{A} and sensitivity curves of the three bluemost GALANTE filters. F420N and F450N have been approximated by gaussians while F348M has been represented by a rectangular filter with a linear atmospheric absorption effect. Note how F348M measures the continuum to the left of the Balmer jump while F420N and F450N are located at the gaps between the Balmer lines.}
\label{fig:filtros_galante}
\end{figure}

\begin{figure}
\includegraphics[width=0.48\textwidth]{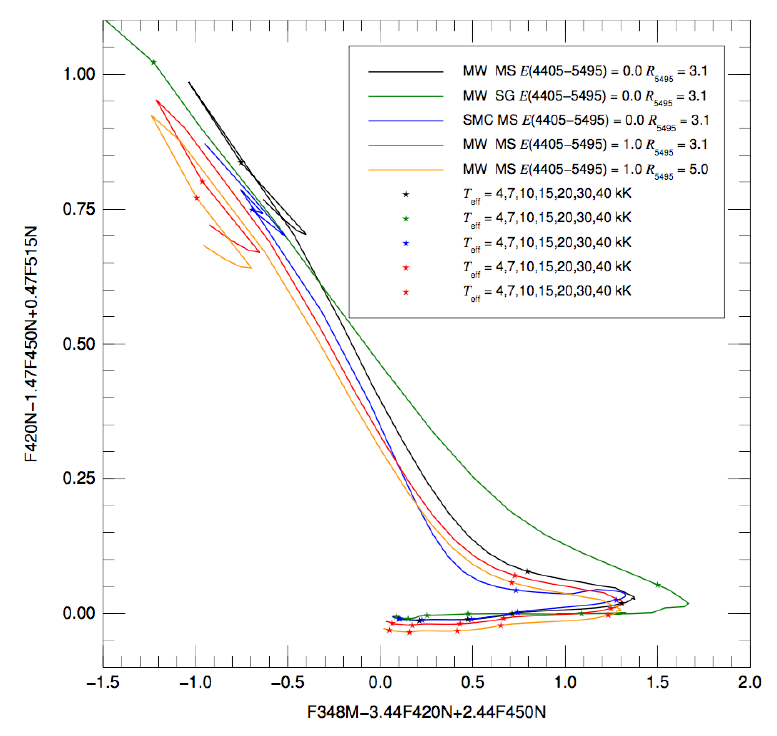}
\centering
\caption{Bracket-like diagram using the GALANTE filter set. It shows how, for a fixed metallicity and extinction law, it is possible to classify the stars by $T_{\rm eff}$ independently of reddening.}
\label{fig:color_color}
\end{figure}

In Figures~\ref{fig:filtros_JPLUS_GALANTE_VEGA} and~\ref{fig:filtros_JPLUS_GALANTE_SDSS} we present the setup of the filters.
Figures~\ref{fig:filtros_JPLUS_GALANTE_VEGA} (a) and (b) show the layout of the J-PLUS and GALANTE filters respectively, leaning on the Vega
spectrum. Note that in this case we only plot the transmission curves of the filters, allowing to the reader a visual location of each one.
Focusing on GALANTE, Figure~\ref{fig:filtros_JPLUS_GALANTE_VEGA} (b) plots the configuration of the selected GALANTE filters, which enable us
to measure the key zones in the spectra to derive physical parameters. In Figure~\ref{fig:filtros_JPLUS_GALANTE_SDSS} we plot the response
functions of the J-PLUS and the GALANTE photometric system filters. In discontinuous grey we represent the response curve of the CCD and
mirror, while discontinuous green is the CCD, mirror and atmospheric transmission at 1.3 air mass curve. The \textit{SDSS} response curves
are represented in continuous black, and the other continuous curves are the GALANTE and J-PLUS filter response curves convolved with CCD,
mirror and atmospheric transmission at 1.3 air mass. Using Figure~\ref{fig:filtros_galante}, we can also explain the function of each GALANTE
filter with greater accuracy. Filter F348M is a \textit{u}-like filter dispose to measure the continuum from the left of the Balmer jump. We
can combine this one with F420N and F450N to measure the Balmer jump on both sides and derive the $T_{\rm eff}$ of stars. Both filters of our design, F420N and F450N, are the most original (compared to other large-scale recent photometric surveys) of the setup. They have been created to fill the gaps between \textit{H$\delta$} and
\textit{H$\gamma$} and between \textit{H$\gamma$} and \textit{H$\beta$} respectively. These are the bluemost wide regions of the
spectrum to the right of the Balmer jump without absorption lines and they provide a measurement of the blue continuum. Another J-PLUS filter
used in GALANTE is F515N. It is a Str\"omgren $y$-like filter positioned in a region free of lines. It can be seen as a V filter for this survey. The next
pair of filters, namely F660N (from J-PLUS) and F665N (own-design), allow us to measure the red continuum, estimate the gravity of hot stars, and flag objects with H$\alpha$ emission. These are two narrow filters centered on \textit{H$\alpha$}. Specifically, F660N includes the line and F665N just the
continuum. The last filter is F861M (from J-PLUS). This is an intermediate filter in the Calcium triplet. It will be used as a detection
filter, to obtain the maximum number of stars. We can also use a combination of some GALANTE filters, for a fixed metallicity and extinction
law,  to obtain effective temperature $T_{\rm eff}$ independently of reddening. Figure~\ref{fig:color_color} shows a GALANTE colour-colour
diagram using F348M, F420N, F450N, and F515N for stars from 4000 K to 40\,000 K with different colour excess $E(4405-5495)$ and extinction type $R_{5495}$.

\subsection{CHARACTERIZATION}

With our GALANTE configuration, we can describe the system using a response curve defined by the three different transmission curves. Figure~\ref{fig:filtros_JPLUS_GALANTE_VEGA} shows the graphic representation of the optical GALANTE+J-PLUS photometric systems. In order to characterize the GALANTE filter set, we need to describe the filter with some quantitative parameters such us: their isophotal wavelengths, wavelength-weighted average, frequency-weighted average, effective wavelength, root mean square, effective band width, and flux sensitivity. Thus, we define the total response of a photometric system (S$_{\lambda}$) by

\begin{equation}
S_{\lambda}=T_{t}(\lambda) T_{f}(\lambda) T_{a}(\lambda)
\label{transmision}
\end{equation}
Here \textit{T$_{t}$} is the mirror+detector throughput; the filter response transmission is \textit{T$_{f}$}; and the atmospheric curve transmission at 1.3 airmass is \textit{T$_{a}$}.

If the SED received at the the detector, \textit{E$_{\lambda}$}, is continuous, and the total response \textit{S$_{\lambda}$} is also continuous and not negative over a range of wavelengths \citep{1974ASSL...41.....G}, using Equation~\ref{transmision} and the mean value theorem, we can say that \textit{$\lambda_{i}$} exists and has the form of:

\begin{equation}
E_{{\lambda}_i}\int_{\lambda_{a}}^{\lambda_{b}} S_{\lambda} d\lambda = \int_{\lambda_{a}}^{\lambda_{b}}E_{\lambda}S_{\lambda} d\lambda
\label{energia}
\end{equation}

If we rearrange this equation:

\begin{equation}
E_{{\lambda}_i}=\langle{E_{\lambda}}\rangle=\frac{\int_{\lambda_{a}}^{\lambda_{b}}E_{\lambda}S_{\lambda} d\lambda}{\int_{\lambda_{a}}^{\lambda_{b}} S_{\lambda} d\lambda}
\label{energia_promedio}
\end{equation}
where \textit{$\lambda_{i}$} is the \textit{isophotal wavelength} and $\langle{E_{\lambda}}\rangle$ is the
mean value of the intrinsic flux above the atmosphere \citep[see also][]{2005PASP..117.1459T}. We show the isophotal
wavelengths for Vega using GALANTE+J-PLUS filters in Table~\ref{tab:parametros_GALANTE}. These values were
obtained from Equation~\ref{energia_promedio} using the Vega spectrum provided by \citet{2007ASPC..364..315B} and available at ftp://ftp.stsci.edu/cdbs/calspec/alpha\_lyr\_stis\_003.fits, for the optical range.

The isophotal wavelength depends on the Spectral Energy Distribution (SED); thus, for each kind of star it will be different for the same filter. Then we need to define several photometric parameters, as shown in Table~\ref{tab:parametros_fotometricos}, depending only on the photometric system.

{
\renewcommand{\arraystretch}{2.5}
\begin{table*}
\caption{Representative Photometric Parameters.}
\label{example}
\begin{center}
\begin{tabular}{clc}
\hline\hline\noalign{\smallskip}
Parameter & Description & Formula  \\
\noalign{\smallskip}
\hline
\noalign{\smallskip}
$\lambda_{m}$  &  Wavelength-weighted average	& $\int\frac{{\lambda} S_{\lambda} d\lambda}{S_{\lambda} d\lambda}$	 \\
$\nu_{m}$  &  Frequency-weighted average	& $\int\frac{{\nu} S_{\nu} d(ln\nu)}{S_{\nu} d(ln\nu)}$	 \\
$\lambda_{\rm eff}$  &  Effective wavelength	& $exp \left[ \frac{\int d(ln\nu) S_{\nu} ln{\lambda}}{\int d(ln\nu) S_{\nu}} \right]$	 \\
$\sigma$  &  Root Mean Square (rms) of the filters	& $\sqrt{ \frac{\int d(ln \nu) S_{\nu}\left[ln \left(\frac{\lambda}{\lambda_{\rm eff}} \right) \right]^2}{\int d(ln \nu) S_{\nu}} }$	 \\
$\delta$  &  Effective band width (with n=0) & $2 (2 ln 2)^{\frac{1}{2}} \sigma \lambda_{\rm eff}$	 \\
$Q$  &  Flux sensitivity & $\int d(ln \nu) S_{\nu}$	 \\
\noalign{\smallskip}
\hline
\end{tabular}
\end{center}
\label{tab:parametros_fotometricos}
\end{table*}
\normalsize
}

We calculated all those parameters and they are collected in Table~\ref{tab:parametros_GALANTE} and Table~\ref{tab:parametros_otros} in order to give a precise description of the GALANTE and J-PLUS filters. Note that the GALANTE photometric system is composed of F348M, F420N, F450N, F515N, F660N, F665N, and F861M. In this way, we put all the 7 GALANTE filters together in Table~\ref{tab:parametros_GALANTE}, adopting a different nomenclature, but some of them are the same filters used in J-PLUS. The first column shows their GALANTE names, while the second one shows the J-PLUS nomenclature.

\tiny
\begin{table*}
\caption{Representative Parameters of the filters of the GALANTE Photometric System.}
\label{example}
\begin{center}
\begin{tabular}{ccccccccccc}
\hline\hline\noalign{\smallskip}
\multicolumn{2}{c}{Filter} & m$_{AB}$(Vega) & $E_{{\lambda}_i}$(Vega) & $\lambda_{\rm iso}$ & $\lambda_{\rm m}$ & c$\nu_{m}^{-1}$ &
 $\lambda_{\rm eff}$ & $\sigma$ & $\delta$ &  $Q$ \\
GALANTE & J-PLUS & (mag)  & (erg s$^{-1}$ cm$^{-2}$ {{\AA}}$^{-1}$) & (nm)   & (nm)              & (nm)            & (nm)            &          &       &          \\
\noalign{\smallskip}
\hline
\noalign{\smallskip}
F348M & uJAVA &  1.071	& 3.215e-09	&  354.6   & 354.8 & 348.5 & 355.1 & 0.0422 &  35.3 &  0.0286 \\
F420N &       &  $-$0.238	& 8.626e-09	& 420.5 & 421.5 & 421.6 & 421.9 & 0.0257 & 25.6  &  0.0125 \\
F450N &       &  $-$0.180	& 8.502e-09	& 450.5 & 451.0 & 451.0 & 451.2 & 0.0185 & 19.7  &  0.0116 \\
F515N & J0515 &  $-$0.049	& 4.301e-09	& 515.2 & 515.4 & 515.0 & 515.5 & 0.0115 & 13.9  &  0.0260 \\
F660N & J0660 &  0.318	& 1.981e-09	&  660.4   & 660.1 & 660.0 & 660.1 & 0.0066 & 10.3  &  0.0175 \\
F665N &       & 0.238 	& 1.979e-09	&  664.7   & 665.2 & 665.2 & 665.2 & 0.0174 & 27.2  &  0.0061 \\
F861M & J0861 & 0.560 	& 8.185e-10	&  861.1   & 861.2 & 861.0 & 861.2 & 0.0137 & 27.7  & 0.0296  \\
\noalign{\smallskip}
\hline
\end{tabular}
\end{center}
\label{tab:parametros_GALANTE}
\end{table*}


\begin{table*}
\caption{Representative Parameters of the additional J-PLUS filters.}
\label{example}
\begin{center}
\begin{tabular}{cccccccccc}
\hline\hline\noalign{\smallskip}
Filter & m$_{AB}$(Vega) & $E_{{\lambda}_i}$(Vega)  & $\lambda_{\rm iso}$  & $\lambda_{\rm m}$ & c$\nu_{m}^{-1}$ & $\lambda_{\rm eff}$ & $\sigma$ &  $\delta$ & $Q$ \\
       & (mag)  	             & (erg s$^{-1}$ cm$^{-2}$ {{\AA}}$^{-1}$) & (nm)   & (nm)              & (nm)            & (nm)            &          &     &          \\
\noalign{\smallskip}
\hline
\noalign{\smallskip}
J0378  &  0.419	& 6.164e-09	&  380.1   & 379.0 & 378.5 & 379.1 & 0.0145 & 13.0  &  0.0142 \\
J0395  &  $-$0.027	& 8.052e-09	& 394.8 & 395.1 & 395.0 & 395.1 & 0.0102 & 9.5  &  0.0115  \\
J0410  &  $-$0.165	& 5.472e-09	& 410.2 & 410.4 & 410.0 & 410.5 & 0.0144 & 14.0  & 0.0261  \\
J0430  &  $-$0.125	& 7.067e-09	& 429.7 & 430.4 & 430.0 & 430.4 & 0.0154 & 15.6  &  0.0274 \\
\textit{gSDSS}  &  $-$0.090	& 5.352e-09	& 468.7 & 478.6 & 480.3 & 477.0 & 0.0880 & 105.3  & 0.1108 \\
\textit{rSDSS}  &  0.163	& 2.497e-09	&  618.0   & 627.6 & 625.4 & 622.2 & 0.0652 & 104.6  & 0.1015 \\
\textit{iSDSS}  &  0.404	& 1.301e-09	&  761.1   & 769.7 & 766.8 & 763.2 & 0.0592 & 108.4  & 0.0766 \\
\textit{zSDSS}  &  0.536	& 9.015e-10	&  891.7   & 896.9 & 911.4 & 904.9 & 0.0586 & 119.8  & 0.0354 \\
\noalign{\smallskip}
\hline
\end{tabular}
\end{center}
\label{tab:parametros_otros}
\end{table*}
\normalsize

\section{STANDARD STARS SYSTEM}
\label{section_standards}

To obtain the GALANTE transformation equations, we use two observational catalogues: NGSL \citep{2006hstc.conf..209G} and MAW \citep{2018A&A...619A.180M}. The NGSL library comprises of 378 high
signal-to-noise stellar spectra. This catalogue has all these stars with a good spectral resolution, all of
them homogeneously flux calibrated, covering a large range of spectral types, gravities, and metallicities
(3100 K $\leq$ $T_{\rm eff}$ $\leq$ 32500 K, 0.45 $\leq$ $\log(g)$ $\leq$ 5.4, -2.0 $\leq$ [Fe/H]
$\leq$ 0.5 and  a range of E(B-V) from 0 to 0.75). The wavelength coverage of these spectra is
from 2000~\r{A} to 10\,200~\r{A} at resolution R$\sim$1000.

We also take advantage of the new MAW library of 122 objects observed with HST/STIS spectrophotometry. This selection provides us a library of hot stars with significant extintion plus three M
dwarfs, thus covering a wide range of colours that can be defined by their intrinsic colour or their reddening.

These two catalogues complement each other in the sense of covering the whole range of stellar temperatures observed in an ample grade of extinction, in such a way that the same colour, for example (\textit{g}-\textit{r}), could represent an unreddened intermediate-type star or a very reddened early-type object. This complementarity will be crucial when estimating the transformation equations between GALANTE and \textit{SDSS} systems.

The AB magnitude system defined by \citep{1983ApJ...266..713O} has been chosen to set the magnitudes in the GALANTE photometric system.

\begin{equation}
AB_{\nu}= -2.5 \cdot \log f_{\nu} - 48.60
\label{AB}
\end{equation}
where \textit{f$_{\nu}$} is the flux per unit frequency of an object in erg cm$^{-2}$ s$^{-1}$ Hz$^{-1}$. The constant is extrated setting AB magnitude equal to V magnitude of the Vega flux:

\begin{equation}
48.60 = -2.5 \cdot \log F_{0}
\label{constant}
\end{equation}
being F$_{0}$ = 3.65 x 10$^{-20}$ erg cm$^{-2}$ s$^{-1}$ Hz$^{-1}$ the flux of Vega at $\lambda$ = 5480 \r{A} used by those authors. Thus we can write the AB$_{\nu}$ magnitudes by:

\begin{equation}
AB_{\nu}= -2.5 \log \frac{\int f_{\nu}S_{\nu}d(\log \nu)}{\int S_{\nu}d(\log \nu)} - 48.60
\label{AB_final}
\end{equation}
where \textit{S$_{\nu}$} is the total response of the atmosphere, filter, detector, and mirror transmission. Another way to construct this magnitude is using the \textit{m$_{ST}$} system to derive a \textit{m$_{AB}$} equation as a wavelength function. We adopted the formulation proposed by \citet{2014MNRAS.444..392C}:

\begin{equation}
m_{AB}=-2.5 \cdot \log\dfrac{\int_{\lambda_{i}}^{\lambda_{f}} \lambda f_\lambda S_\lambda d \lambda }{F_0 \cdot c \int_{\lambda_{i}}^{\lambda_{f}}\frac{S_\lambda}{\lambda} d \lambda}
\label{AB_final_wavelength}
\end{equation}
where \textit{c} is the speed of light in \r{A} s$^{-1}$ and \textit{f$_{\lambda}$} being the flux per unit of wavelength in erg cm$^{-2}$ s$^{-1}$ \r{A}$^{-1}$. Equation~\ref{AB_final_wavelength} is used to obtain AB synthetic magnitudes.

The choice of a spectrophotometric library to perform the calibration of a new photometric system is still a topic open to discussion \citep[i.e.][]{2010AJ....139.1242A,2011PASP..123.1442B,2012A&A...538A.143K,2018A&A...619A.180M,2018A&A...617A.138W}. \citet{2018A&A...619A.180M} identify differences of zero-point (ZP) between the MAW and NGSL stellar libraries that can reach up to 0.05 magnitudes for some objects and that on average present an rms of 0.03 magnitudes. However, these differences do not seem to depend on the colour of the stars \citep{2018A&A...617A.138W}. For the first calibration of GALANTE photometry we will use both libraries although there could be spurious differences of up to 0.05 mag between the two catalogues.

\section{GALANTE-SDSS TRANSFORMATION EQUATIONS}
\label{section_trans_eqs}

Since the GALANTE project is an optical photometric survey in the optical range, we can think of another useful and well known optical survey to transform GALANTE AB magnitudes. This survey is the \textit{SDSS} \citep{1996AJ....111.1748F,2002AJ....123.2121S}, based on an optical photometric system composed of five bands (\textit{ugriz}) in the range from 3000~\r{A} to 11\,000~\r{A}. The GALANTE and \textit{SDSS} photometric systems share the same otical wavelength range, thus we will use the NGSL and MAW catalogues to derive transformation equations between both systems.

It is worth noting that both photometric systems are different in several aspects: number of bands and filter bandwidth, as we show in Figure~\ref{fig:filtros_JPLUS_GALANTE_VEGA}. While \textit{SDSS} uses five wide-band filters, GALANTE applies a mix of seven narrow and intermediate-band filter set.

We obtain synthetic photometric GALANTE and \textit{SDSS} AB magnitudes for NGSL and MAW libraries using the response curves shown in Figure~\ref{fig:filtros_JPLUS_GALANTE_SDSS}.

\begin{figure*}
\hspace{1.0cm}
\includegraphics[scale=0.40]{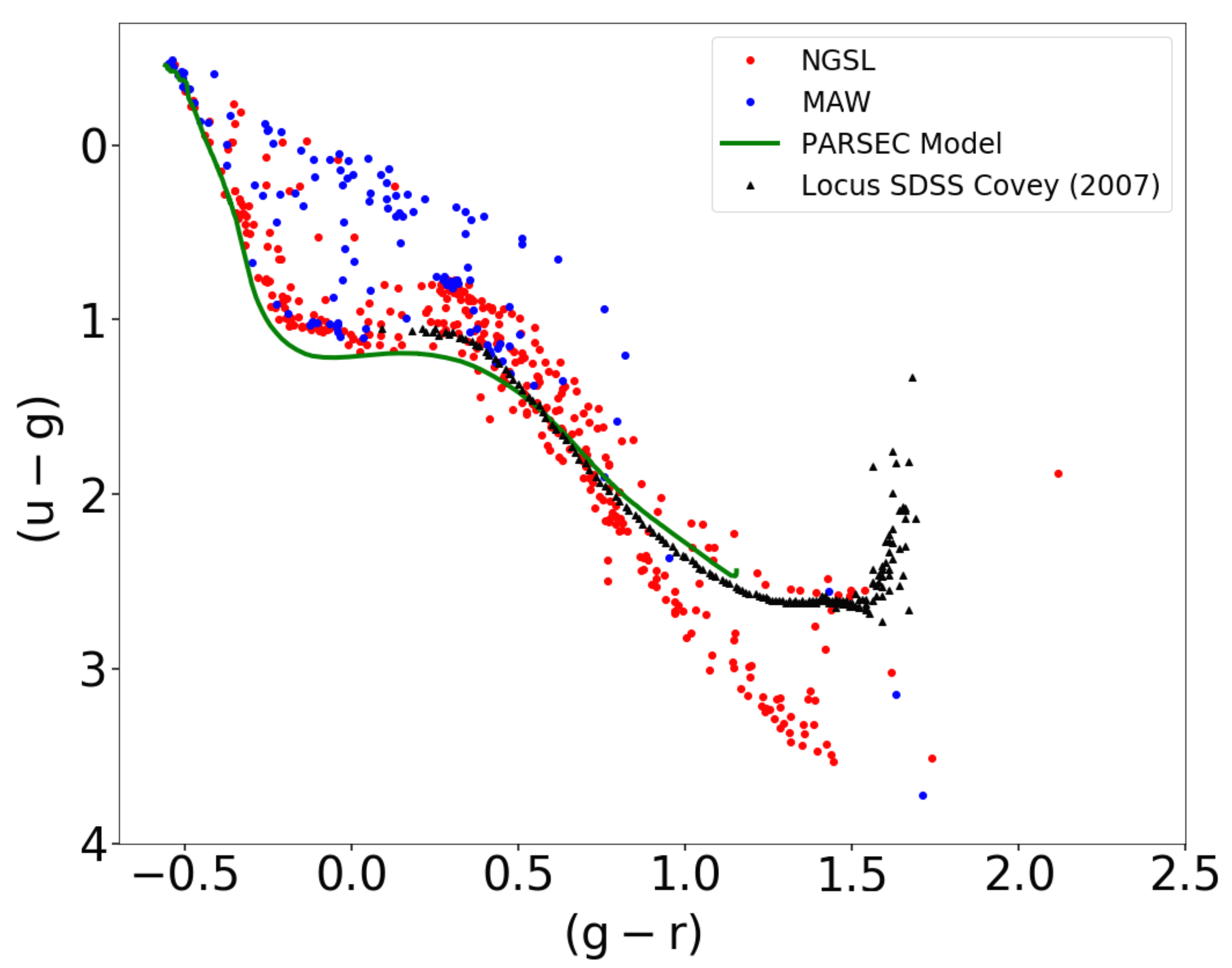}
\centering
\caption{Colour-colour diagram using \textit{SDSS} synthetic photometry for both catalogues. NGSL stars are red dots while MAW stars are shown as blue dots. We draw a 1 Myr theoretical PARSEC curve with solar metallicity and no extinction. Black triangles represent main stellar locus in \textit{SDSS} photometry from \citet{2007AJ....134.2398C}.}
\label{fig:H-R_diagram}
\end{figure*}

Figure~\ref{fig:H-R_diagram} shows a colour-colour diagram for both catalogues. NGSL stars are represented by red dots while MAW stars are shown in blue dots. We also draw a 1 Myr theoretical PARSEC curve with solar metallicity without extinction from http://stev.oapd.inaf.it/cmd, isochrones PARSEC release v1.2S + COLIBRI release PR16 \citep{2017ApJ...835...77M}. Black triangles represent main locus in \textit{SDSS} photometry from \citet{2007AJ....134.2398C}. Looking at both catalogues in this figure one can see that the MAW library covers a good number of reddened high-temperature stars, showing a sparse distribution for dwarf late-type and giant stars, whereas the NGSL library covers a wider range of temperature and gravity but is limited in reddening. Lastly, in order to derive GALANTE transformation equations that are as general as possible, we use both catalogues as a single one to better probe the diagram of Figure~\ref{fig:H-R_diagram}.

We have considered the possibility of modeling these transformations through multilinear fitting with the 4 independent \textit{SDSS} colours (\textit{u-g}, \textit{g-r}, \textit{r-i}, and \textit{i-z}). However, a multilinear analysis where the independent variables are not really stochastically independent can introduce significant biases. For this reason, we first analysed the covariance matrix of both catalogues, which are shown below (Equation~\ref{NGSL_corr} and~\ref{MWEILER_corr}), where we see a high correlation between these 4 colours. Therefore, we have decided to use only a linear fitting to a colour. Equation~\ref{linear_fitting} shows the general case, where \textit{i} indicates the GALANTE bands, \textit{k} the \textit{SDSS} band, and \textit{j} specifies the \textit{SDSS} colour used. To select the best fit, we have considered all the possible combinations of \textit{SDSS} colours, choosing the one that shows a lower BIC parameter.

\begin{equation}
NGSL_{corr} = \begin{pmatrix} 1.000 & 0.918 & 0.772 & 0.797\\ 0.918 & 1.000 & 0.893 & 0.908\\ 0.772 & 0.893 & 1.000 & 0.991\\ 0.797 & 0.908 & 0.991 & 1.000\end{pmatrix}
\label{NGSL_corr}
\end{equation}

\begin{equation}
MAW_{corr} = \begin{pmatrix} 1.000 & 0.842 & 0.810 & 0.826\\ 0.842 & 1.000 & 0.887 & 0.931\\ 0.810 & 0.887 & 1.000 & 0.983\\ 0.826 & 0.931 & 0.983 & 1.000\end{pmatrix}
\label{MWEILER_corr}
\end{equation}

\begin{equation}
Gal_i - \textit{SDSS}_k =  c_{ijk} \cdot (\textit{SDSS}_{j} - \textit{SDSS}_{j+1}) + d_{ijk}
\label{linear_fitting}
\end{equation}

We obtain the transformation equations between the \textit{SDSS} and the GALANTE photometric systems (and vice versa) using the following procedure. To do this, we use AB GALANTE synthetic magnitudes as dependent and the \textit{SDSS} synthetic magnitudes as independent variables. We obtain magnitudes in the GALANTE photometric system from \textit{SDSS} using the statistical package \textsl{Statsmodels} in \textbf{Python} for this linear fitting. This algorithm provides a goodness of fit parameter called Bayesian Information Criterion (BIC), defined as

\begin{equation}
BIC = -2 \cdot ln\textit{L} + \textit{f} \cdot ln(\textit{n})
\label{BIC}
\end{equation}
where \textit{n} is the sample size of the catalogue, \textit{f} is the number of free parameters, and \textit{L} is the likelihood. This parameter has been chosen to select the best solution, choosing that fitting with a lower BIC for each \textit{SDSS} colour. Equation~\ref{Coefficients_SDSS_GALANTE} summarizes the transformation equations from \textit{SDSS} DR12 to GALANTE photometry.

\begin{align}
  \begin{split}
& F348M - \textit{u} =  0.149;\ rms = 0.067\\
& F420N - \textit{g} = 0.317 \cdot (\textit{u-r}) - 0.182;\ rms = 0.068 \\
& F450N - \textit{g} = 0.125 \cdot (\textit{g-i});\ rms = 0.027  \\
& F515N - \textit{g} = -0.300 \cdot (\textit{g-r}) - 0.032;\ rms = 0.028\\
& F660N - \textit{r} = -0.134 \cdot (\textit{g-z}) + 0.040;\ rms = 0.019\\
& F665N - \textit{r} = -0.138 \cdot (\textit{g-i}) + 0.010;\ rms = 0.009\\
& F861M - \textit{z} = 0.047 \cdot (\textit{r-z}) + 0.005;\ rms = 0.008\\
\label{Coefficients_SDSS_GALANTE}
  \end{split}
\end{align}

\begin{figure*}
\includegraphics[scale=0.19]{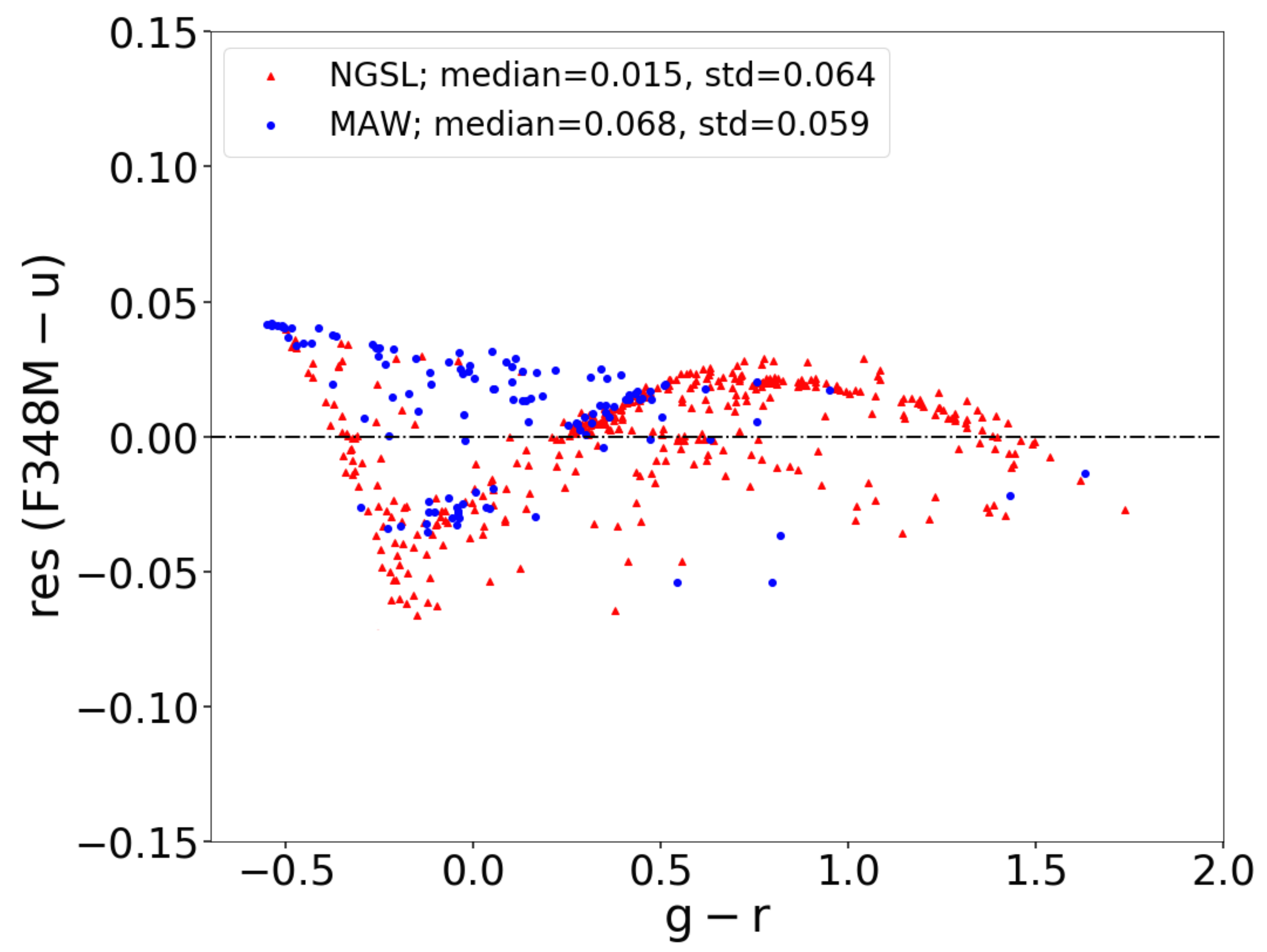}
\includegraphics[scale=0.19]{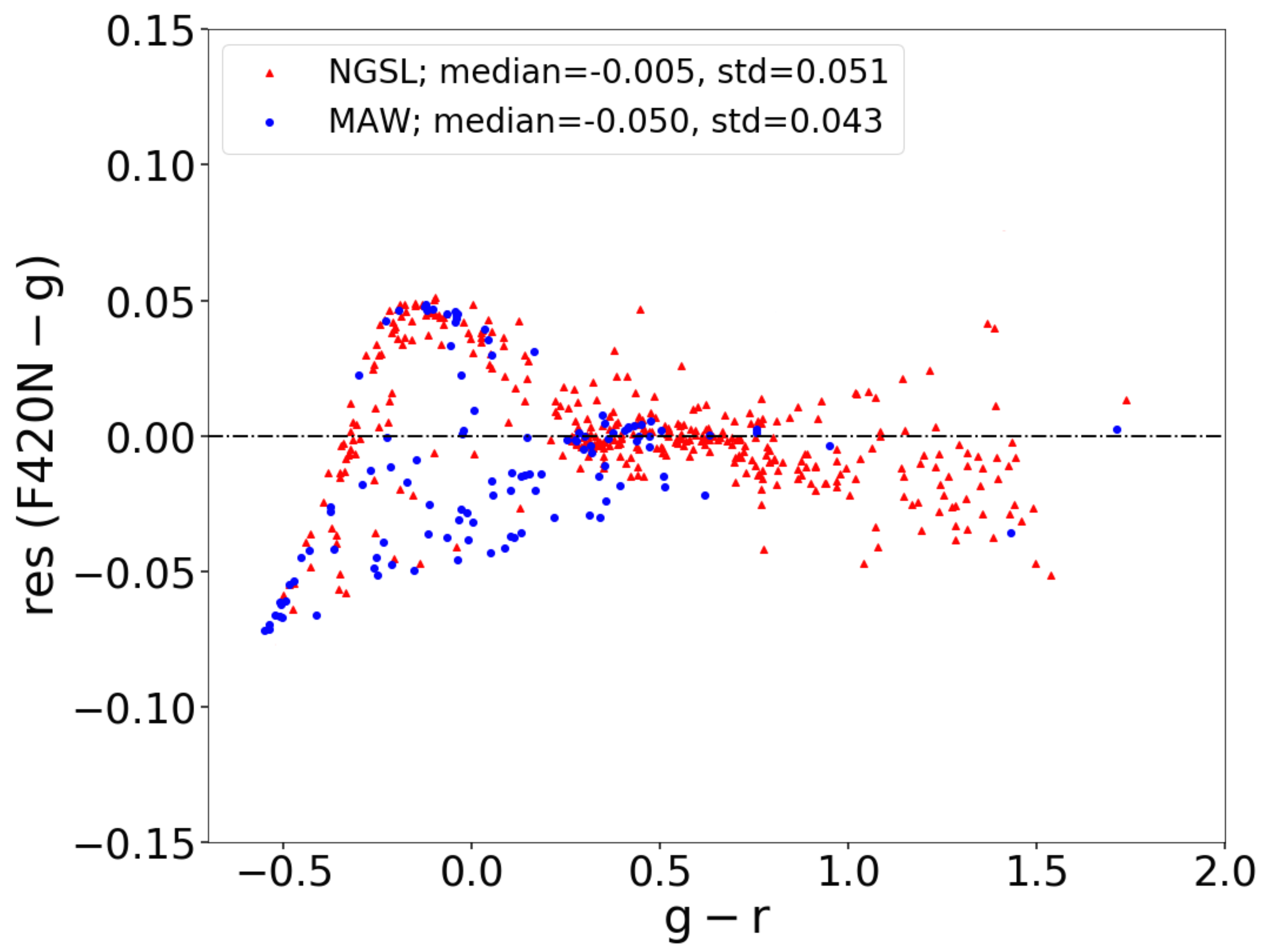}
\includegraphics[scale=0.19]{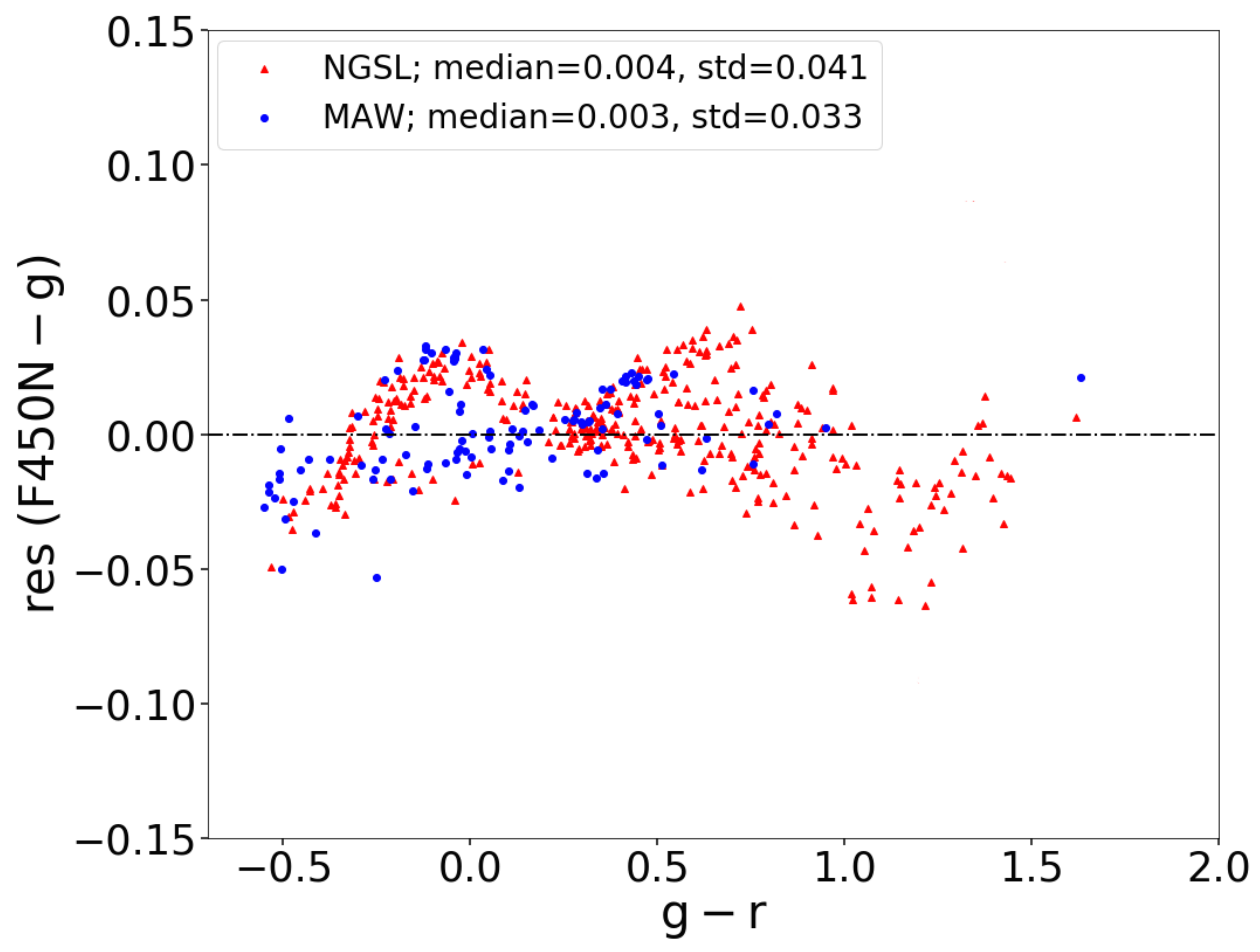}
\includegraphics[scale=0.19]{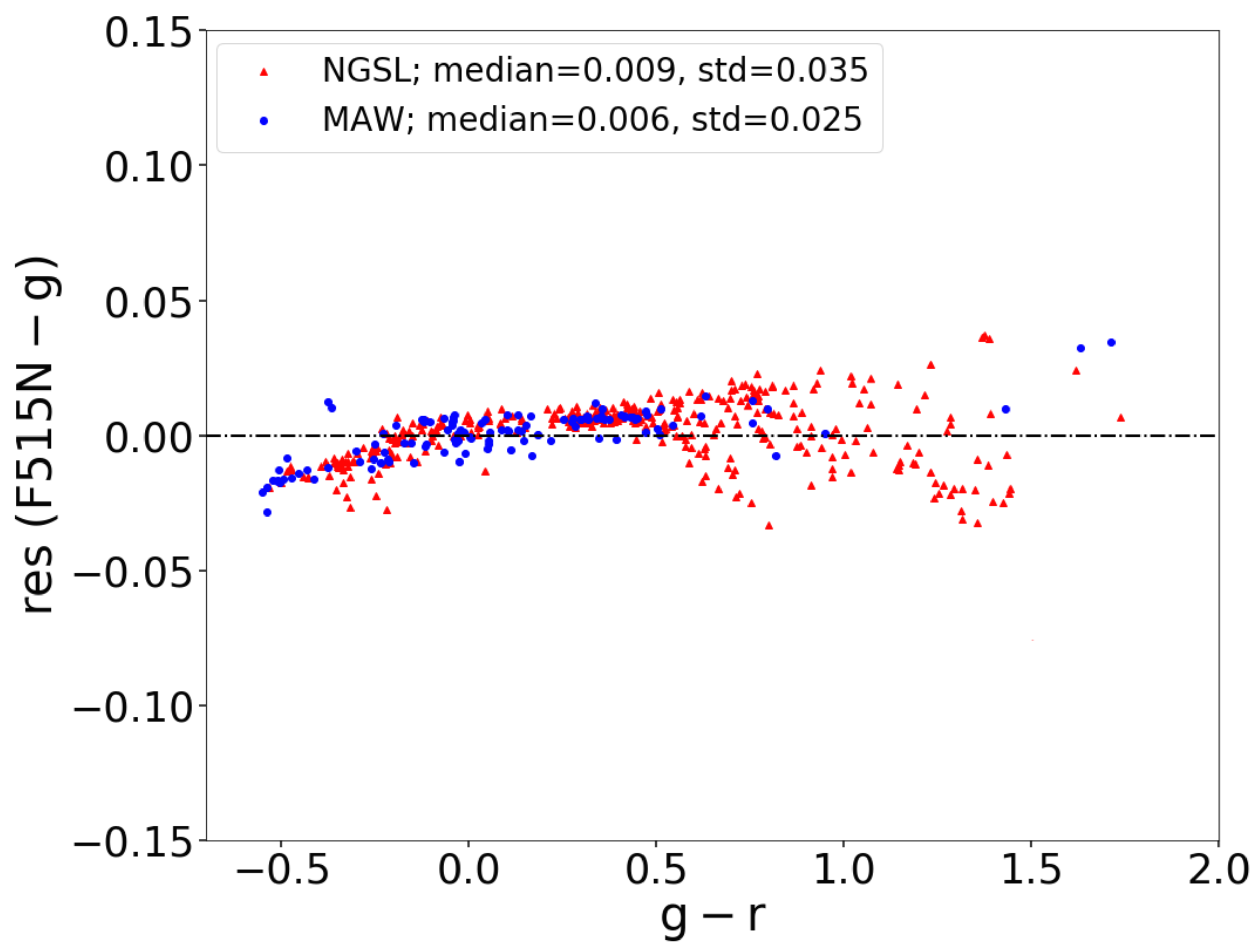}
\includegraphics[scale=0.19]{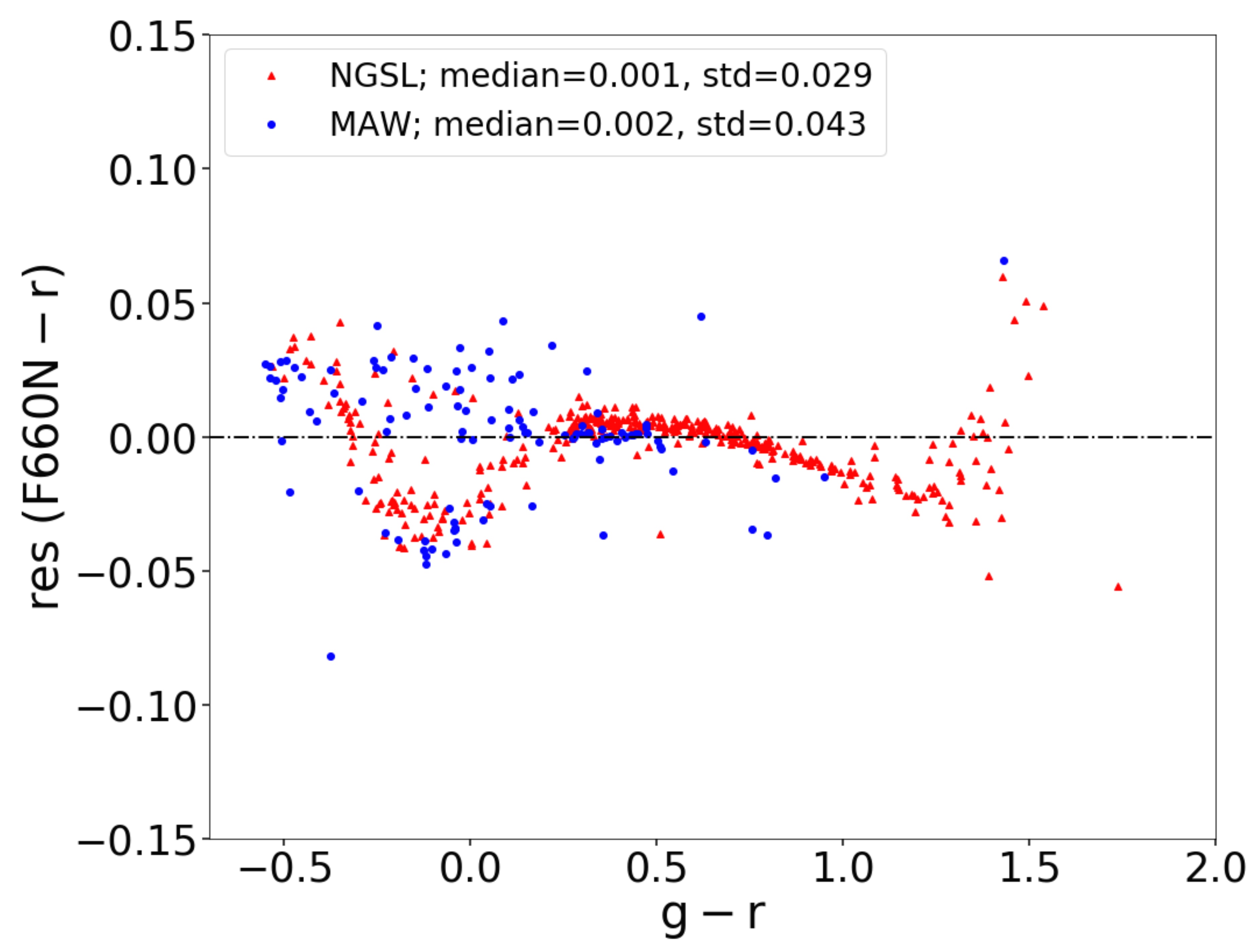}
\includegraphics[scale=0.19]{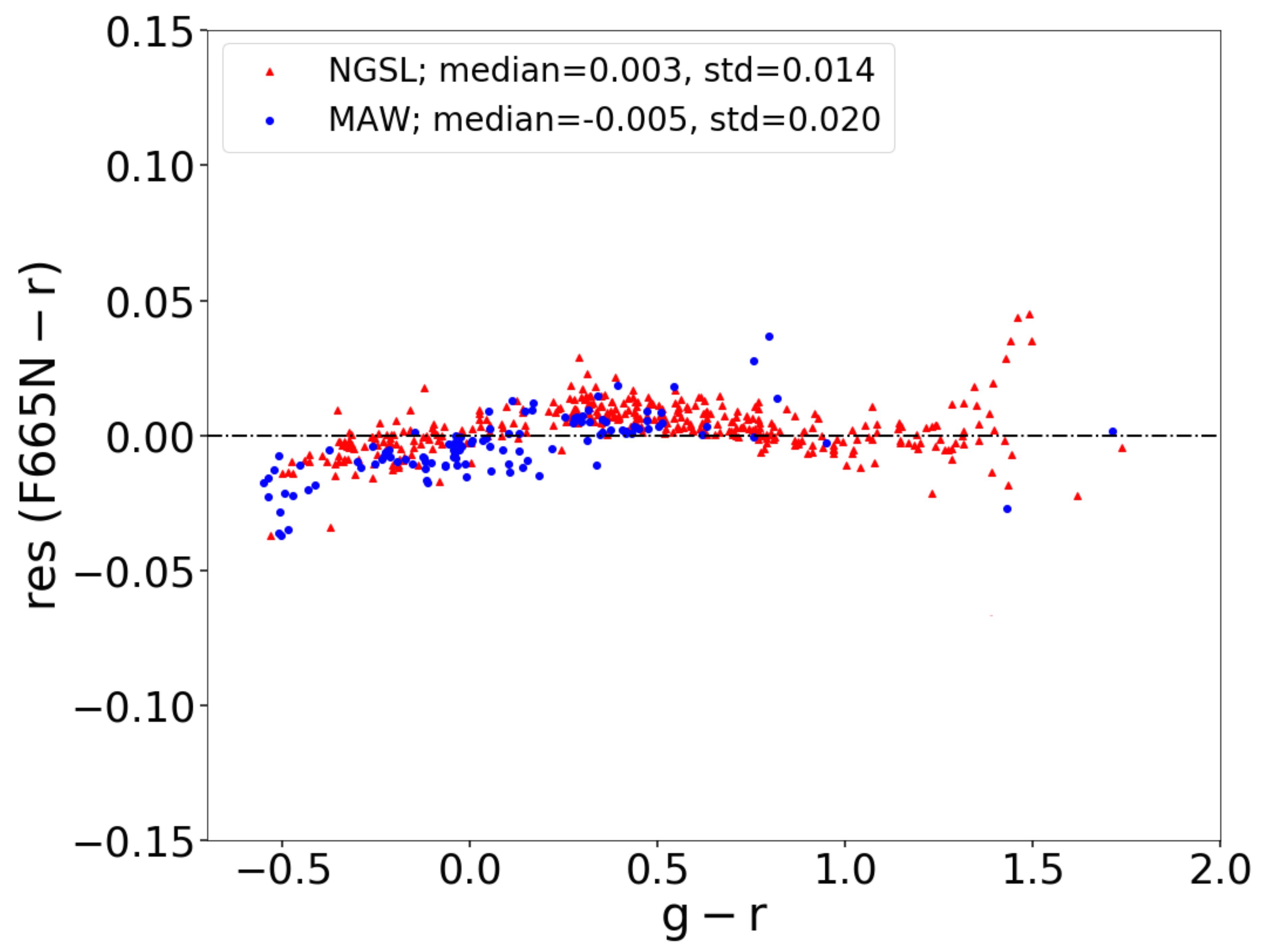}
\includegraphics[scale=0.19]{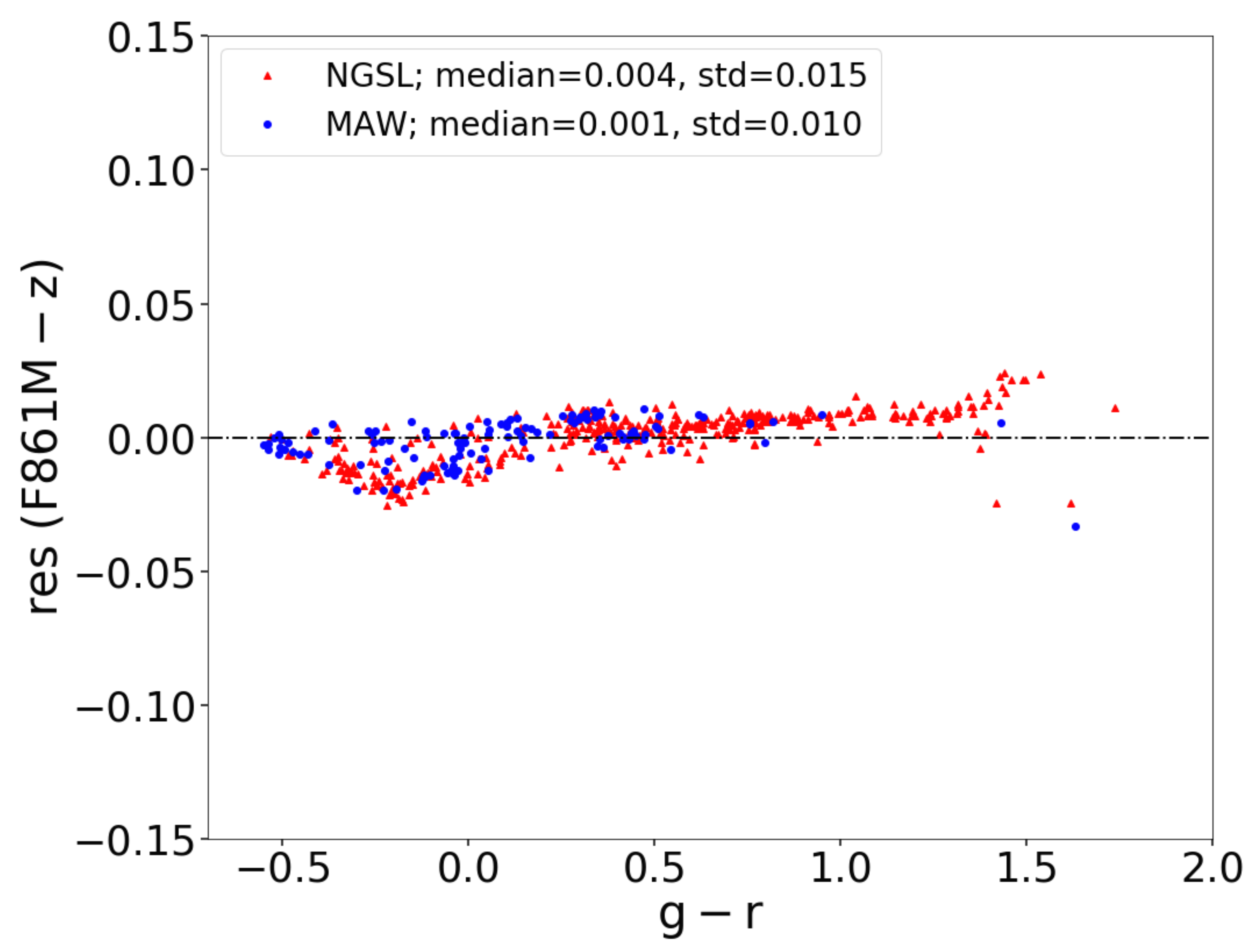}
\centering
\caption{Residual fitting errors for both catalogues from Equation~\ref{Coefficients_SDSS_GALANTE}. NGSL stars are plotted in red dots and MAW stars are shown in blue dots.}
\label{fig:residuo_g_r}
\end{figure*}

 After this fitting we plot the residuals versus (\textit{g}-\textit{r}) in Figure~\ref{fig:residuo_g_r}. The residuals have been split for each library. For (\textit{g}-\textit{r})<0.3, the high-reddened hot stars (blue points) are well differentiated from the distribution of red points representing less-reddened intermediate-type stars, especially for the F348M, F420N, F450N, and F660N filters.

If we take a look at these figures, F348M-\textit{u}, and F420N-\textit{g} residuals are more scattered than for the other filters, which was expected due to the bands being closer to the Balmer jump. The rms of these residuals is always lower than 6$\%$ for both libraries.

\begin{figure*}
\includegraphics[scale=0.19]{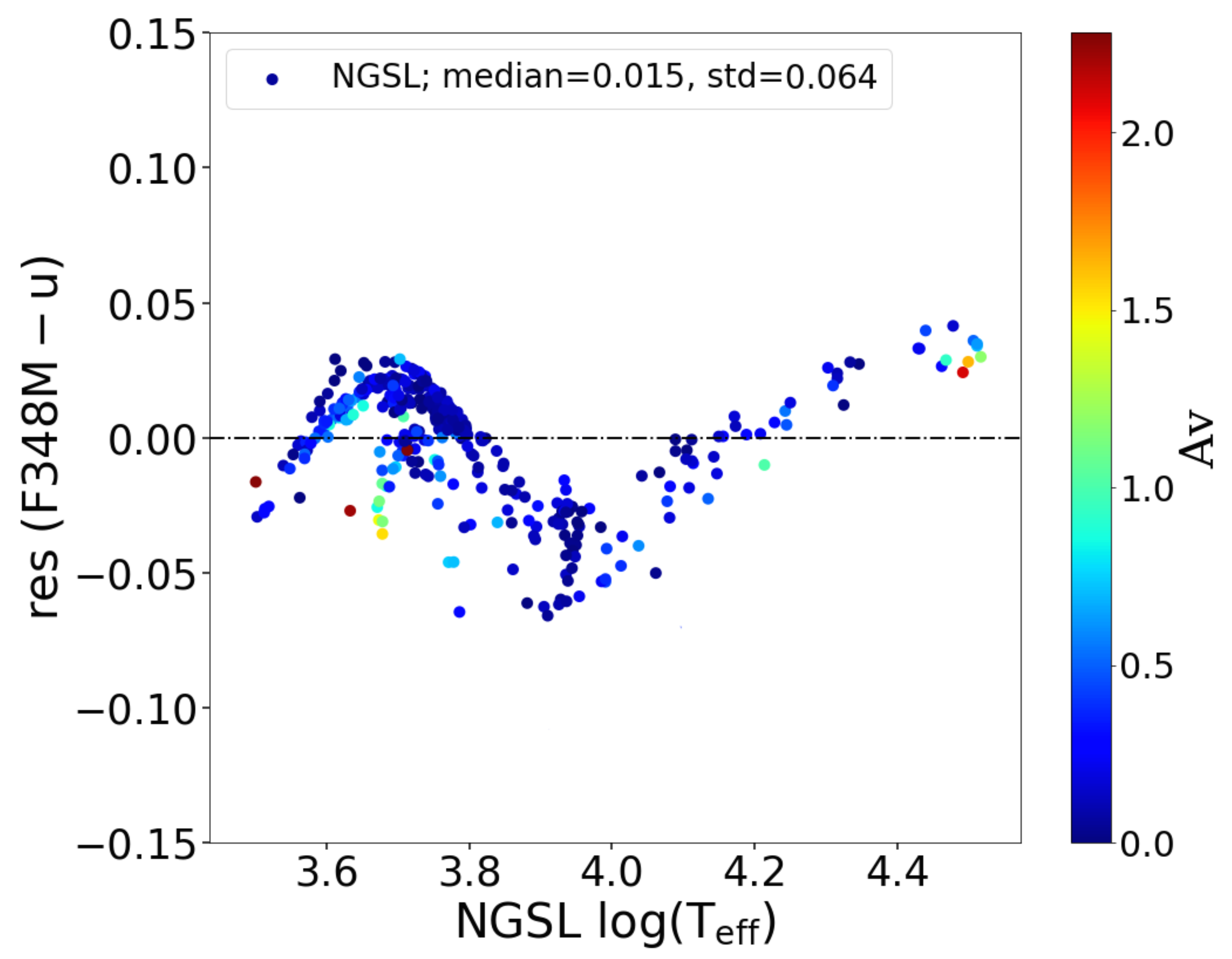}
\includegraphics[scale=0.19]{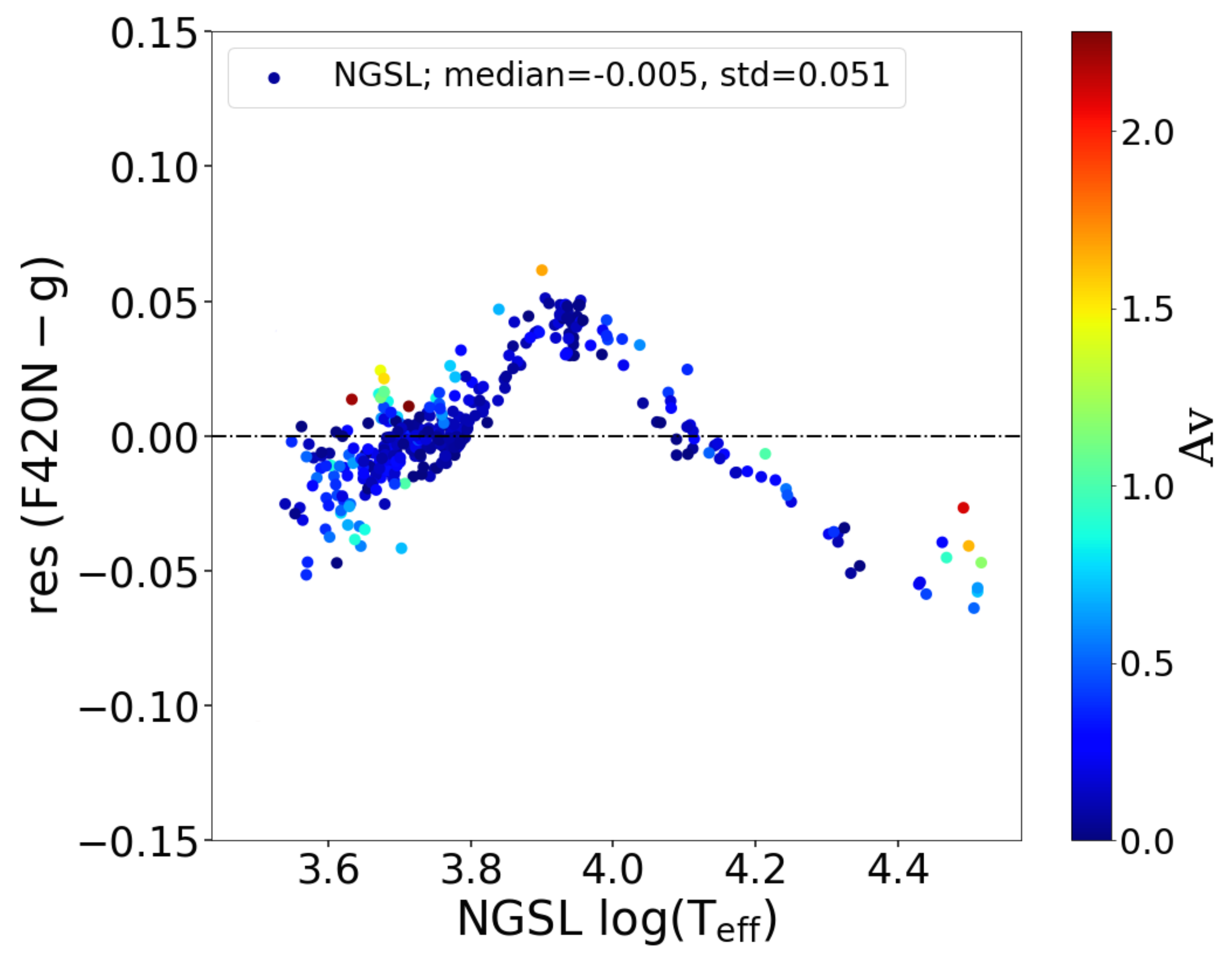}
\includegraphics[scale=0.19]{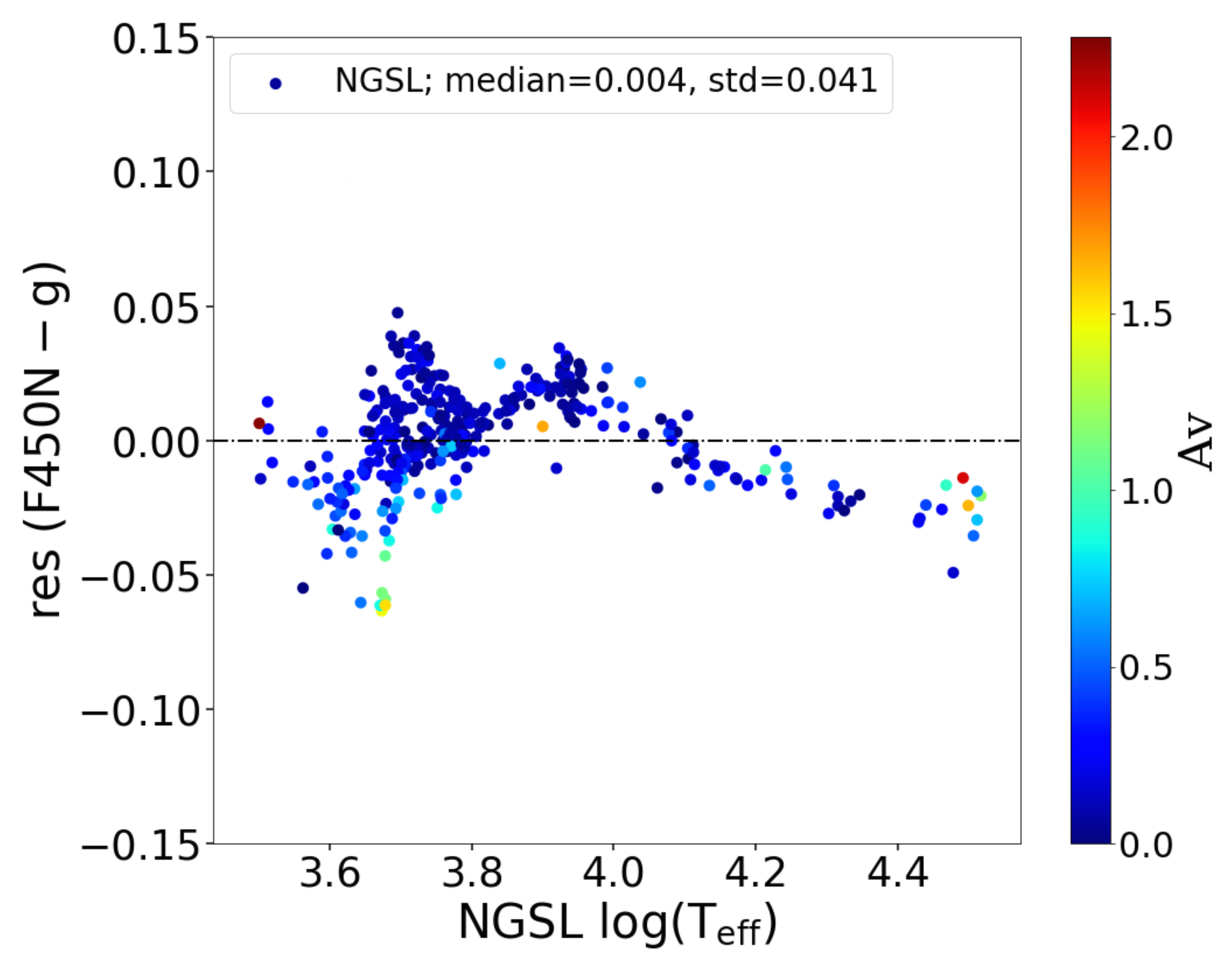}
\includegraphics[scale=0.19]{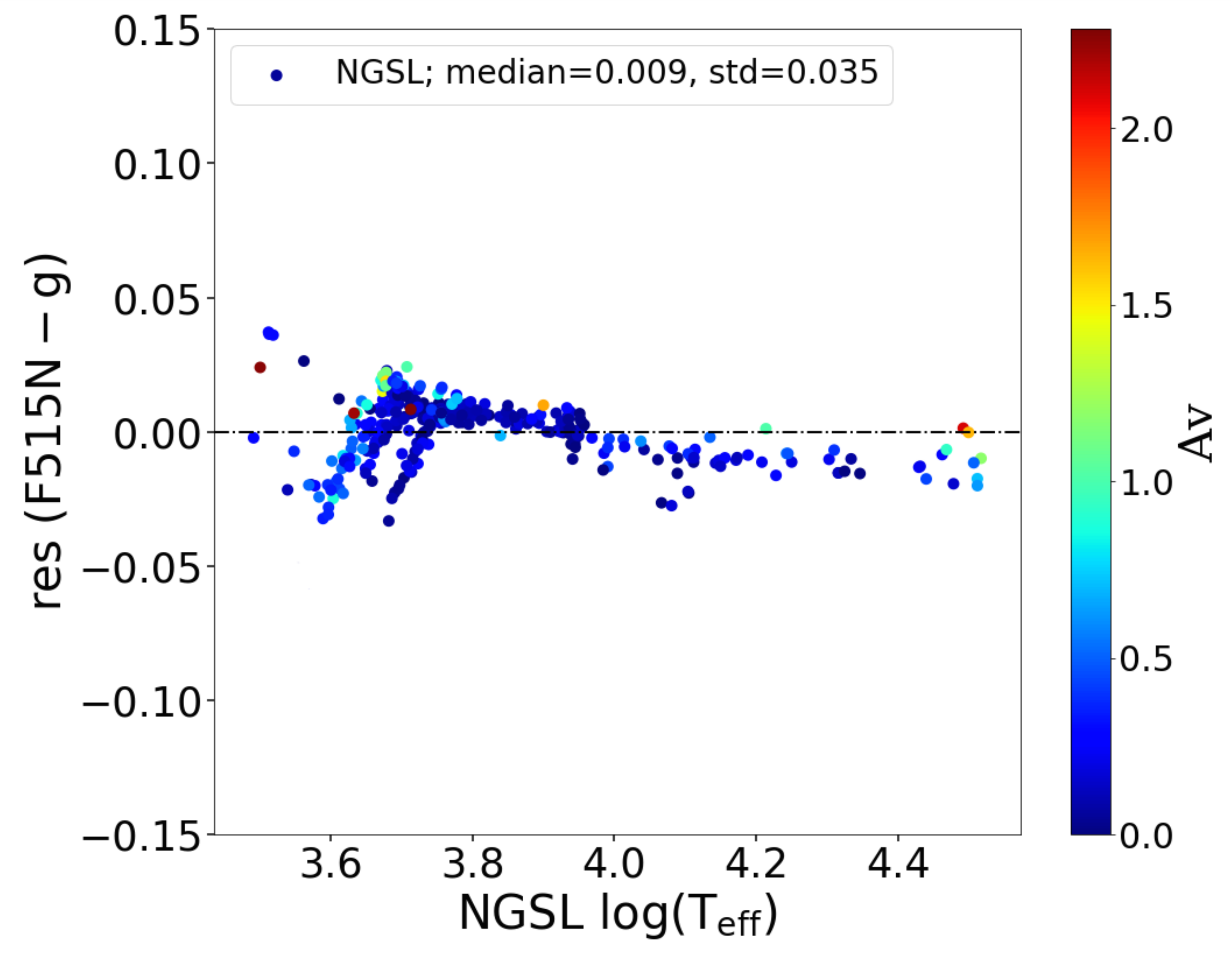}
\includegraphics[scale=0.19]{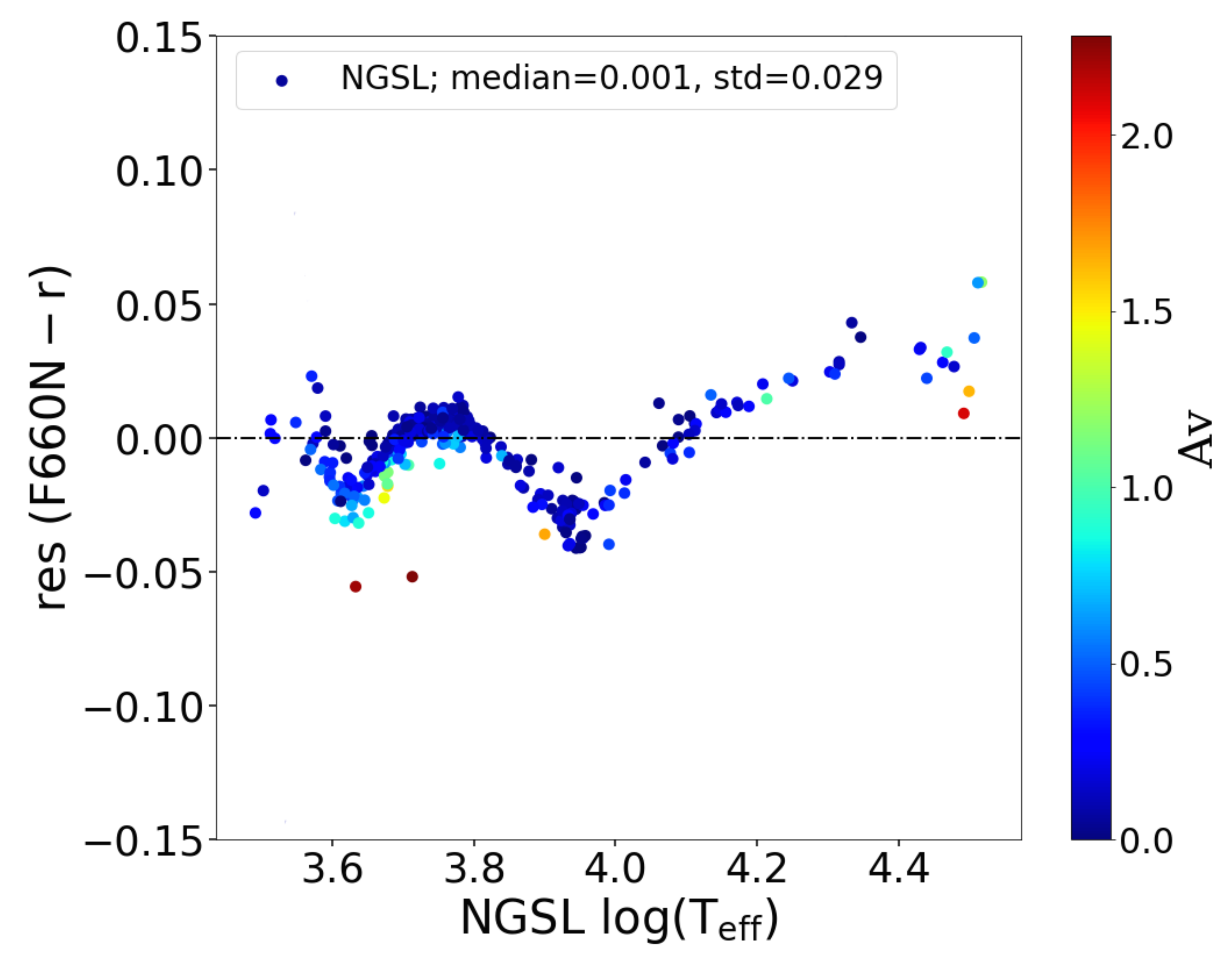}
\includegraphics[scale=0.19]{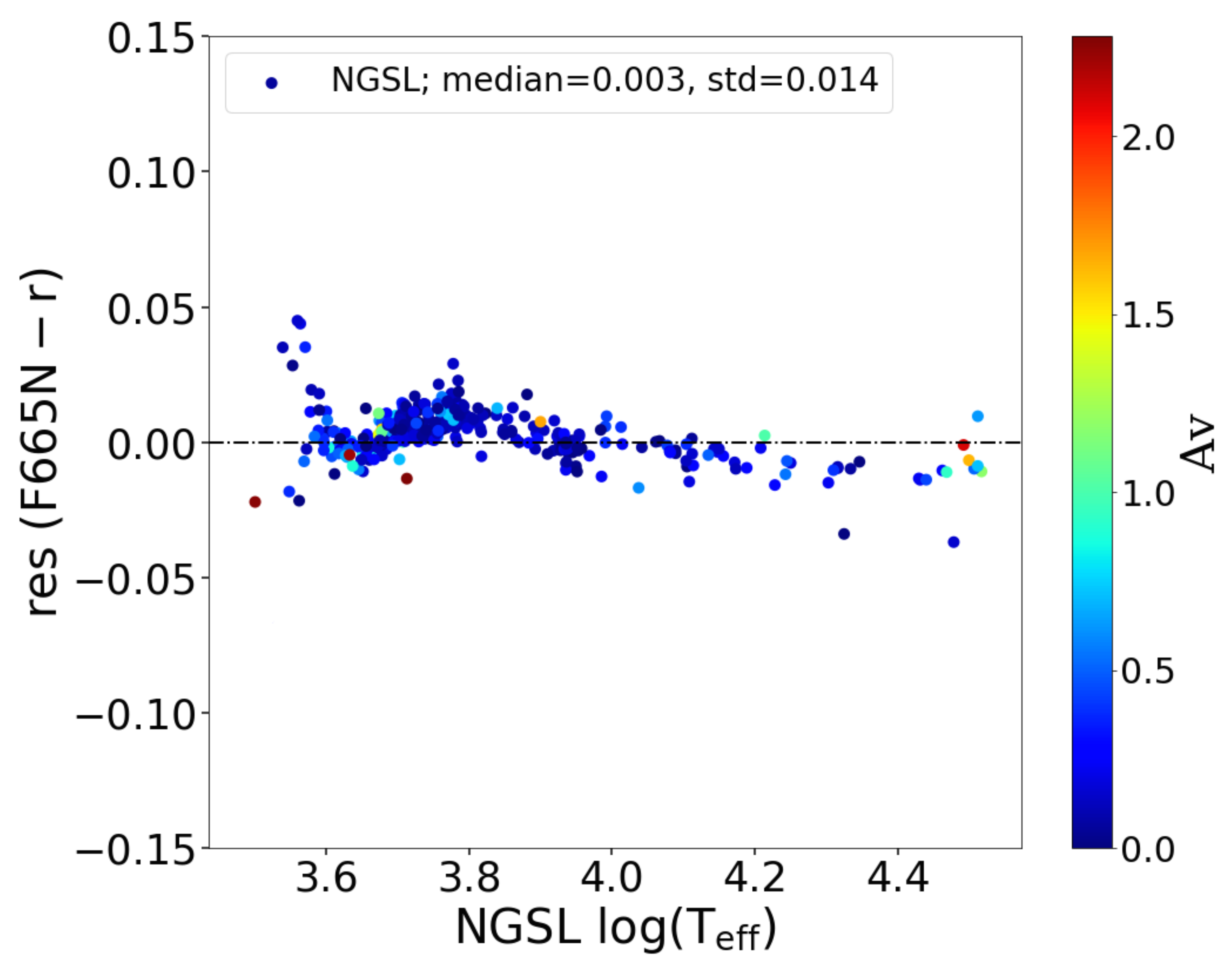}
\includegraphics[scale=0.19]{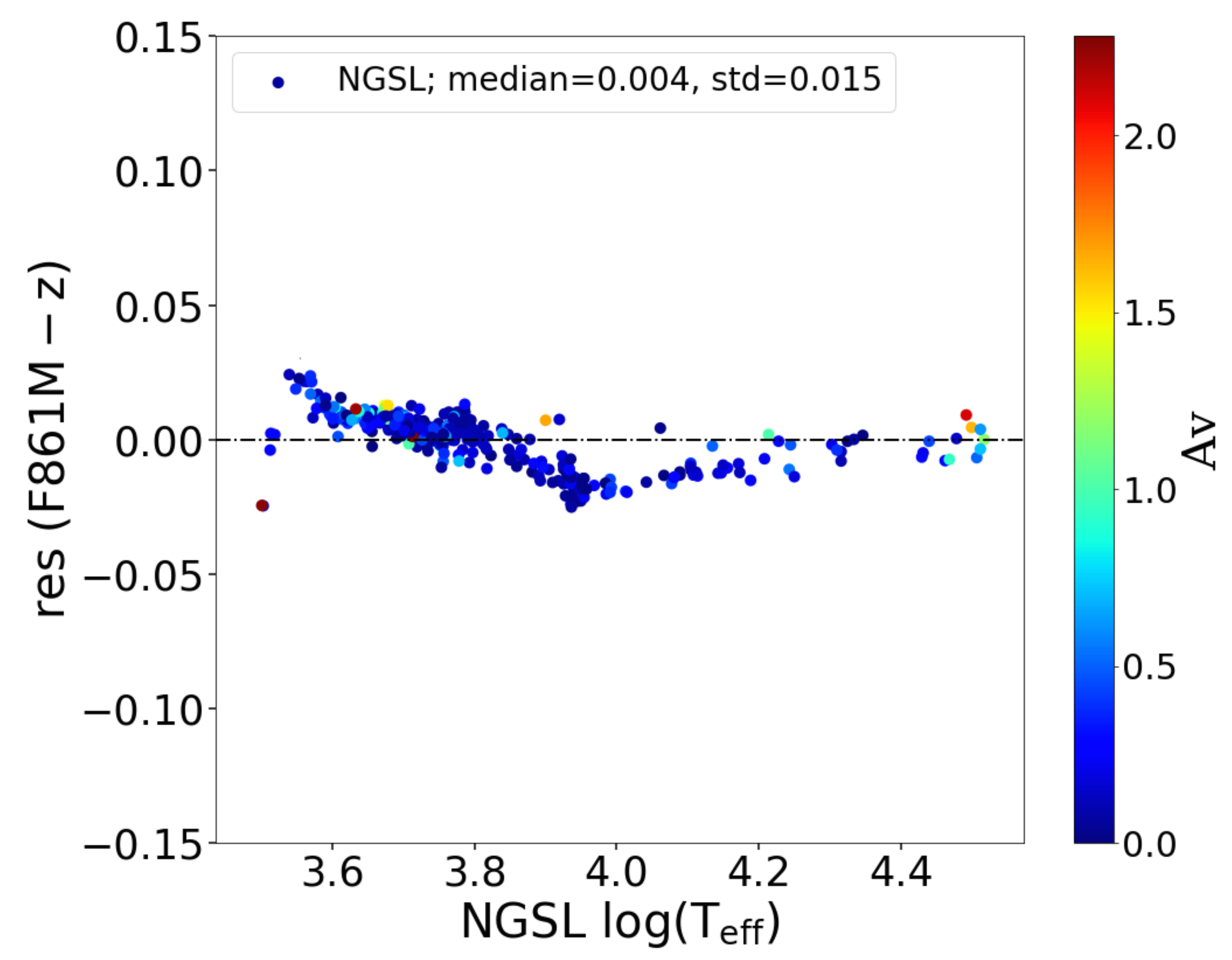}
\centering
\caption{These plots show the residual errors from the fitting depending on the effective temperature and colour excess. The colour map scale represents A$_{v}$. Both stellar parameters are extracted from the NGSL library.}
\label{fig:residuo_Teff}
\end{figure*}

The relationship of NGSL residuals with temperature and reddening is shown more clearly in Figure~\ref{fig:residuo_Teff}. Here we have represented fitting residuals versus tabulated NGSL temperature together with its absorption in the visible range (A$_{v}$) marked by different colours ({https://archive.stsci.edu/prepds/stisngsl/}, \citet{2016ASPC..503..211H}). As can be seen, differences due to reddening are not very marked, since in all cases we are limited to values lower than 0.7. However, fitting these transformations by a single colour generates substructures in the residuals, such as those observed in filters F348M, F420N, F450N and F660N. Nonetheless, these transformation equations allow us to establish a first ZP of the GALANTE photometry, which will be very dependent on the \textit{SDSS} data quality in the regions to be calibrated.

We also estimated the inverse transformation equations from GALANTE to \textit{SDSS} following the same procedure. The results are shown in Equation~\ref{Coefficients_SDSS_GALANTE_inversos}.

\begin{align}
  \begin{split}
& \textit{u} - F348M = - 0.149;\ rms = 0.067\\
& \textit{g} - F515N = 0.591 \cdot (F450N - F515N) + 0.025;\ rms = 0.018\\
& \textit{r} - F665N = 0.151 \cdot (F450N - F660N) - 0.013;\ rms = 0.013\\
& \textit{i} - F861M = 0.205 \cdot (F515N - F861M) - 0.055;\ rms = 0.022\\
& \textit{z} - F861M = -0.068 \cdot (F660N - F861M) - 0.005;\ rms = 0.013\\
\label{Coefficients_SDSS_GALANTE_inversos}
  \end{split}
\end{align}

\section{CALIBRATION OF THE GALANTE PHOTOMETRY}
\label{calibration_using_SDSS}

\subsection{CALIBRATION USING SDSS}
In this section, we will calibrate the GALANTE photometry for a small region of Cyg OB2, applying transformation equations from Equation~\ref{Coefficients_SDSS_GALANTE}. These data have been obtained over the past 3 years through requests for open observation times from the Javalambre observatory. Data has been previously reduced by the CEFCA team, and aperture photometry has been obtained at the IAA.

We select DR8 and DR12 as possible sources for obtaining GALANTE ZPs through the transformation equations. DR8 has been selected because a previous Cyg OB2 study by \citet{2012ApJS..202...19G} was based on this release, while our transformation is based on DR12. Thus we now want to compare both releases in order to see the ZP differences we can obtain using both catalogues.

\begin{figure*}
\includegraphics[scale=0.19]{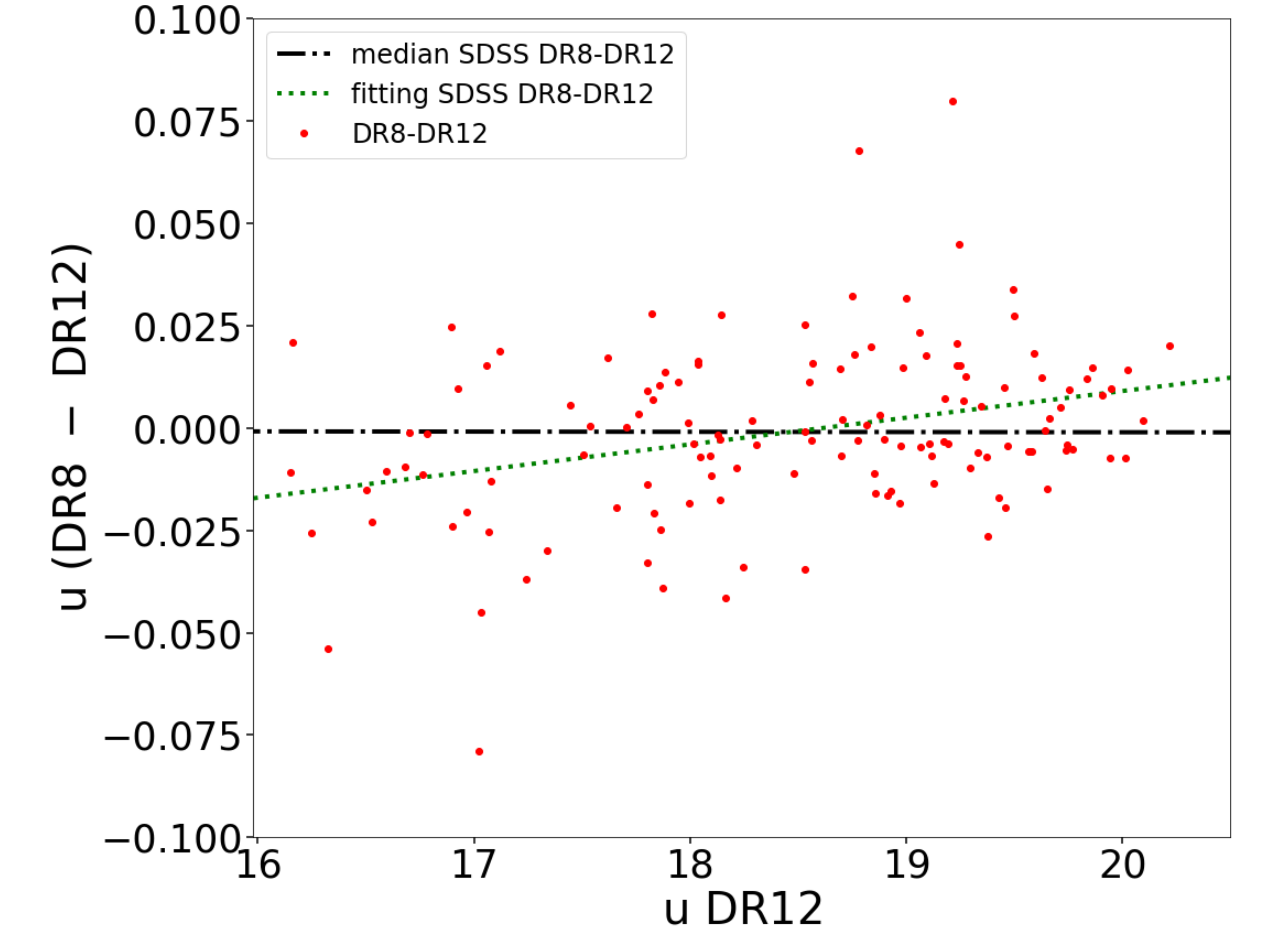}
\includegraphics[scale=0.20]{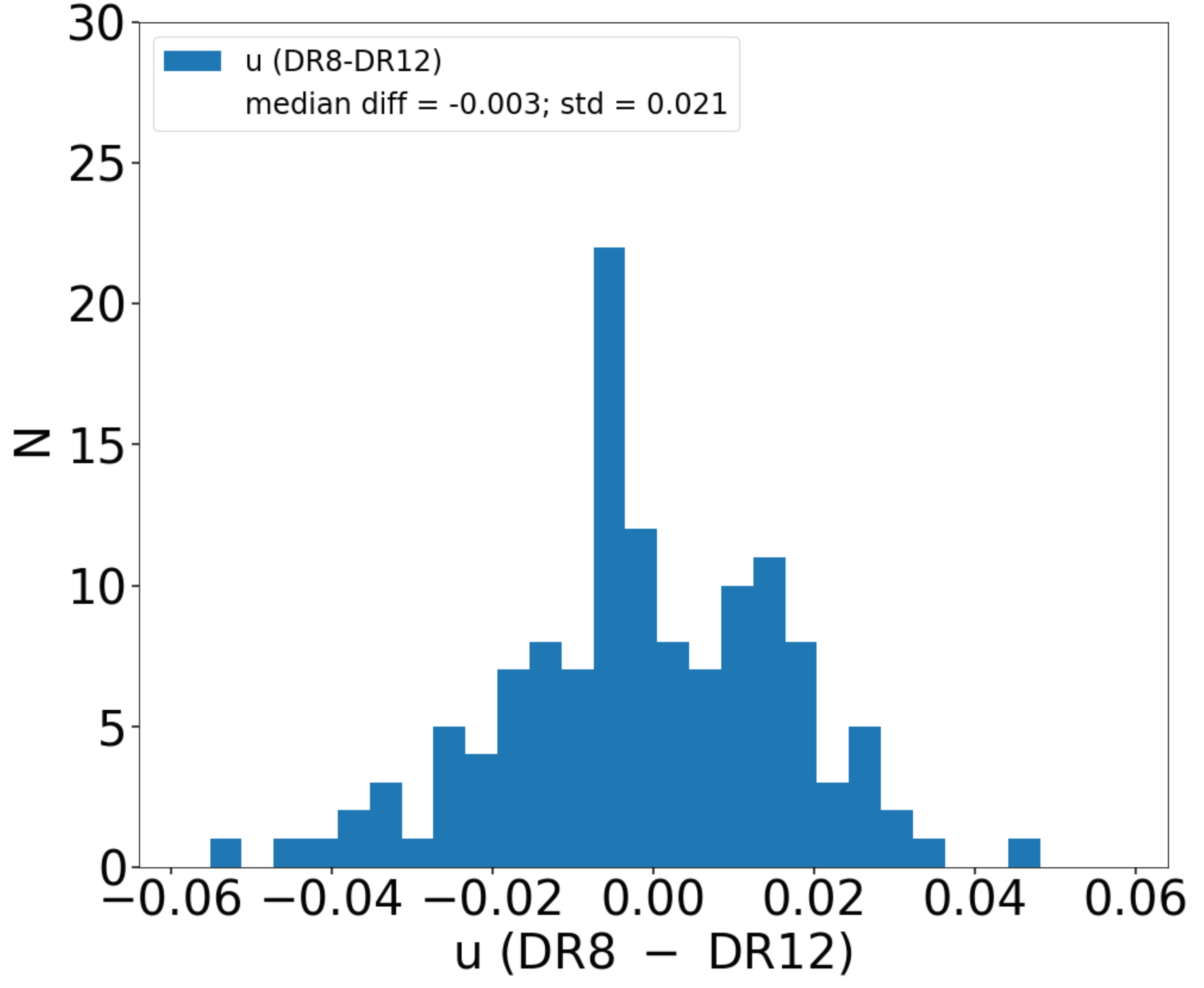}
\includegraphics[scale=0.19]{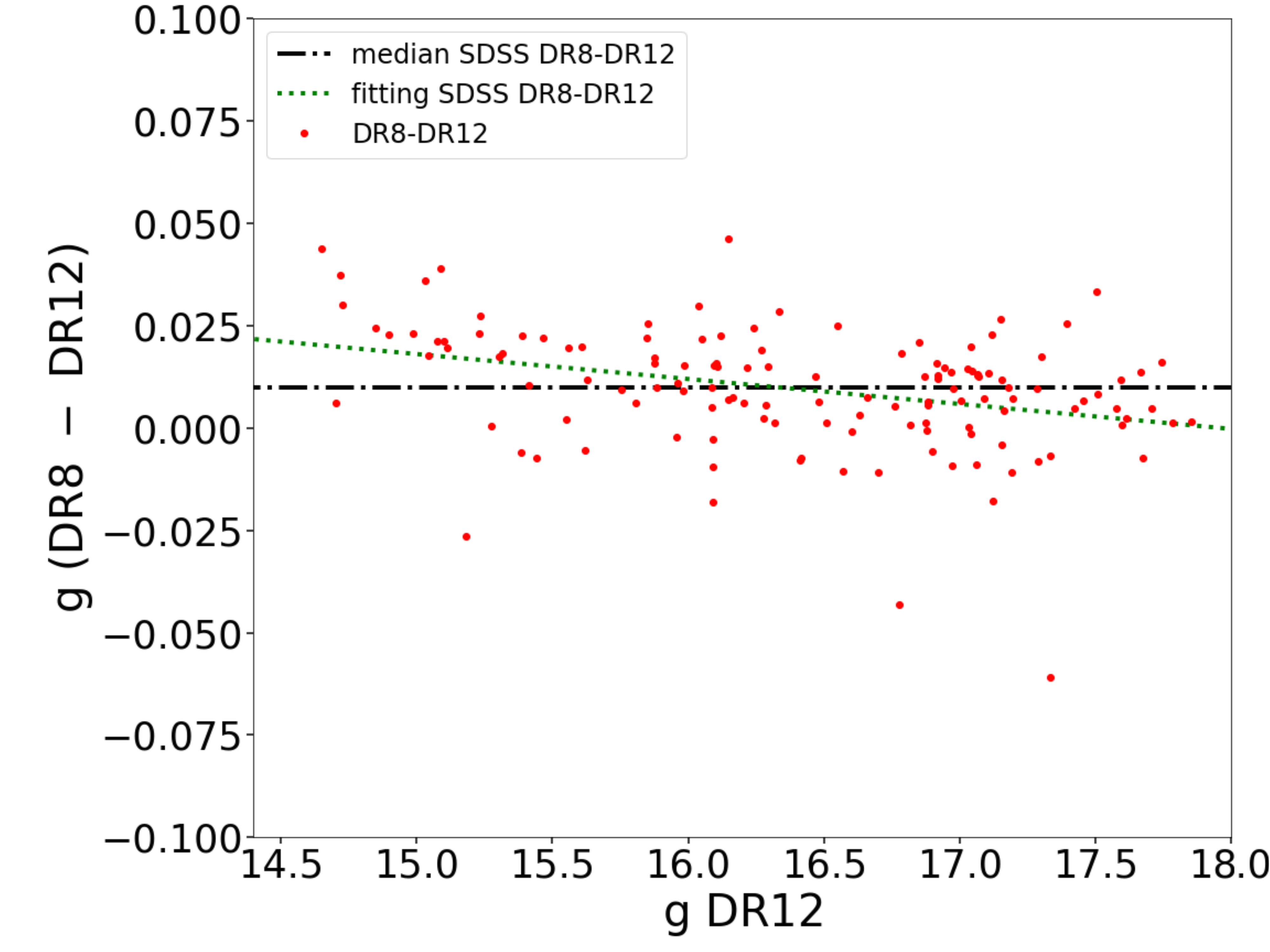}
\includegraphics[scale=0.195]{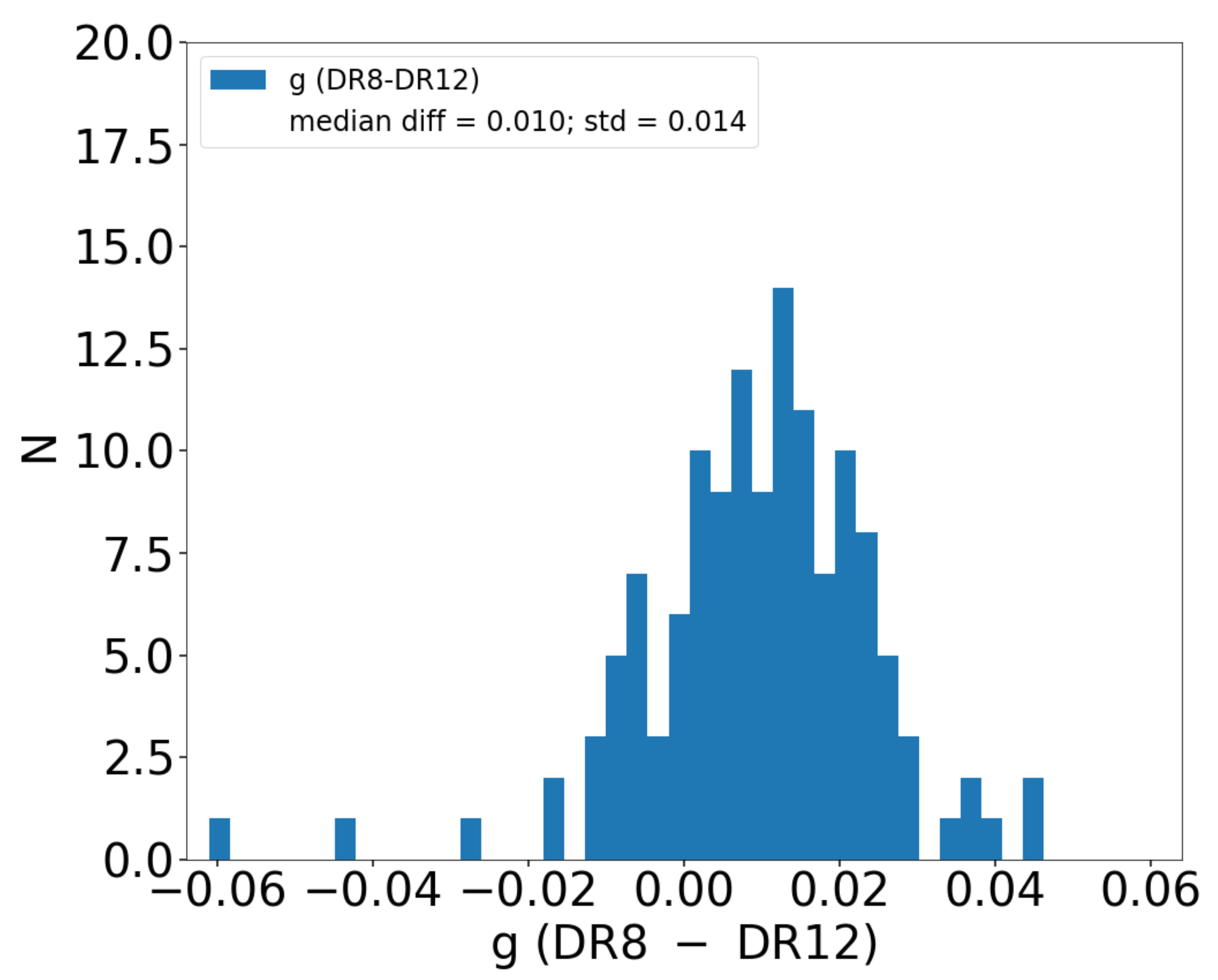}
\includegraphics[scale=0.185]{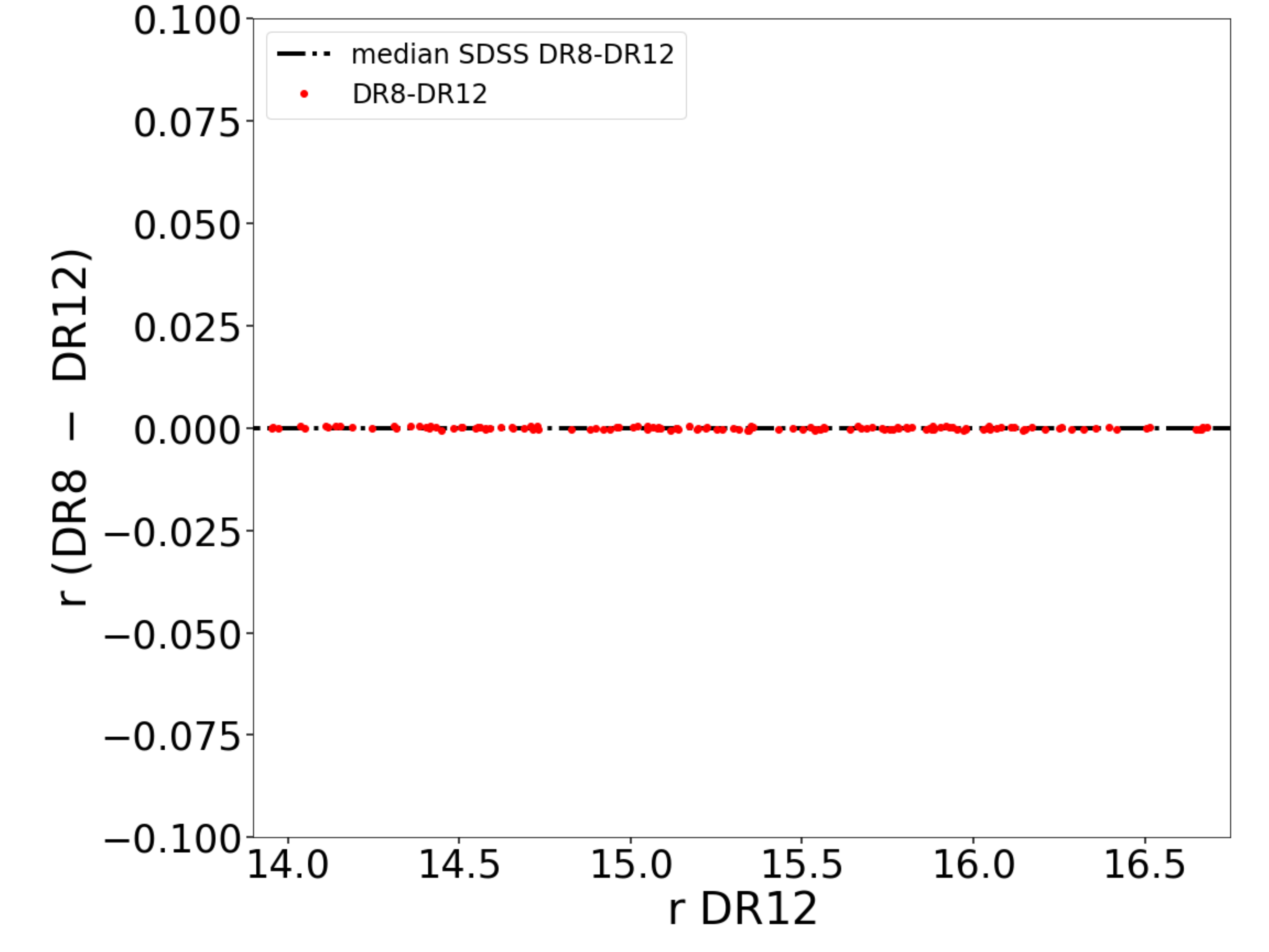}
\includegraphics[scale=0.185]{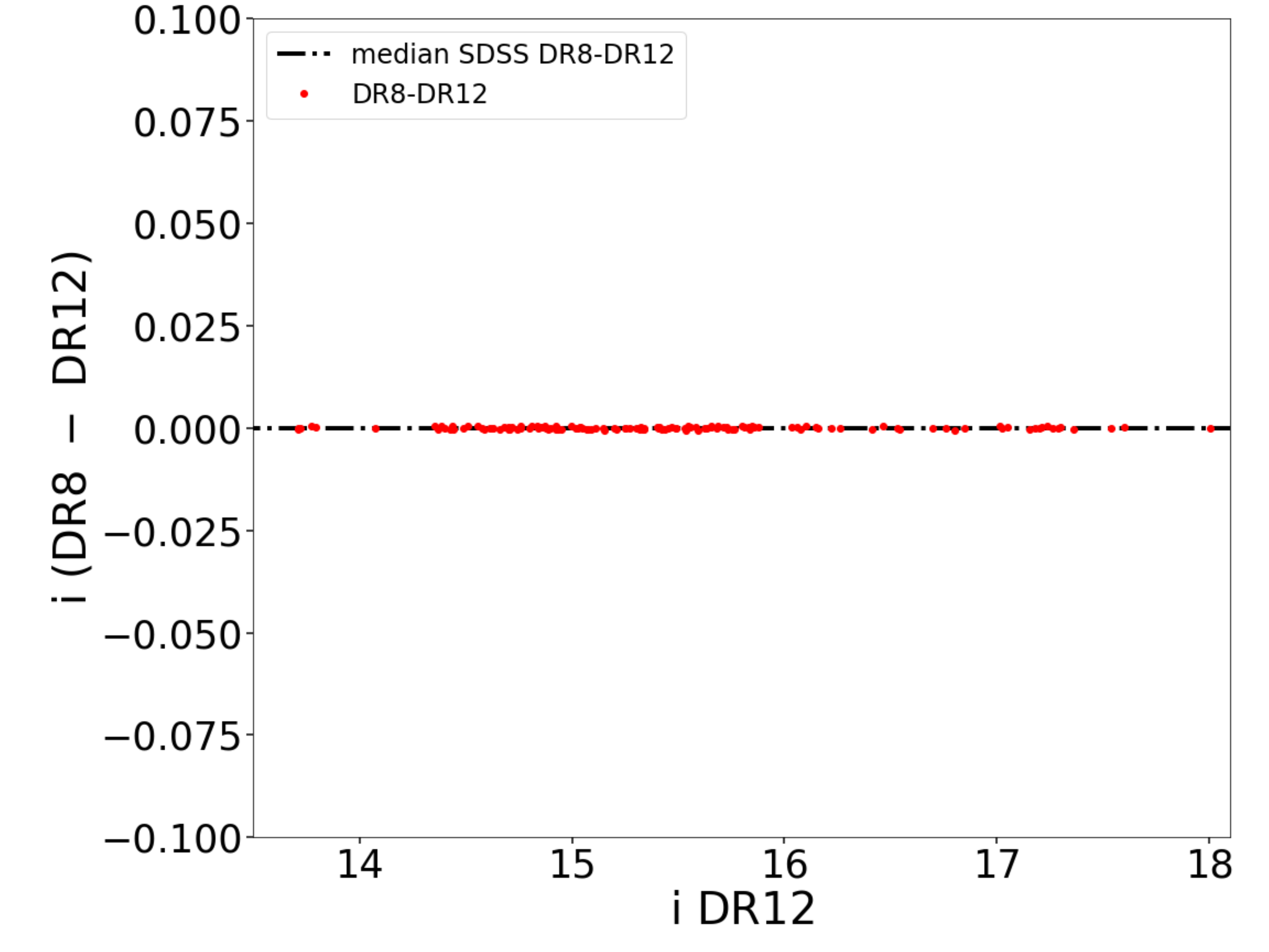}
\includegraphics[scale=0.19]{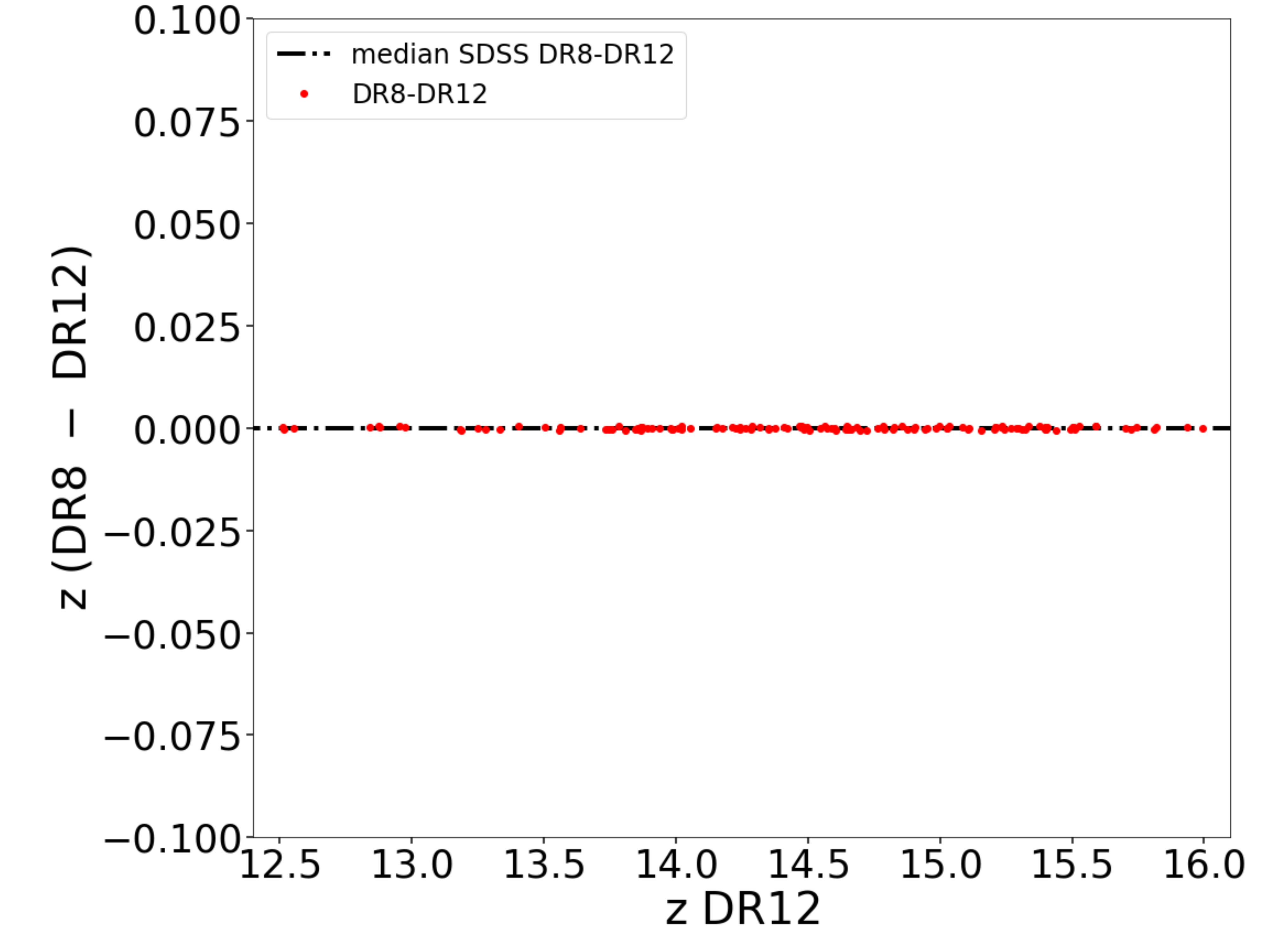}
\centering
\caption{These figures represent differences for stars observed with the Javalambre observatory T-80 telescope in Cyg OB2 and their \textit{SDSS} photometry taken directly from DR8 and DR12.}
\label{fig:diff_DR8_DR12}
\end{figure*}

 Firstly, we directly compare \textit{SDSS} photometry in both data releases taking common stars observed with the Javalambre Auxiliary Survey Telescope (hereinafter T-80 telescope). This crossmatch gives a total of 130 stars in Cyg OB2. Figure~\ref{fig:diff_DR8_DR12} represents the differences between both releases.

Figure~\ref{fig:diff_DR8_DR12} shows a significant difference in \textit{u} and \textit{g} bands for both releases, showing a light magnitude equation (green dots line) for these two bands. This effect was already known by the \textit{SDSS} team and appeared to be due to the ubercalibration procedure described by \citet{2008ApJ...674.1217P}. DR12 was already corrected of this effect. Differences in \textit{r}, \textit{i}, and \textit{z} bands are shown to be lower than a few thousandths of magnitudes and without any evidence of systematic effect. Nonetheless, even though we are observing an equation magnitude between both releases, differences between magnitudes show median values below one hundred, as seen in the histograms of Figure~\ref{fig:diff_DR8_DR12}.

\begin{figure*}
\includegraphics[scale=0.19]{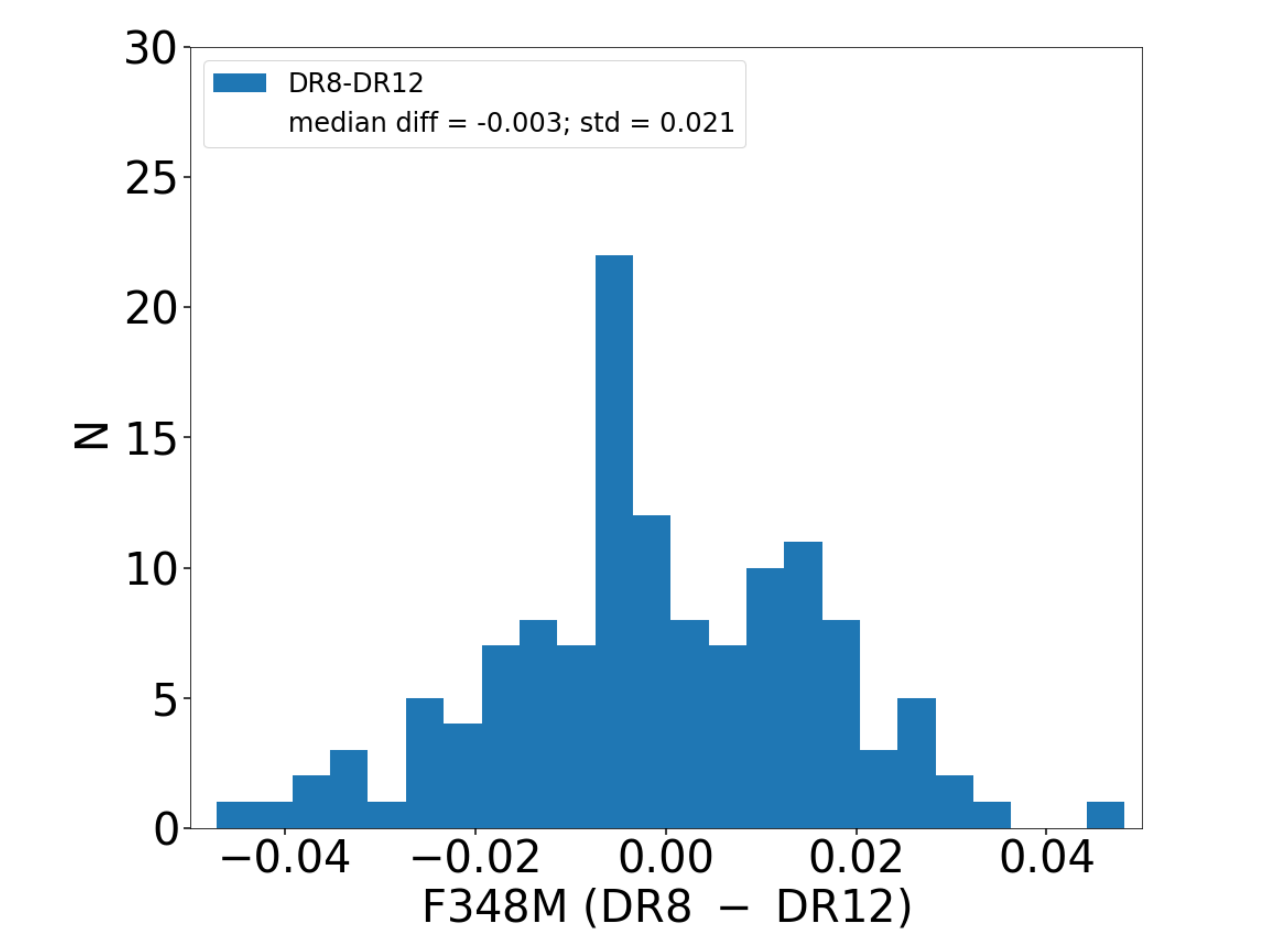}
\includegraphics[scale=0.19]{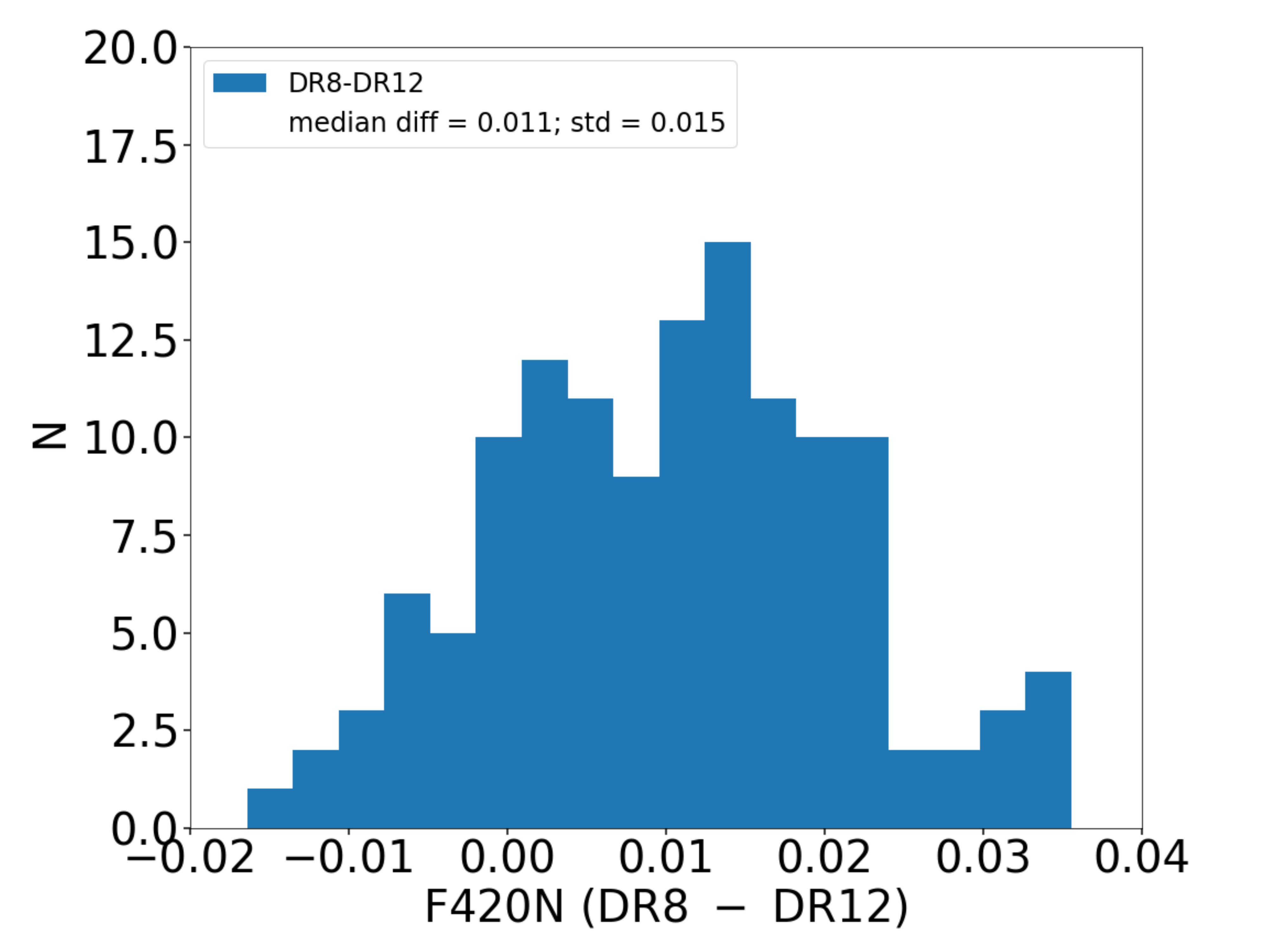}
\includegraphics[scale=0.19]{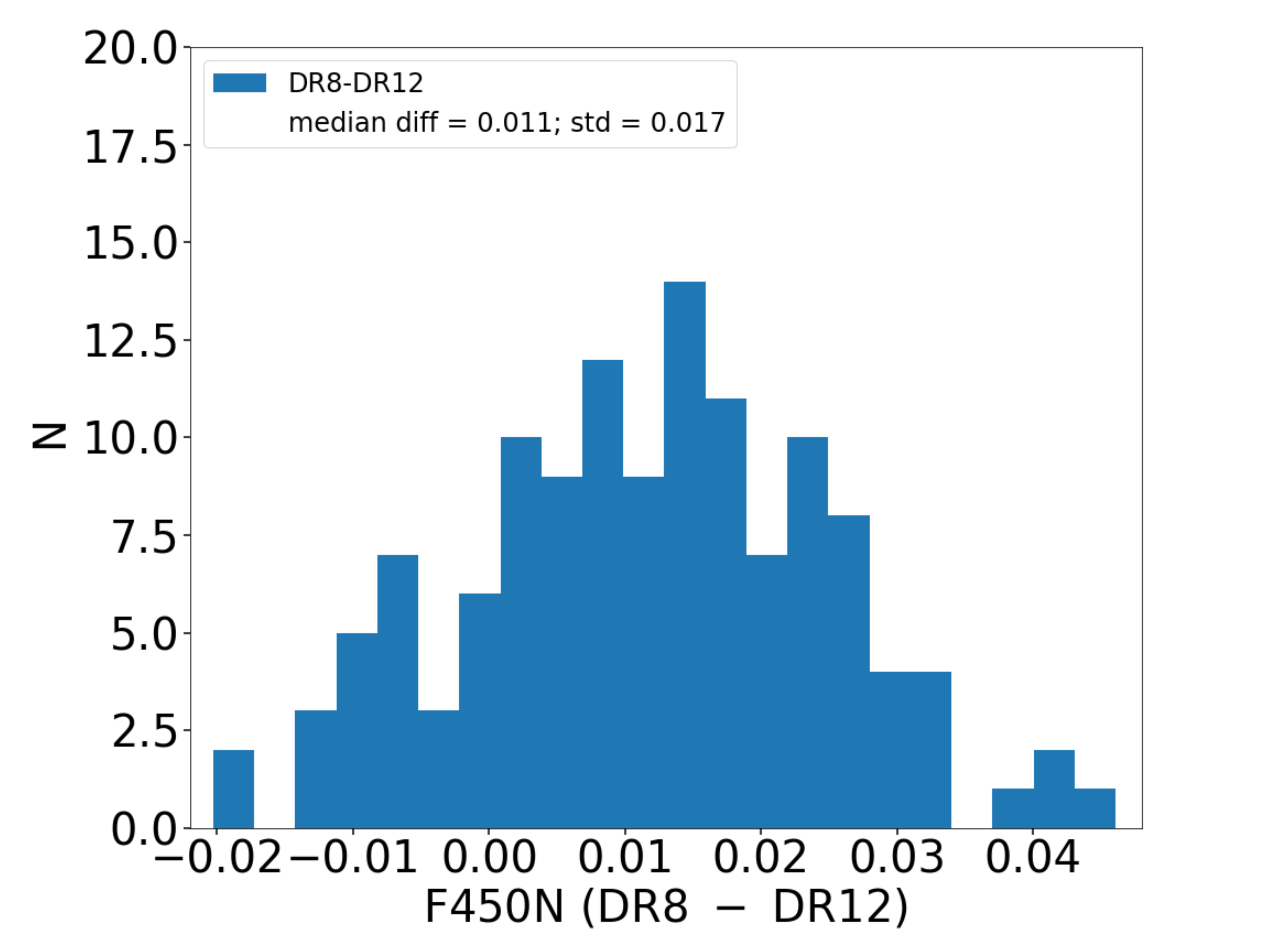}
\includegraphics[scale=0.19]{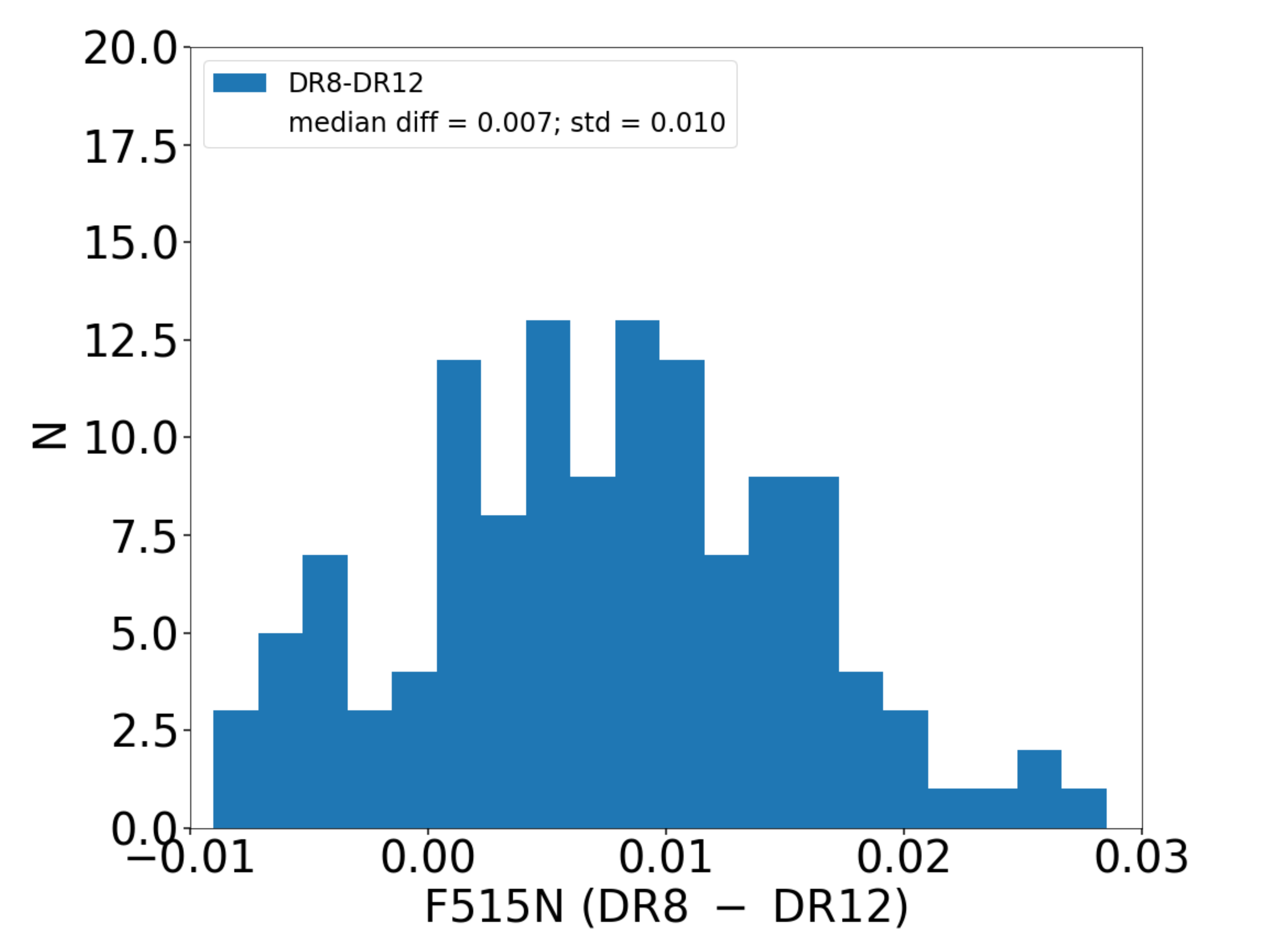}
\includegraphics[scale=0.19]{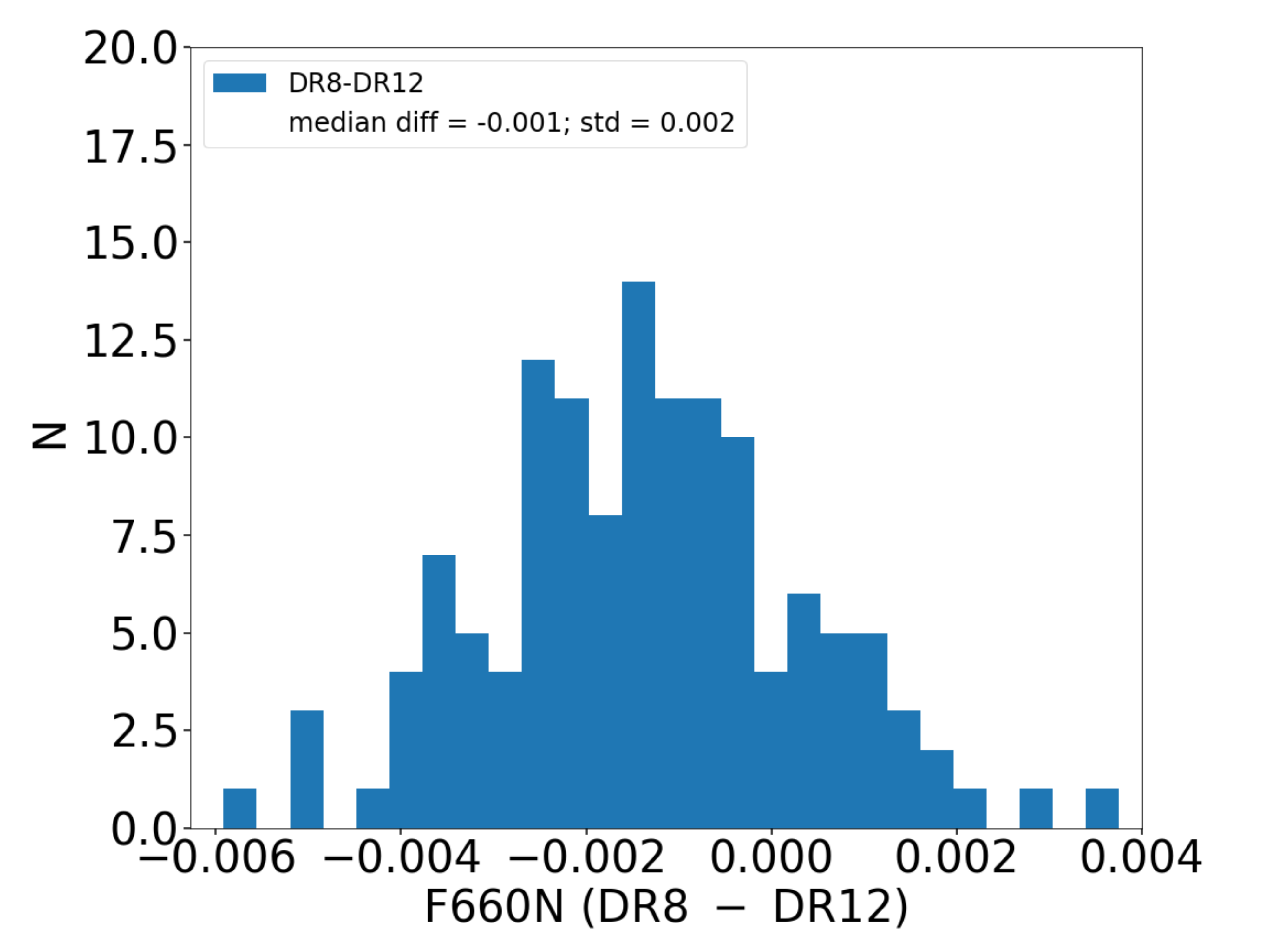}
\includegraphics[scale=0.19]{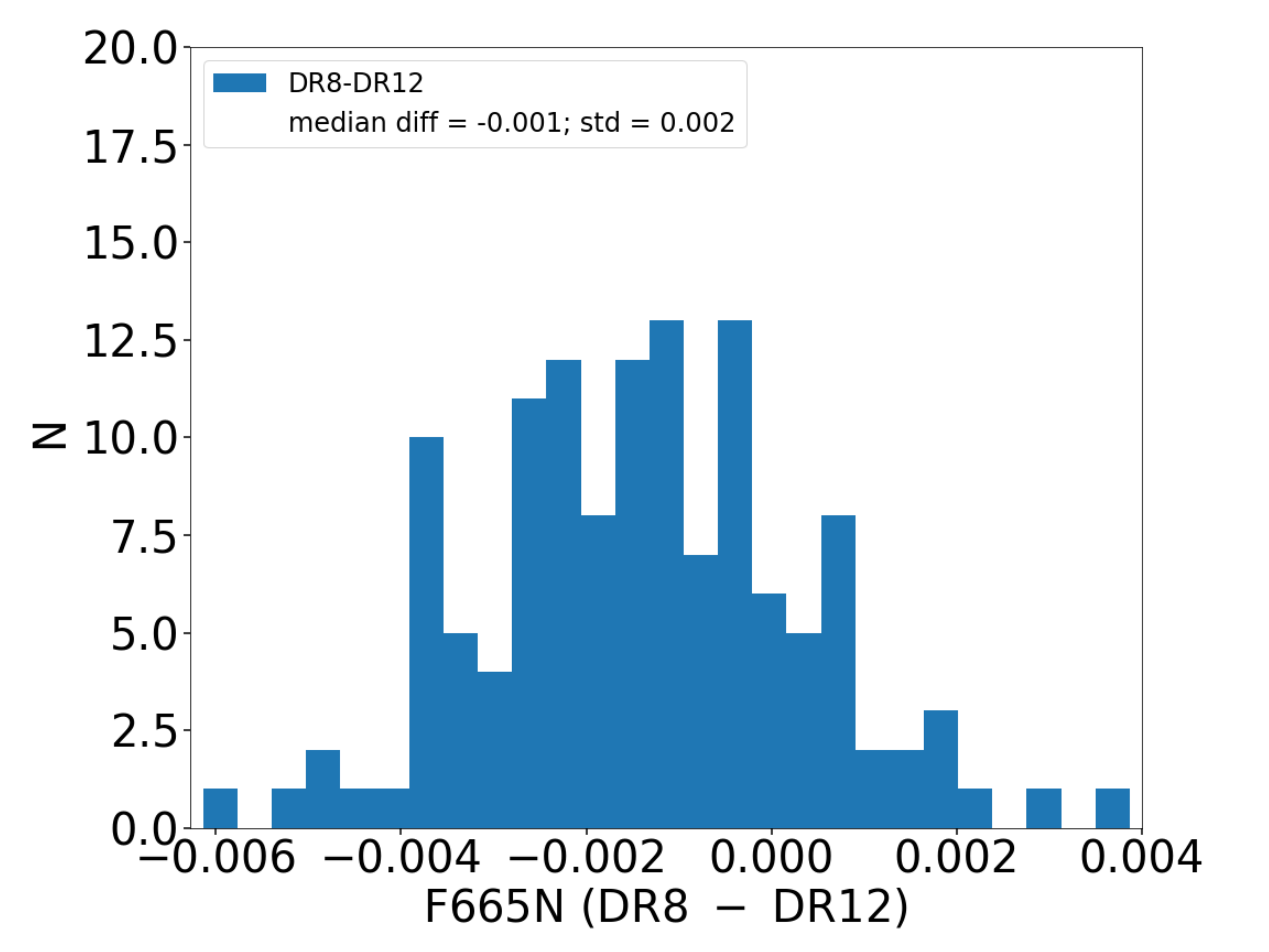}
\includegraphics[scale=0.19]{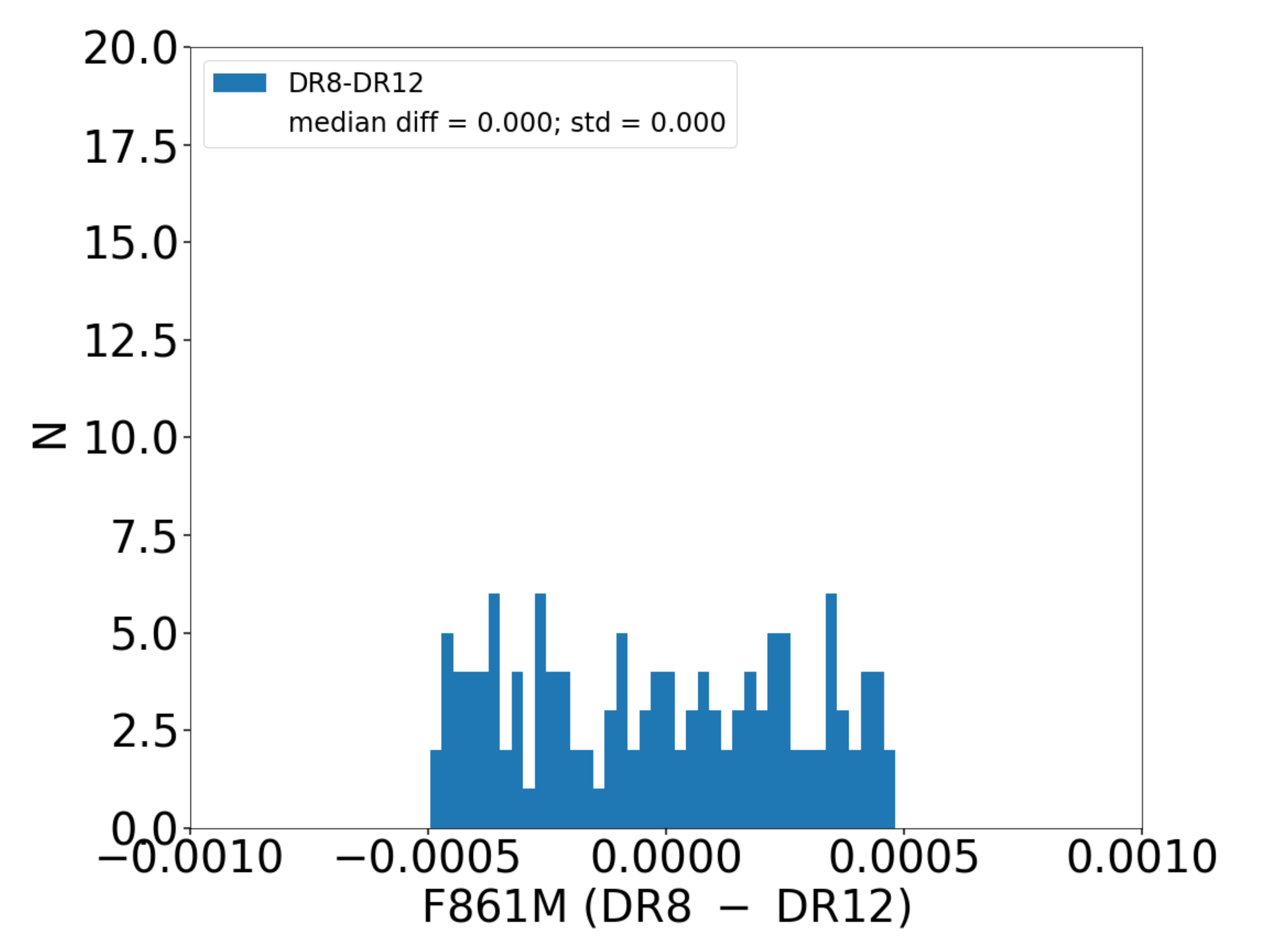}
\centering
\caption{GALANTE photometry histograms for differences between \textit{SDSS} DR8 and DR12 using a small region of Cyg OB2 observed by the T-80 telescope.}
\label{fig:histog_diff_DR8_DR12}
\end{figure*}

After this analysis, we decided to obtain GALANTE photometry ZPs based on both releases and comparing them afterward. Using both releases, we obtain the difference between GALANTE photometries shown in Figure~\ref{fig:histog_diff_DR8_DR12}, where we can see that the median and rms of these distributions are below 0.02 magnitudes for all bands. The final ZPs for each GALANTE band are shown in Table~\ref{tab:ZPs_DR8_DR12}. Thus, we choose to calibrate GALANTE photometry with \textit{SDSS} data release 12 before to analyze the calibration procedure with the new \textit{RefCat2} catalogue.

\tiny
\begin{table}
\caption{GALANTE photometric ZPs from \textit{SDSS} DR8 and DR12.}
\begin{center}
\begin{tabular}{ccccc}
\hline\hline\noalign{\smallskip}
Band      &   ZP DR8     &    error DR8    &    ZP DR12     &    error DR12 \\
\noalign{\smallskip}
\hline
\noalign{\smallskip}
F348M      &   26.046     &    0.004    &    26.058     &    0.004  \\
F420N      &   24.365     &    0.006    &    24.349     &    0.005  \\
F450N      &   24.496     &    0.005    &    24.482     &    0.005  \\
F515N      &   26.443     &    0.004    &    26.433     &    0.004  \\
F660N      &   25.905     &    0.004    &    25.907     &    0.004  \\
F665N      &   24.057     &    0.006    &    24.058     &    0.006  \\
F861M      &   26.427     &    0.003    &    26.427     &    0.003  \\
\noalign{\smallskip}
\hline
\end{tabular}
\end{center}
\label{tab:ZPs_DR8_DR12}
\end{table}
\normalsize

\subsection{CALIBRATION USING RefCat2}
\label{calibration_using_RefCat2}

\begin{figure*}
\includegraphics[scale=0.195]{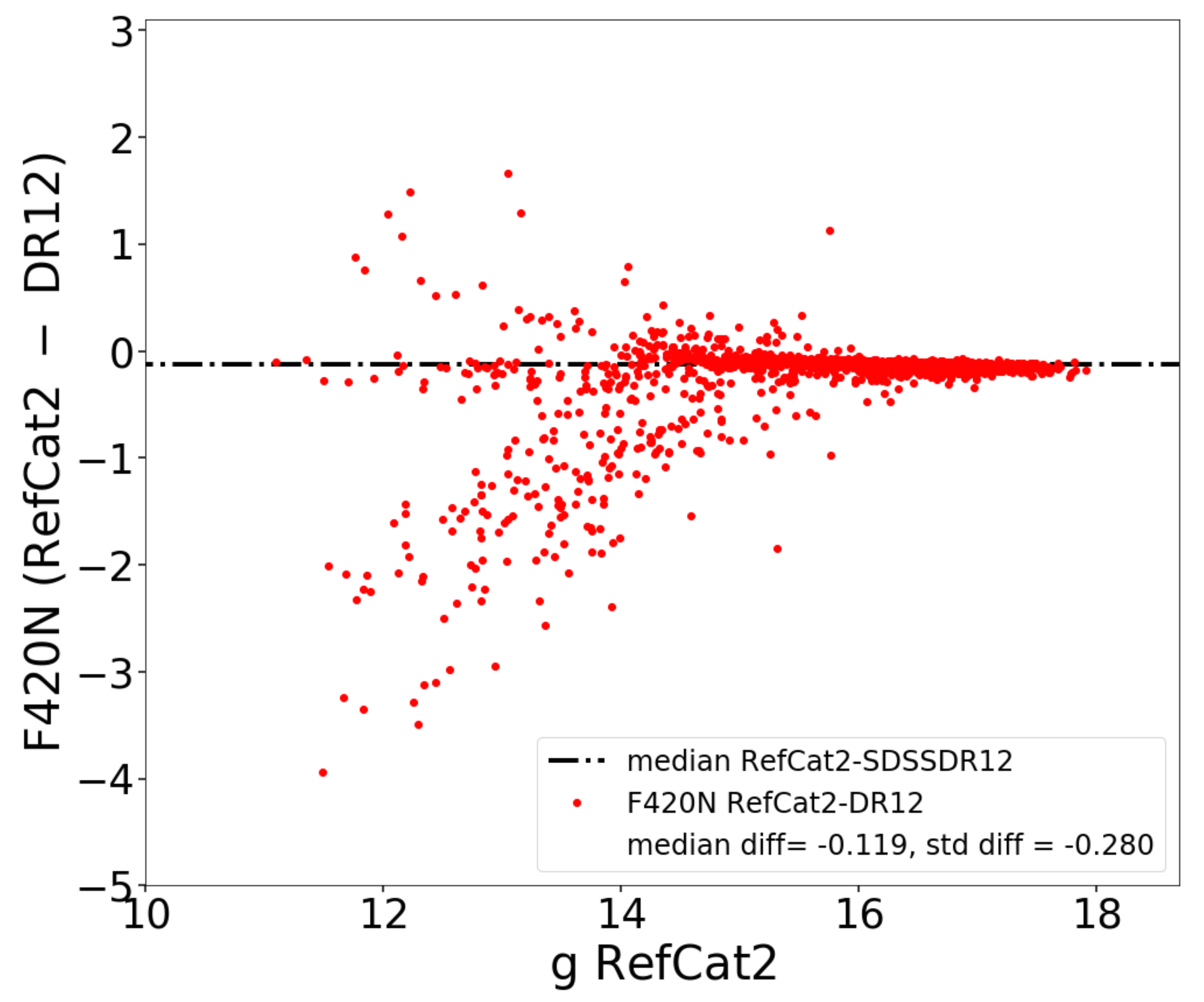}
\includegraphics[scale=0.195]{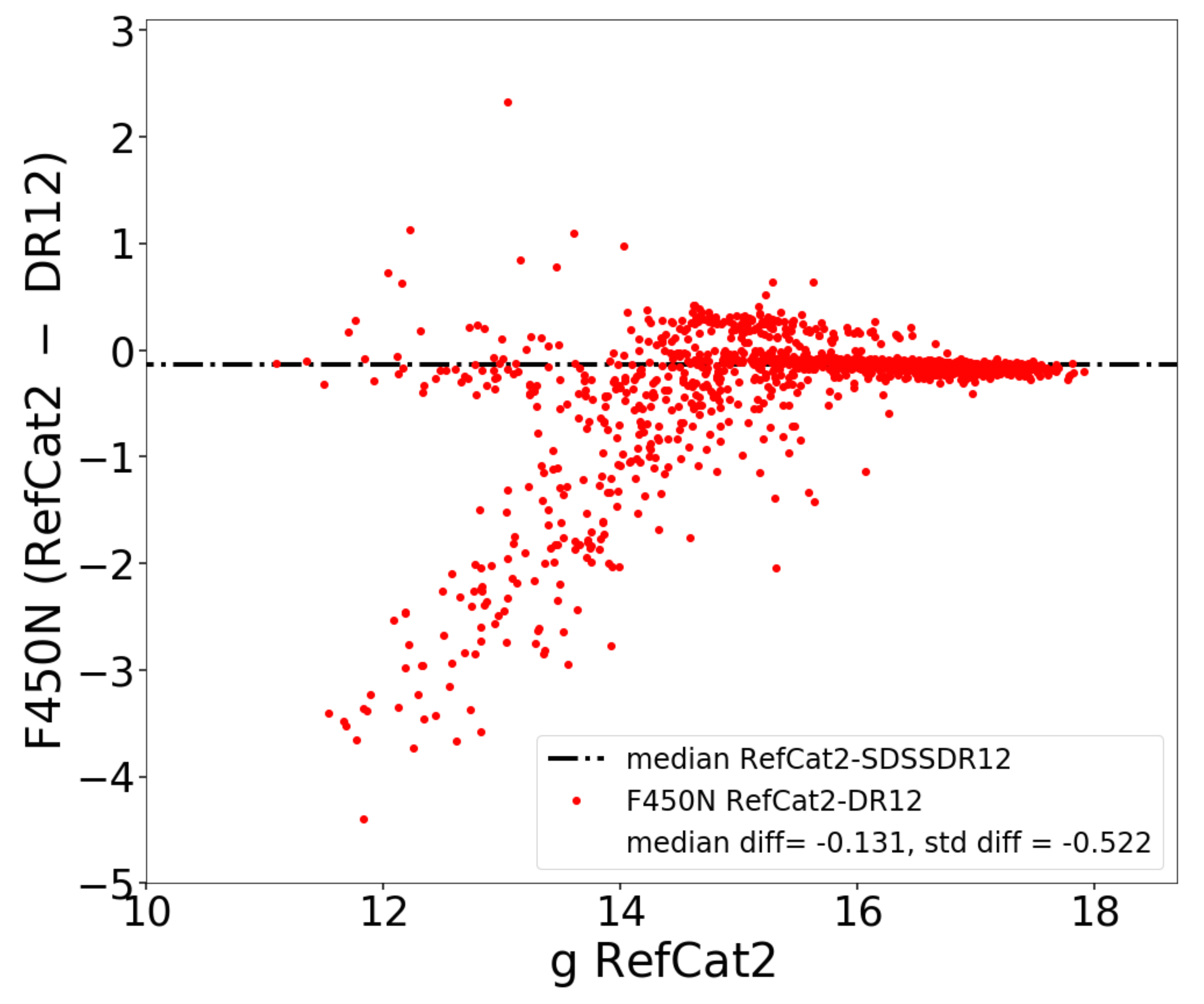}
\includegraphics[scale=0.195]{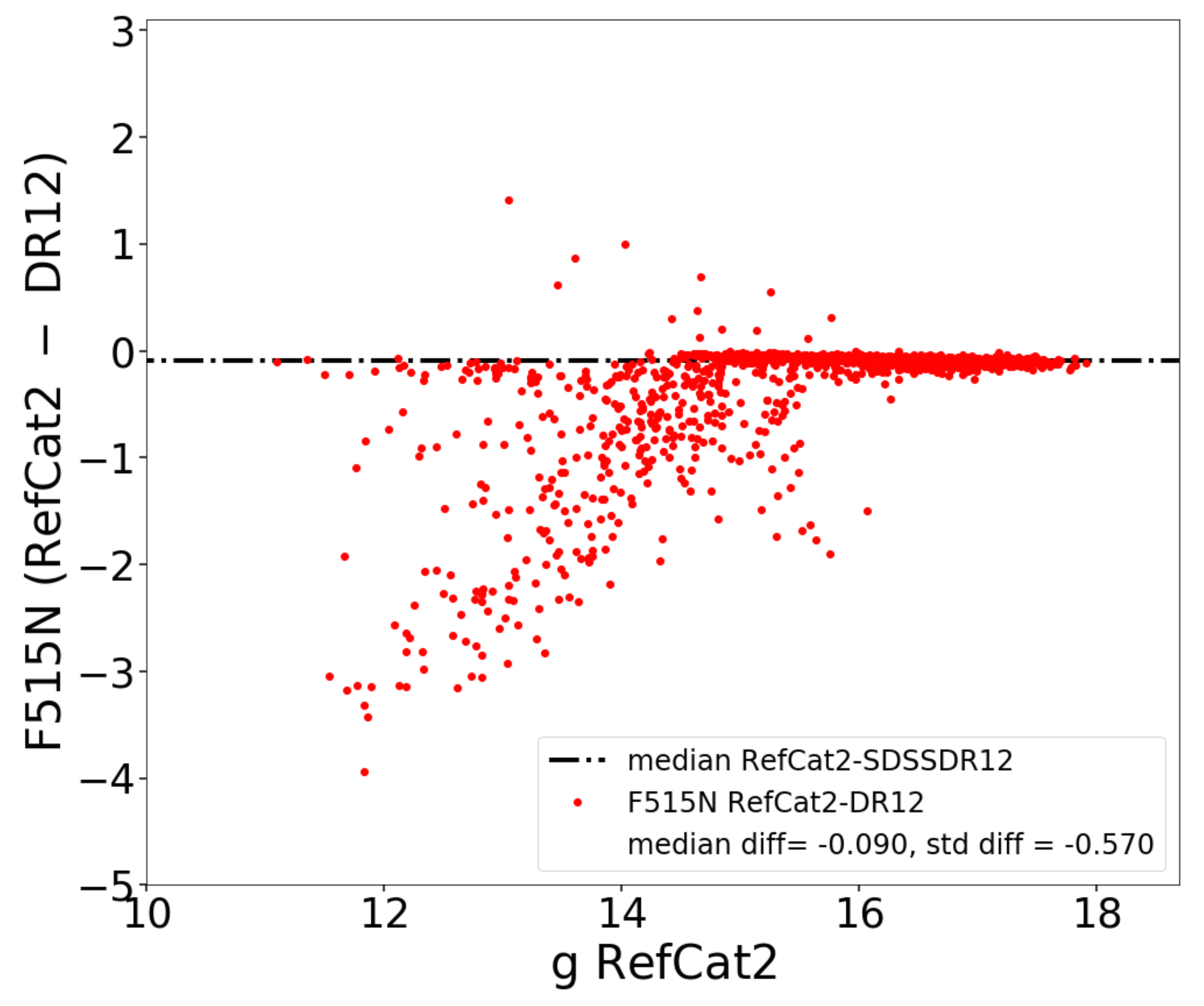}
\includegraphics[scale=0.195]{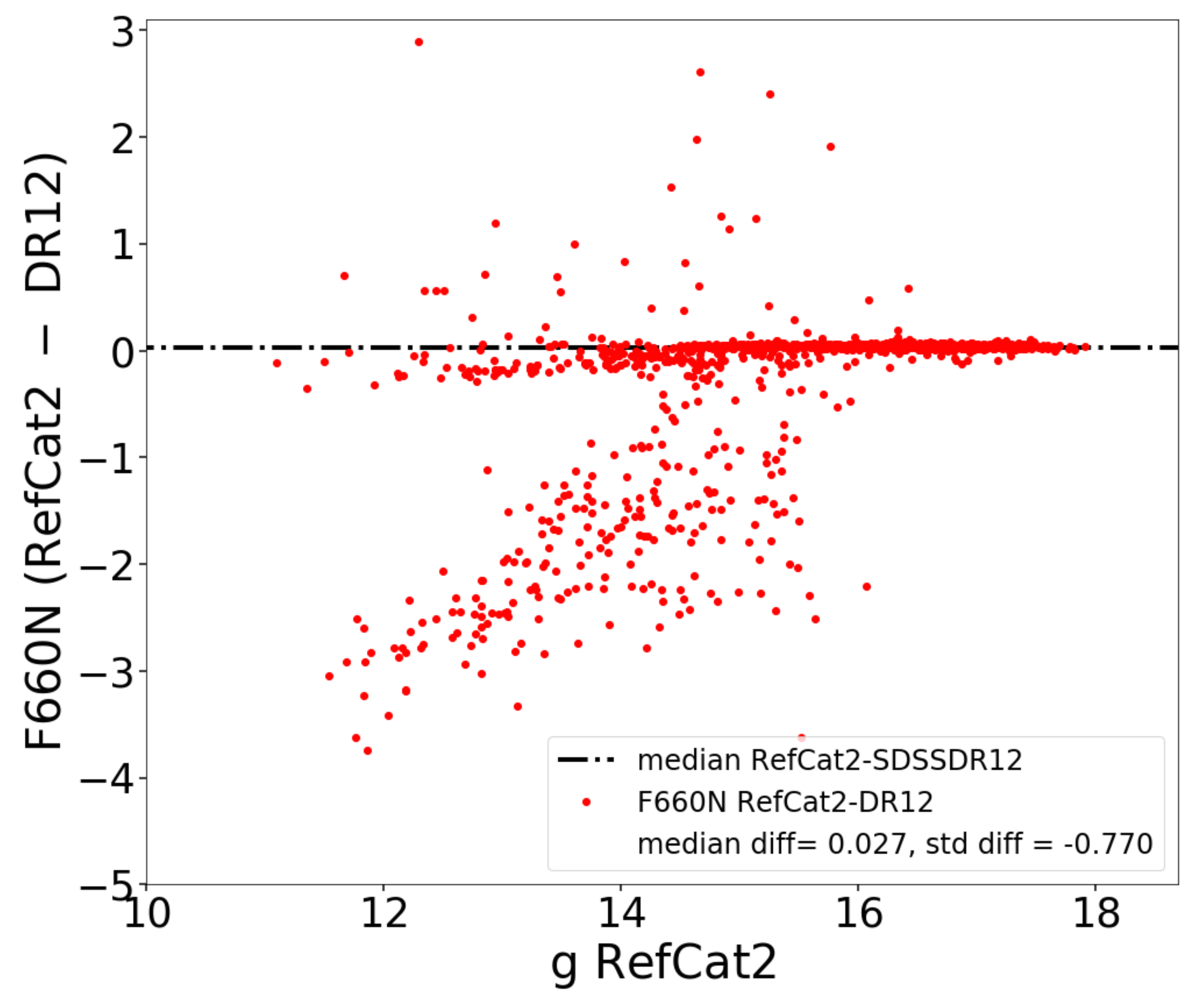}
\includegraphics[scale=0.195]{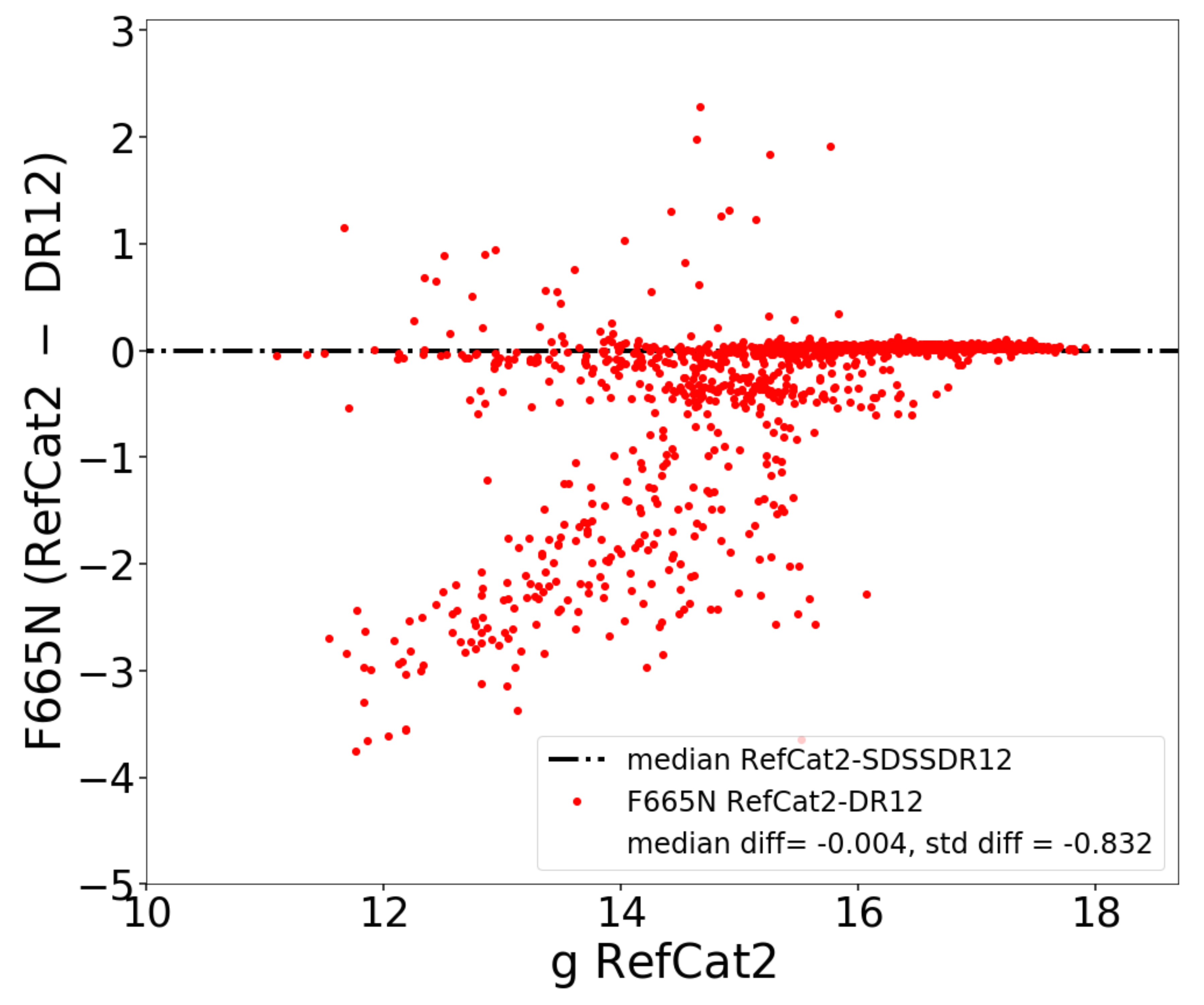}
\includegraphics[scale=0.195]{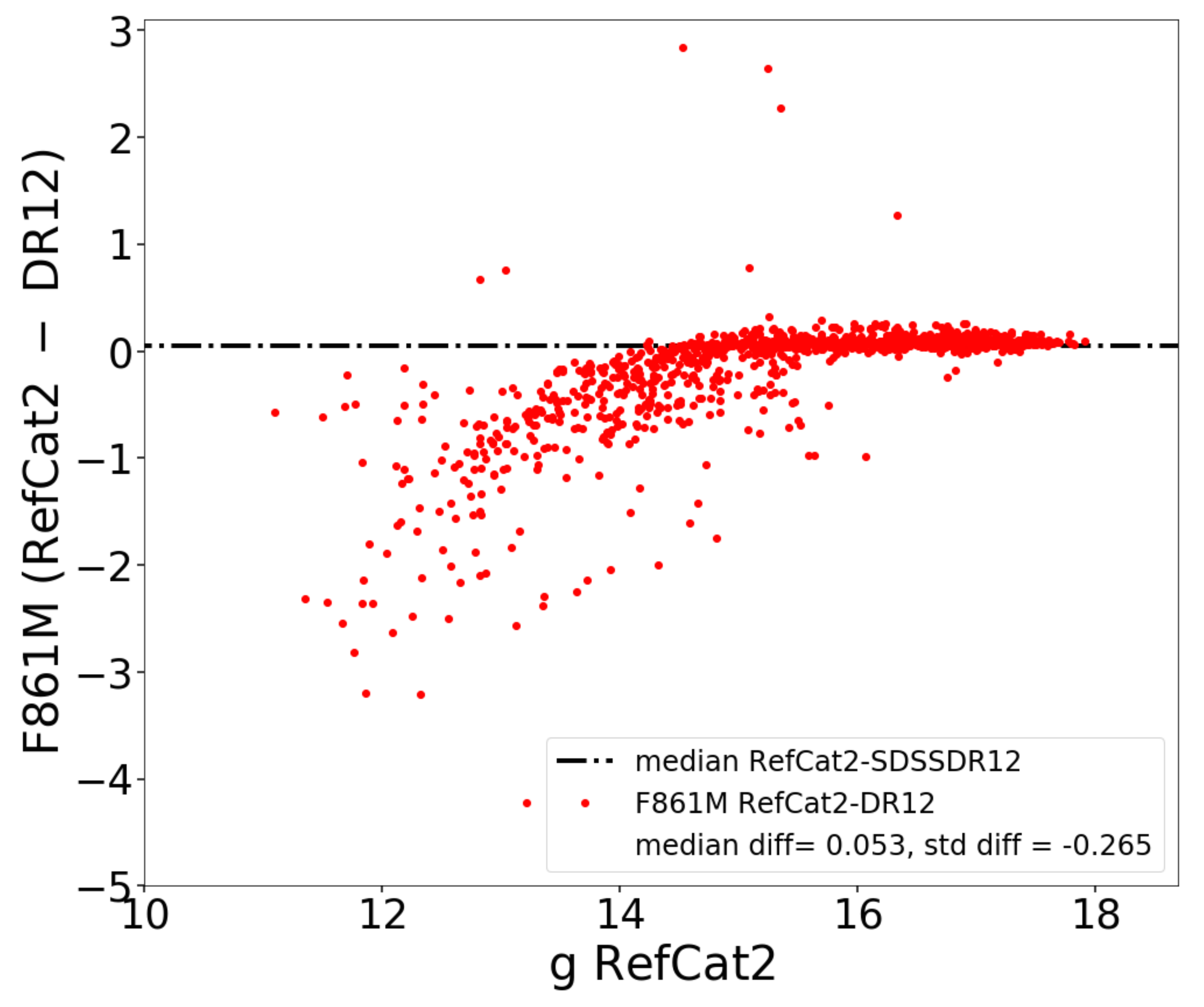}
\centering
\caption{These figures represent differences for stars observed with the Javalambre observatory T-80 telescope in Cyg OB2 and their \textit{RefCat2} and \textit{SDSS} DR12 photometry taken directly from both libraries.}
\label{fig:diff_RefCat2_DR12}
\end{figure*}

While we were writing this work, \citet{Tonry2018} published a new catalogue (\textit{RefCat2}) with \textit{griz} photometry of 993 million stars to m<19. According to the authors, \textit{RefCat2} has an internal precision of 0.02 mag for stars in the Galactic disk and is free of systematic effects. In accordance with these premises and at refferee's suggestion, we decided to obtain the GALANTE ZPs using this library, however, due to the lack of a \textit{u} band in this catalogue, we have to continue using  \textit{u} from \textit{SDSS} DR12. Attending to both libraries, we compare differences between these ZPs in Figure~\ref{fig:diff_RefCat2_DR12}.

\begin{figure*}
\includegraphics[scale=0.19]{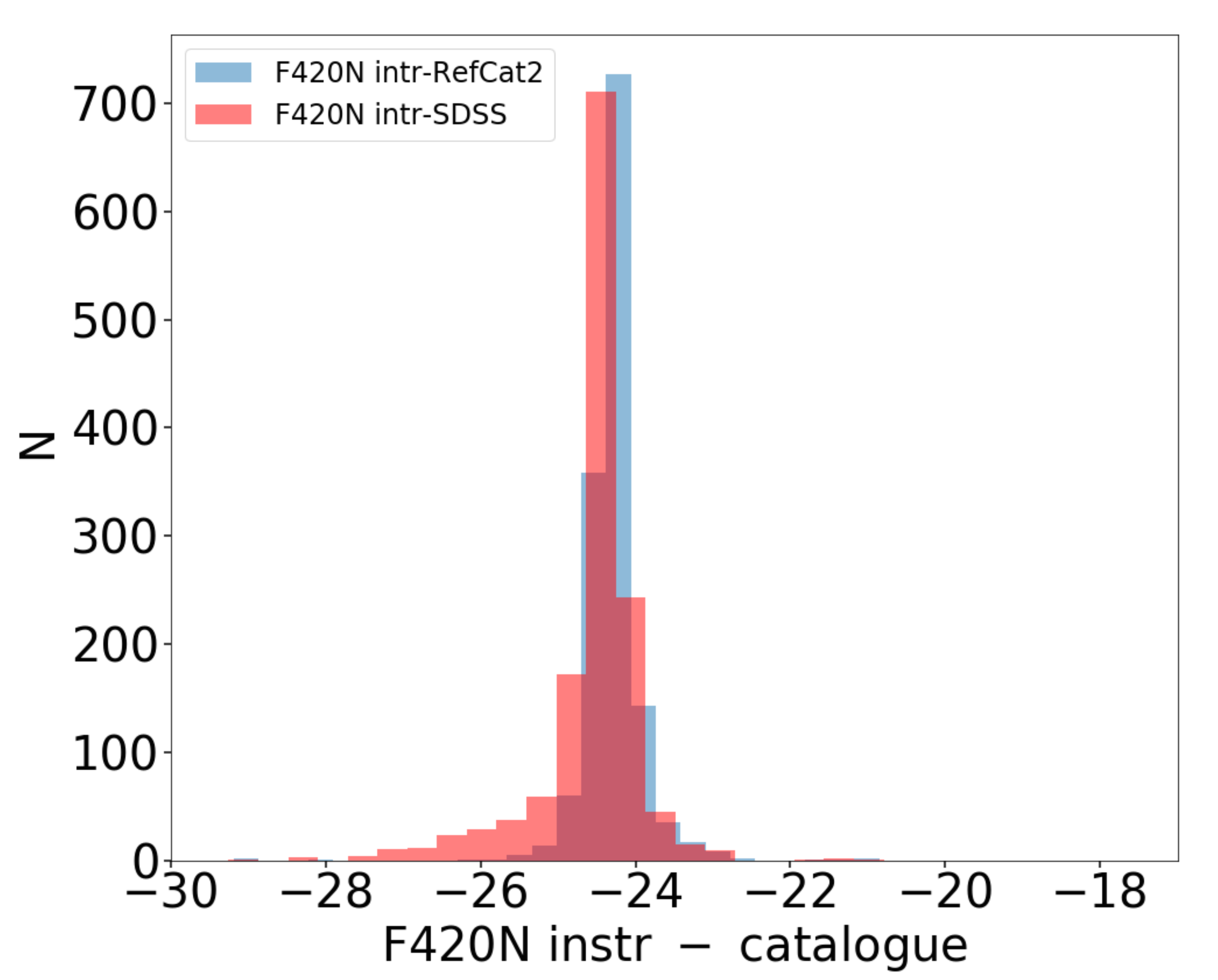}
\includegraphics[scale=0.19]{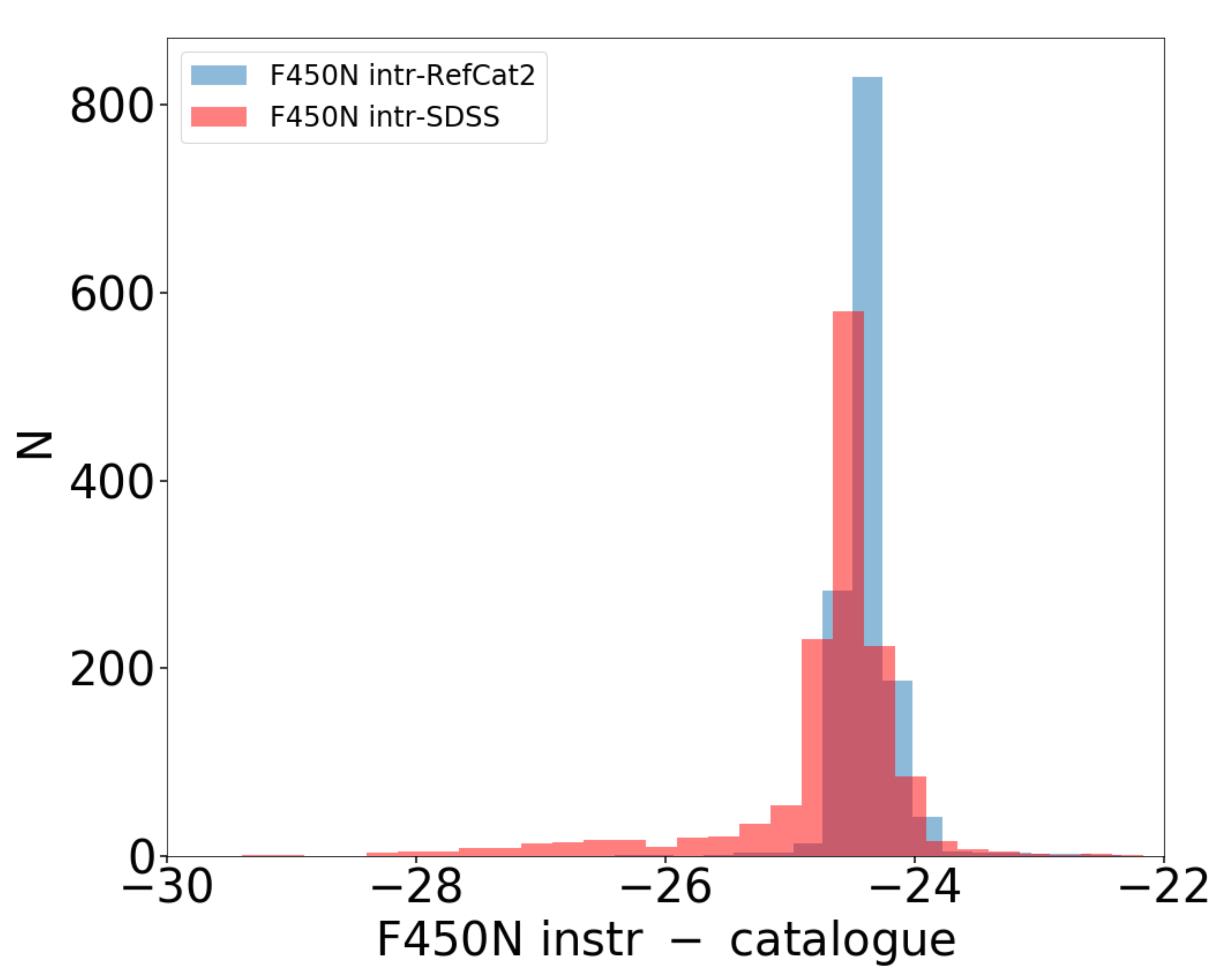}
\includegraphics[scale=0.19]{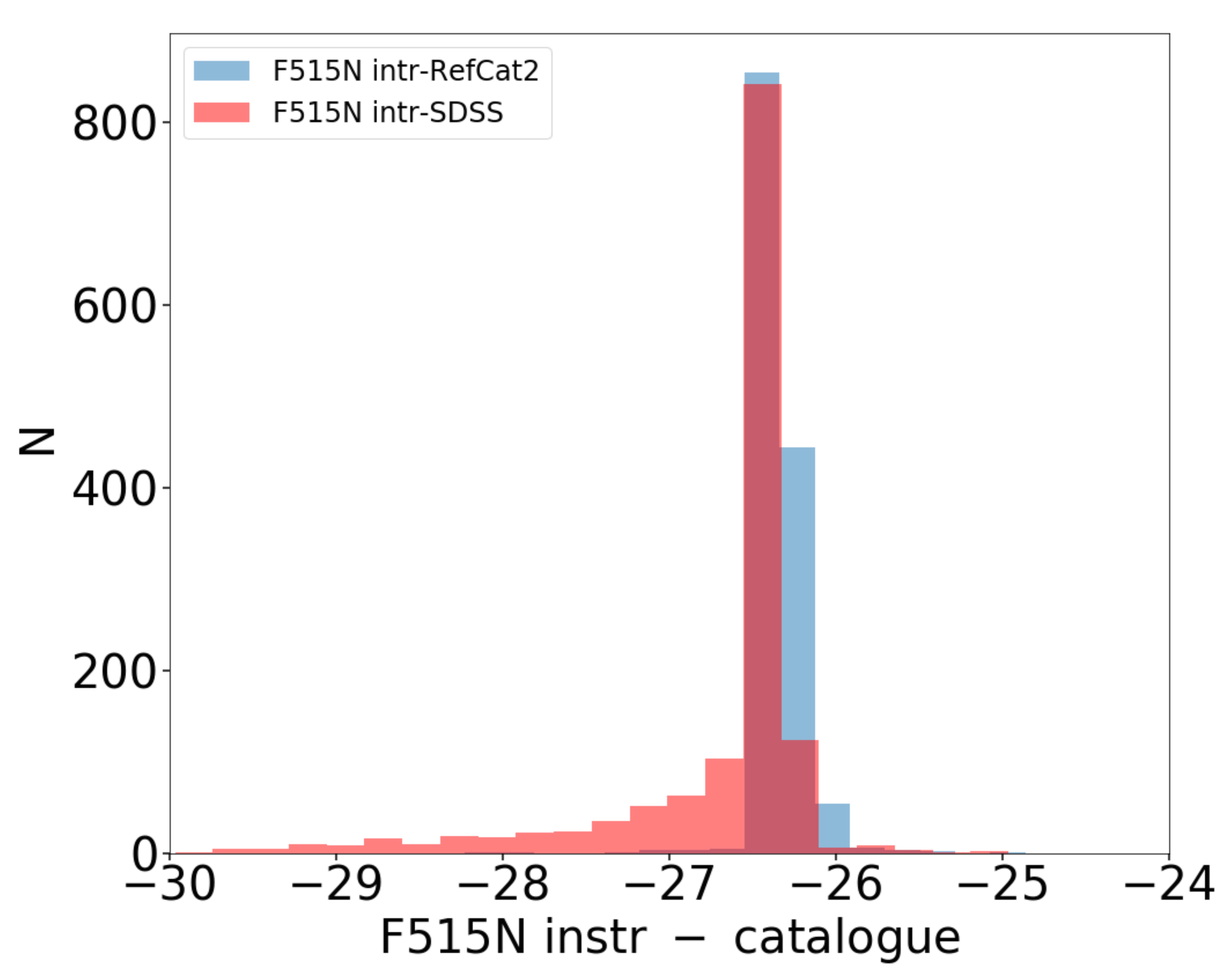}
\includegraphics[scale=0.19]{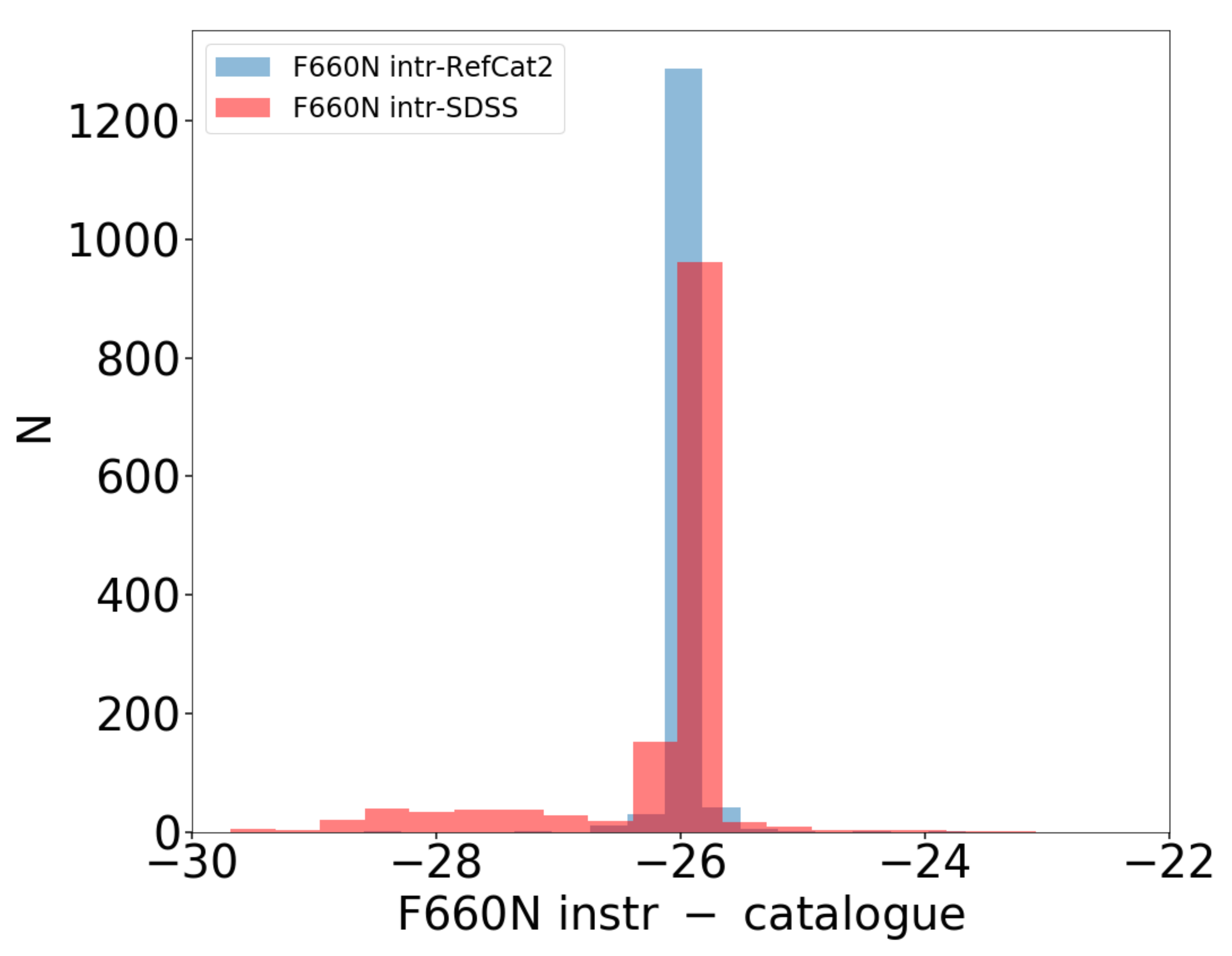}
\includegraphics[scale=0.19]{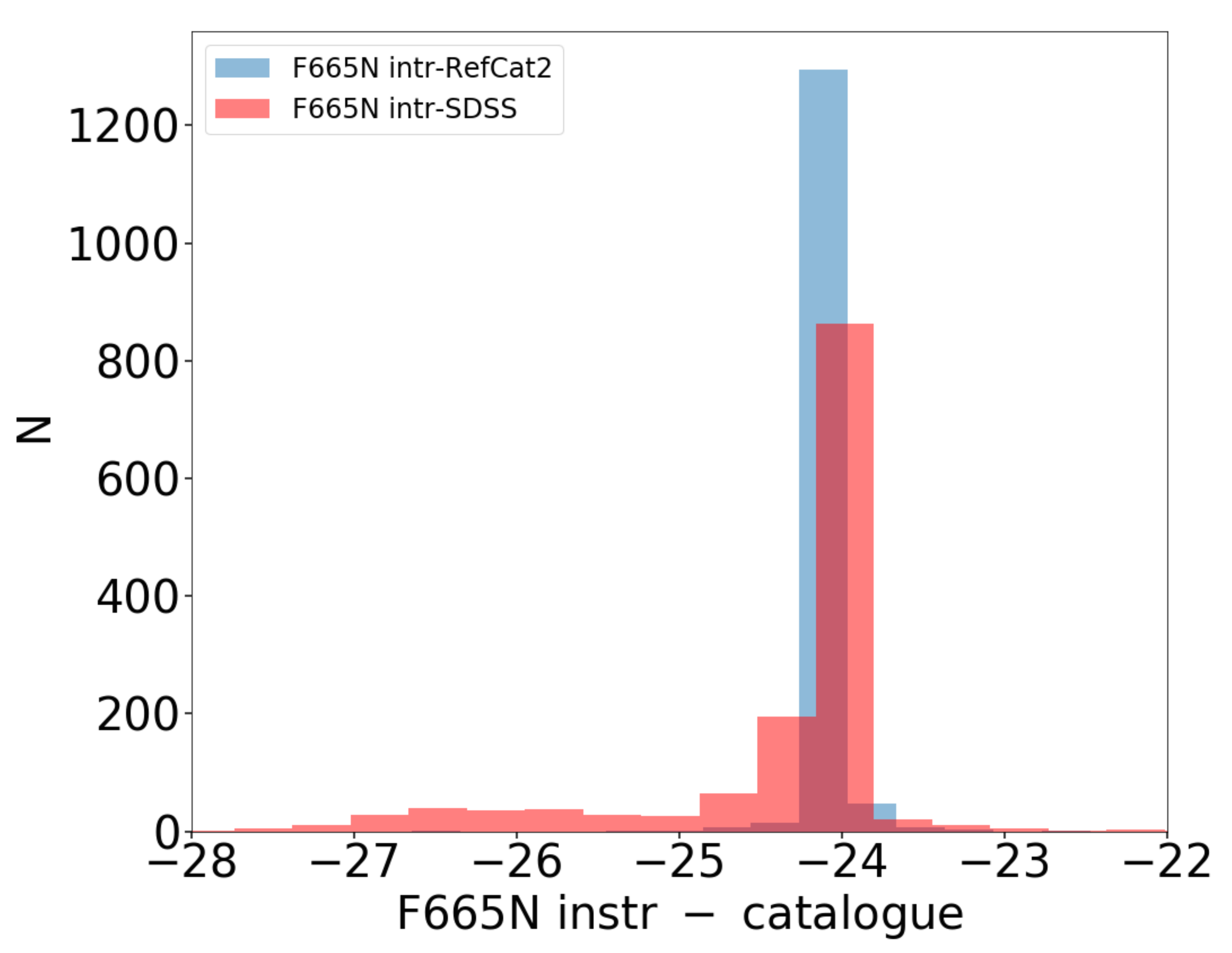}
\includegraphics[scale=0.19]{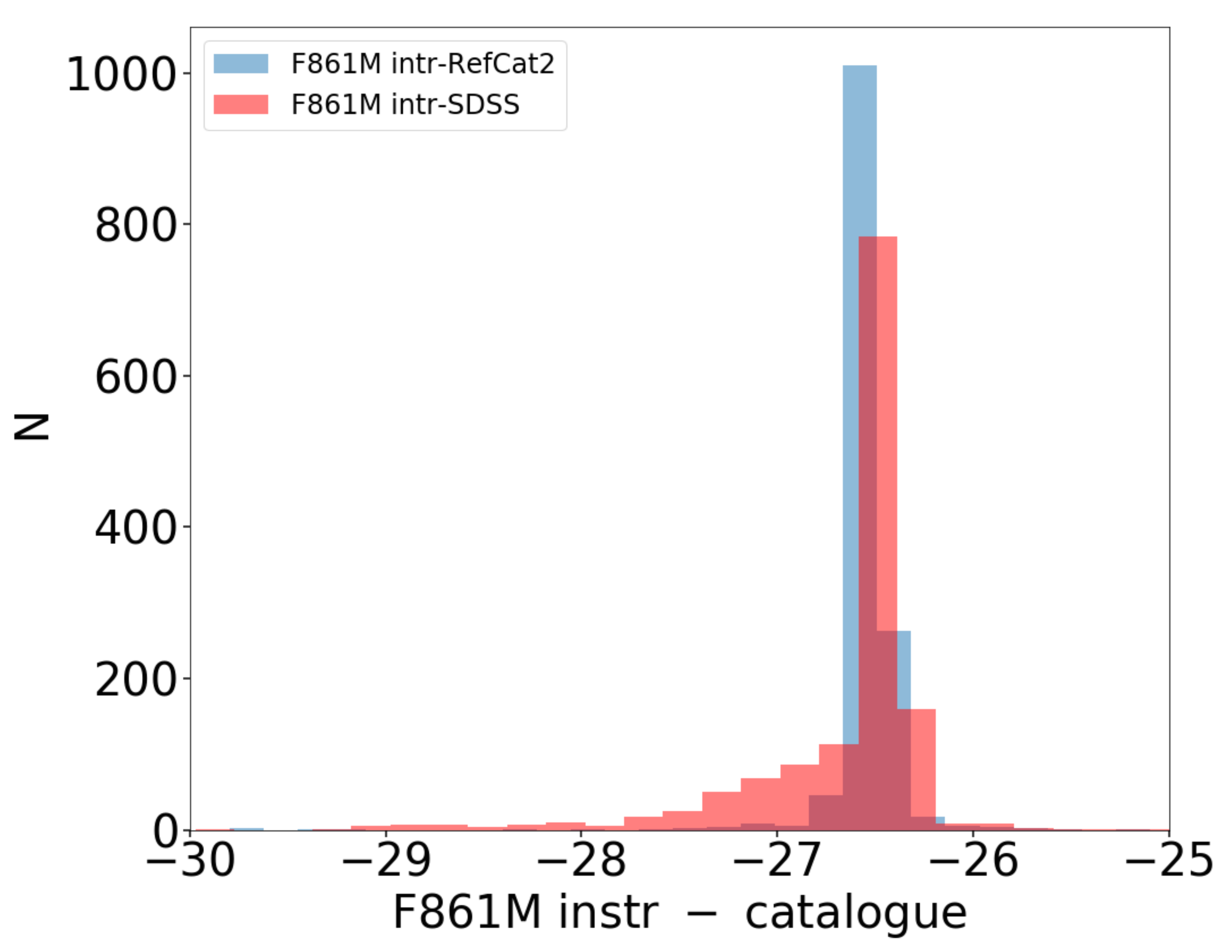}
\centering
\caption{GALANTE photometry histograms for differences between instrumental AB magnitudes minus \textit{RefCat2} (in blue) and \textit{SDSS} DR12 (in red) using a small region of Cyg OB2 observed by the T-80 telescope.}
\label{fig:histog_diff_instrumental_RefCat2_DR12}
\end{figure*}

A clear fact to validate this catalogue is that, if these stars present lower systematic errors than those of \textit{SDSS} DR12, this should be translated into the ZP distribution (instrumental - catalogue) for each filter. Figure~\ref{fig:histog_diff_instrumental_RefCat2_DR12} represents GALANTE photometry histograms for these ZP differences comparing with \textit{RefCat2} (in blue) and \textit{SDSS} DR12 (in red). Two facts are clearly observed: \textit{RefCat2} ZPs are more peaked than \textit{SDSS} based ones, and they are also more symmetrical, without the large skews present in the \textit{SDSS} distribution.

Tails that we observe in the ZPs from \textit{SDSS} DR12, suggest the presence of systematic errors for the brightest stars in \textit{SDSS} DR12, as seen in Figure~\ref{fig:diff_RefCat2_DR12}. After this analysis, and considering that \textit{RefCat2} is an all-sky survey, we consider \textit{RefCat2} our base catalogue to obtain the preliminary calibration of the GALANTE photometric system.

\begin{figure}
\includegraphics[width=0.48\textwidth]{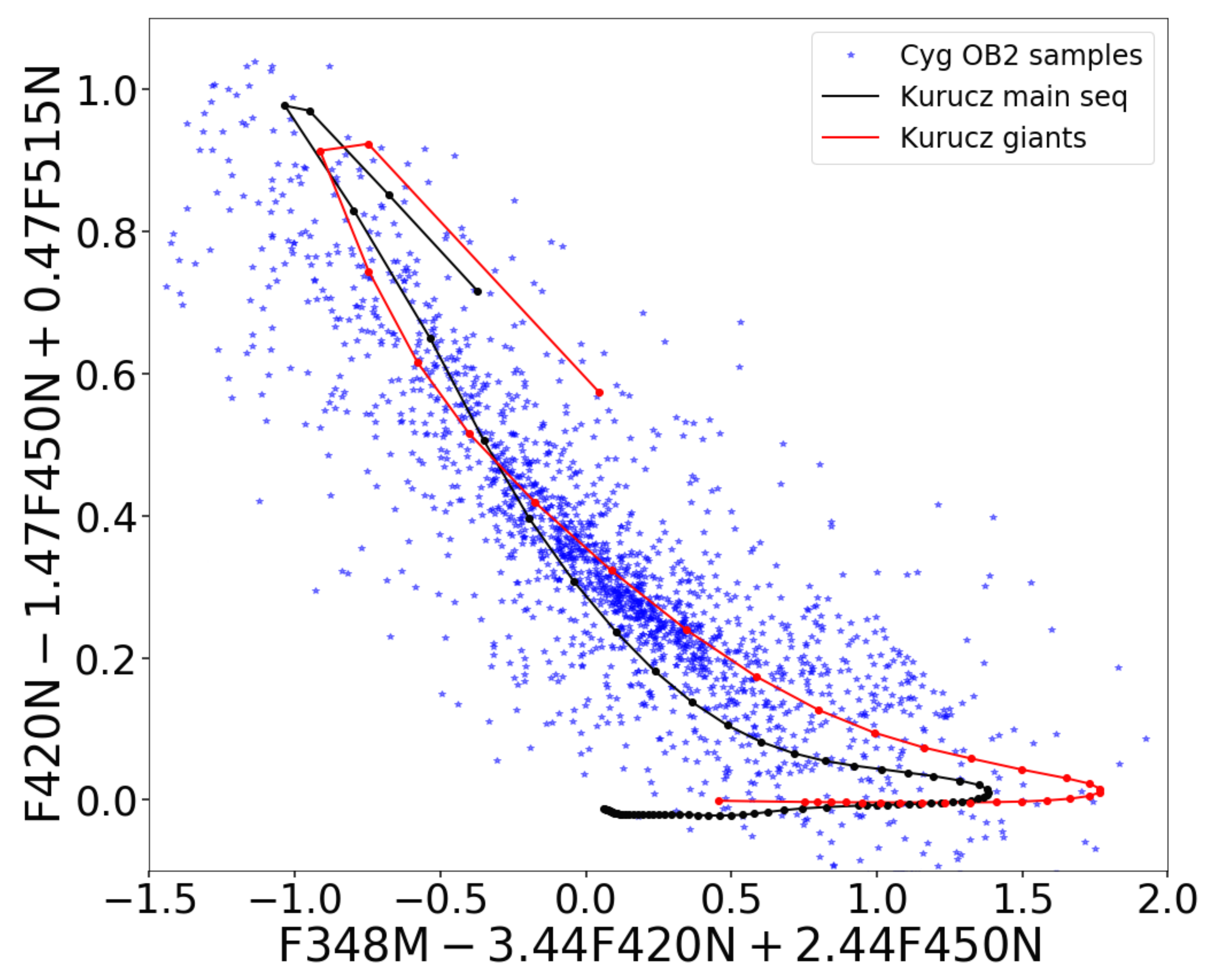}
\centering
\caption{Bracket diagram using GALANTE filter set similar to Figure~\ref{fig:color_color}. It shows only tracks with solar metallicity models with log(g)=4.5 (black line) and log(g)=2.5 (red line). We represent Cyg OB2 calibrated samples for this work by blue star-shapes.}
\label{fig:color_color_CYGOB2}
\end{figure}

As an example of the obtained results using \textit{RefCat2}, we have drawn the bracket diagram of Figure~\ref{fig:color_color} for our Cyg OB2 field. The results are shown in Figure~\ref{fig:color_color_CYGOB2}, where the main and giant sequences by Kurucz as listed by \citet{1997A&A...318..841C} are also overplotted. We selected \textit{Kurucz ODFNEW/NOVER} theoretical spectra to obtain the GALANTE synthetic photometry. Kurucz's library provides a coverage of 3500 K $\leq$ $T_{\rm eff}$ $\leq$ 50\,000 K (in steps of 200 K below $T_{\rm eff}$ = 13\,000 K, and 1000 K otherwise), 0.0 $\leq$ $\log (g)$ $\leq$ 5.0 (in steps of 0.5 dex) and -2.5 $\leq$ [Fe/H] $\leq$ 0.5 (in steps of 0.2 and 0.5). These low-resolution spectra are sampled from 90.9 \r{A} to 160 $\mu$m, which includes the GALANTE wavelength range. Solar metallicity, reddening free lines (main sequence: black, and giant: red) are shown.

We observe that most stars are arranged between the two main lines (main sequence and giant stars) and that the most scattered ones could be dispersed due to high variable reddening and/or ZP error propagation. However, the observed stellar distibution in this plot produces confidence in our ZP choice as a preliminary calibration of the GALANTE photometry. We will compare this calibration with a more sophisticated one in a following paper where we will present the GALANTE photometry of Cyg OB2.

\section{Conclusions}
\label{section_conclusions}
We present the characterization of the GALANTE photometric system as defined by the convolution of four different response functions: atmosphere, filter, detector, and mirror. The set of primary standard stars to
define the GALANTE photometric system is composed of the 378 stars from NGSL library, which cover a wide range of spectral types, luminosity classes and metallicities. We enhance it with the MAW library, which provides
a sample of 122 stars. Both libraries provide 500 GALANTE spectrophotometric standard stars. Using this compilation to obtain the GALANTE and \textit{SDSS} synthetic photometries in the AB system, we derive transformation equations from \textit{SDSS} photometry to the GALANTE photometric system (and vice versa) using the BIC parameter as a figure of merit for selecting the best model.

Figure~\ref{fig:residuo_g_r} shows that a simple \textit{SDSS} model (band+colour; Equation~\ref{Coefficients_SDSS_GALANTE}) is able to shape the GALANTE system up to an rms level of 0.06 magnitudes in the worst case. It is worth noting that our compiled final library covers a wide range of effective temperature, log(g), metallicity, and, most importantly, reddening. This means that our transformations, despite the noise, can be considered as unbiased. This point is better observed in Figure~\ref{fig:residuo_Teff}, where only NGSL stars are plotted and their dependence on temperature and reddening show a rippled way but limited to an amplitude lower than 0.05 magnitude.

We use Equation~\ref{Coefficients_SDSS_GALANTE} to obtain the GALANTE photometric ZPs in a field of Cyg OB2 observed using the Javalambre observatory. The ZPs have been obtained from \textit{SDSS} DR8 and DR12 releases. Despite the \textit{u} and \textit{g} magnitude equations between both data releases, the ZP difference is always below 0.01 with an rms of the order of 0.02. During our work, a new \textit{griz} catalogue called \textit{RefCat2} from \citet{Tonry2018} was published. However, the observed stellar distribution of this new catalogue, covering the whole sky, represents an excellent choice for the preliminary calibration of the GALANTE photometry, if, as its authors proclaim, the average internal precision is 0.02 magnitudes for the stars of the Galactic disk and is free of systematic effects. We have made a comparison between the ZPs obtained from \textit{SDSS} DR12 and \textit{RefCat2}, finding that the distribution of the ZPs derived from \textit{RefCat2} shows a better behavior, with more centralized values and shorter tails (see Figures~\ref{fig:diff_RefCat2_DR12} and~\ref{fig:histog_diff_instrumental_RefCat2_DR12}). For this reason we adopted \textit{RefCat2} as the base catalogue for GALANTE calibration. The application of the \textit{RefCat2} ZPs to our photometry is shown in Figure~\ref{fig:color_color_CYGOB2}, where the Kurucz's fiducial lines corresponding to main sequence and giant stars of solar metallicity are also overplotted. The agreement between the observed photometry and the theoretical tracks means that we are confident in this ZP calibration. In our next paper, we will compare this calibration with another more complex procedure for a final analysis of the GALANTE photometry of the Cyg OB2 association.

\section*{Acknowledgements}

We would like to thank the referee, Prof. Bessell, for his invaluable comments and suggestions, helping us
to improve the presentation of our results. We thank the Centro de Estudios de F\'isica del Cosmos de Arag\'on (CEFCA) team in Teruel and Javalambre for supporting us in this project, giving us the opportunity to use non-J-PLUS useful nights to develop the GALANTE survey. We also want to recognize the work of the NGSL team \citep{2006hstc.conf..209G,2016ASPC..503..211H} whose observational catalogue has been of great help for our work.
This research made use of Python ({\tt \href{http://www.python.org}{http://www.python.org}}); Numpy \citep{2011CSE....13b..22V}; Scipy \citep{Jones:2001aa}; and Matplotlib \citep{Hunter:2007}, a suite of open-source python modules that provides a framework for creating scientific plots.
We also acknowledge the use of STILTS and TOPCAT tools \citet{2005ASPC..347...29T}.
A.L.-G, E.J.A, and J.M.A. acknowledge support from the Spanish Government Ministerio de Ciencia, Innovaci\'on y Universidades though grants
AYA2013-40\,611-P and AYA2016-75\,931-C2-1/2-P. A.L.-G and E.J.A also acknowledge support from the State Agency for Research of the Spanish MCIU through the "Center of Excellence Severo Ochoa" award for the Instituto de Astrof\'isica de Andaluc\'ia (SEV-2017-0709).




\bibliographystyle{mnras}
\bsp	
\label{lastpage}

\bibliography{/Users/antoniolorenzogutierrez/Desktop/MNRAS_paper_GALANTE/actualizado_04_02_2019_aplicadas_correcciones_propuestas_Jesus_correcciones_Emilio_Jesusdenuevo_FINAL_pdf_Bib_JAMES_MAIZ_enviadorevista_CORRECCIONES_Reviewer_SUBIDOFINAL_arXiv/bibliography.bib}
\end{document}